\documentclass{article}

\usepackage{arxiv}

\usepackage[utf8]{inputenc} 
\usepackage[T1]{fontenc}    
\usepackage{hyperref}       
\usepackage{url}            
\usepackage{booktabs}       
\usepackage{amsfonts}       
\usepackage{nicefrac}       
\usepackage{microtype}      
\usepackage{lipsum}
\usepackage{graphicx}
\graphicspath{ {./images/} }

\usepackage{float}
\usepackage{xcolor}
\usepackage{color,soul}
\usepackage{subfigure}
\usepackage{dblfloatfix}
\usepackage{booktabs}
\usepackage[numbers]{natbib}
\usepackage[utf8]{inputenc}
\usepackage[english]{babel}

\title{Predicting Short-term Mobile Internet Traffic from Internet Activity using Recurrent Neural Networks}

\author{
  Guto Leoni Santos \\
  Centro de Informática\\
  Universidade Federal de Pernambuco (UFPE)\\
  Recife, Brazil \\
  \texttt{gls4@cin.ufpe.br} \\
  \And
  Pierangelo Rosati \\
  Business School\\
  Dublin City University (DCU)\\
  Dublin, Ireland \\
  \texttt{pierangelo.rosati@dcu.ie} \\
  \And
  Theo Lynn \\
  Business School\\
  Dublin City University (DCU)\\
  Dublin, Ireland \\
  \texttt{theo.lynn@dcu.ie} \\
  \And
  Judith Kelner \\
  Centro de Informática\\
  Universidade Federal de Pernambuco (UFPE)\\
  Recife, Brazil \\
  \texttt{jk@gprt.ufpe.br} \\
  \And
  Djamel Sadok \\
  Centro de Informática\\
  Universidade Federal de Pernambuco (UFPE)\\
  Recife, Brazil \\
  \texttt{jamel@gprt.ufpe.br} \\ 
  \And
  Patricia Takako Endo \\
  Universidade de Pernambuco (UPE)\\
  Caruaru, Brazil \\
  \texttt{patricia.endo@upe.br} \\ 
}

\begin{document}
\maketitle
\begin{abstract}
Mobile network traffic prediction is an important input in to network capacity planning and optimization. Existing approaches may lack the speed and computational complexity to account for bursting, non-linear patterns or other important correlations in time series mobile network data. We compare the performance of two deep learning architectures - Long Short-Term Memory (LSTM) and Gated Recurrent Unit (GRU) - for predicting mobile Internet traffic using two months of Telecom Italia data for the metropolitan area of Milan. K-Means clustering was used \emph{a priori} to group cells based on Internet activity and the Grid Search method was used to identify the best configurations for each model. The predictive quality of the models was evaluated using root mean squared error. Both Deep Learning algorithms were effective in modeling Internet activity and seasonality, both within days and across two months. We find variations in performance across clusters within the city. Overall, the LSTM outperformed the GRU in our experiments.
\end{abstract}


\section{Introduction}
\label{sec:introduction}
At the beginning of 2020, it was estimated that 67\% of the global population (5.2bn people) subscribed to mobile services \cite{gsma20}. According to industry forecasts, we are about to enter in to a period of unprecedented mobile data growth driven by the Internet of Things (IoT). By 2023, machine-to-machine (M2M) connections that support a broad range of IoT applications will represent about 50\% (14.7 billion) of total global devices and connections; and 45\% of all networked devices will be mobile-connected \cite{cisco20}. The combination of this influx of new mobile-connected devices, faster broadband speeds, greater video consumption, and the capabilities of 5G networks is dramatically increasing mobile data traffic \cite{cisco20}. The nature and scale of this growth poses significant challenges to mobile network providers including the management of complexity, scalability, Quality of Service (QoS), Quality of Experience (QoE), and privacy, all against the backdrop of constrained budgets and intense competition \cite{aguzzi2013definition}.

This massive surge in demand for mobile broadband requires solutions that satisfy QoS and QoE requirements with minimum service delay and within budget constraints. Capacity planning is the process of adjusting the capacity across the network in response to changing or predicted demands \cite{tikunov2007traffic}. Mobile network traffic modeling and prediction is an important input in to multiple capacity planning tasks including network design, performance evaluation, control, and network optimization \cite{ma2020survey,narejo2018an}. Network traffic has been characterized as a time series with non-linear and chaotic characteristics and is correlated over both long and short time frames \cite{narejo2018an}. There is a well-established literature that focuses on trend, seasonality and anomaly prediction at the network-level and the cell-level to guide mobile network investments and optimization \cite{ma2020survey,bui2017survey,li2017learning}. Much of the extant research focuses on traffic prediction across that while presenting good forecast results, may have unacceptable training times, turnaround times, lack computational complexity and may not therefore account for characteristics such as bursting, non-linear patterns or other important correlations \cite{narejo2018an,tran2019cellular}. 

To address this shortcoming not only requires a shift from coarse to fine prediction but the adoption of novel techniques that can accommodate high dimensionality and provide a satisfactory solution quickly. In the last decade, we have seen a number of applications that can benefit from short-term or fine mobile network traffic prediction including opportunistic scheduling \cite{huang2017study}, multimedia optimization \cite{samulevicius2015most}, and energy efficiency \cite{pollakis2016anticipatory}. More recently, a combination of deep learning (DL) techniques and new data sources have emerged that show promising results in mobile traffic prediction  \cite{narejo2018an,hua2018traffic,qiu2018spatio}.

This paper compares the performance of two recurrent neural networks (RNNs), Long Short-Term Memory (LSTM) and Gated Recurrent Unit (GRU), for predicting mobile Internet traffic. Using two months (62 days) of Telecom Italia spatio-temporal log data for the metropolitan area of Milan, we constructed twelve clusters each with its own time series. We identify the best configuration of each model using the Grid Search method \cite{syarif2016svm} and use root mean squared error (RMSE) to evaluate the performance of the models. The models satisfactorily learned the pattern of Internet activity and seasonality, both within days and across the two months. We find variations in performances based on geographic clusters within the city. Overall the LSTM performed better than the GRU in most cases. We also compare the mean absolute error (MAE) results for our models with extant research using similar data \cite{chen2018deep} and find both of our models significantly outperform previous works.  

The rest of this paper is structured as follows. In Section \ref{sec:background}, we provide an overview of RNNs and clustering algorithms. The Section \ref{sec:related-works} presents some related works. Section \ref{sec:methodology} details the mobile Internet traffic data set and data pre-processing, the configuration of the LSTM and GRU models, and the metrics used to evaluate model performance. Section \ref{sec:results} presents and discusses our results and analysis. We conclude this article with a brief summary of the main contributions of the article and propose future directions for research in Section \ref{sec:conclusion}.


\section{Background}
\label{sec:background}
\subsection{Deep Learning Neural Networks}
\label{sec:deep-learning-neural-networks}
In the last five years, deep neural networks, ‘deep learning’, have increased in prominence in research and practice. DL is a sub-branch of machine learning (ML) that, at a high level, ``enables an algorithm to make predictions, classifications or decisions based on data, without being explicitly programmed'' \cite{zhang2019deep}. DL addresses the limitations of single-layer neural networks by using multiple layers to transform their input into higher-dimensional representations and then into the output. The emergence of DL is largely driven by increased computational power through heterogeneous processors such as GPUs, the availability of large data sets for training, and advances in optimization algorithms \cite{kraus2020deep}.

In contrast to traditional ML and neural networks, DL can cater for high dimensionality in data thus enabling DL networks to model highly complex non-linear relationships between variables \cite{kraus2020deep}. As such, it is particularly suitable to the mobile and wireless networking domain which is characterized by massive volumes of high velocity unlabeled heterogeneous data \cite{zhang2019deep}. In addition, DL can significantly reduce operational and capital expenditure by reducing or eliminate the time and effort required by valuable and scarce human resources in feature extraction, and reduce computational and memory requirements through multi-task learning \cite{zhang2019deep}. However, DL is not without its limitations. It hides its internal logic to the user thereby sacrificing accuracy for interpretability with practical and ethical consequences \cite{guidotti2019a}. Other limitations include vulnerabilities to adversarial and privacy attacks \cite{Ansari2020}, computational demands unsuitable to small-form computing in edge networks, and the time taken to find optimal configurations, particularly for highly-parameterised data and multi-step network prediction \cite{ma2020survey,zhang2019deep}.

Common DL architectures include Multi-layer Perceptron (MLP), Restricted Boltzmann Machines (RBM), auto-encoded (AE), convolutional neural networks (CNNs), and recurrent neural networks (RNNs) \cite{zhang2019deep}. These can be differentiated by the data structures that they target, and their respective tuning parameters \cite{kraus2020deep}. For example, MLP targets feature vectors of fixed length and are tuned by the activation function setting and the number of layers and units \cite{kraus2020deep}. In contrast, CNNs target high-dimensional data with local dependencies and are tuned by the number and width of convolutional kernels or filters \cite{kraus2020deep}. Because both MLP and CNN assume that all inputs are independent of each other, they are not suited to modeling sequential data, where sequential correlations exist between samples. RNNs specifically target sequential data, like time series data flows from mobile networks.  As such, we focus on the use of RNNs in this paper.
 
\subsection{Recurrent Neural Networks}
\label{sec:recurrent-neural-networks}

\subsubsection{LSTM}
\label{subsubsec:bkg-lstm}
\begin{figure}[h!]
\centering
\includegraphics[width=0.8\columnwidth]{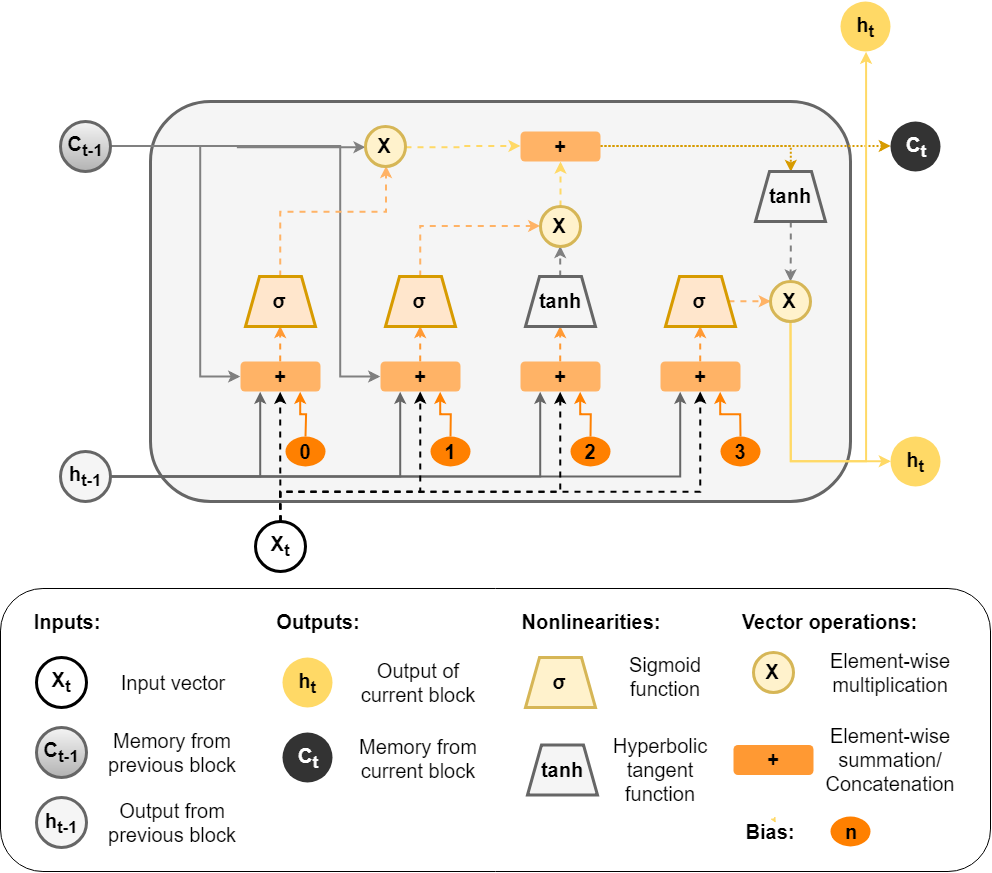}
\caption{Example of a LSTM block (adapted from \cite{lstmExample2016block})}
\label{fig:lstm-block-example}
\end{figure}

As discussed earlier, traditional ML, MLPs and CNNs, typically target input vectors with fixed dimensions. The formalization for sequential data is fundamentally different and thus MLP and CNN are not suitable for time series data. RNN architectures were specifically designed to model sequential data by producing output via recurrent connections (cells) between hidden units \cite{zhang2019deep}. These recurrent connections are the \emph{memory} that stores the previous data allowing RNNs to learn the temporal dynamicity of the sequential data \cite{ordonez2016deep}. In RNNs, the output of the current timestamp is influenced by the output of the previous timestamps, which is critical when the sequence of events or data is important to determine the outcome of a problem. 
Despite being designed to model sequential data, early RNNs suffer from long time dependencies resulting from both vanishing and exploding gradient problems \cite{hochreiter1998vanishing} that negatively impacted training using the Back-Propagation Through Time (BPTT) algorithm. To overcome this limitation, a variation of traditional RNNs was proposed, Long Short-Term Memory (LSTM), which introduces the concept of gates to mitigate gradient problems \cite{ordonez2016deep,jozefowicz2015empirical, greff2017lstm}. Figure \ref{fig:lstm-block-example} presents the basic schema of an LSTM unit. 

The LSTM unit state updates through specific gate operations: write (input gate), read (output gate), or reset (forget gate). These operations consist of component-wise multiplications and apply different functions in the input data as shown in the following equations \cite{ordonez2016deep}:
\begin{equation}
    i_{t} = \sigma_{i}(W_{xi}x_{t}+W_{hi}h_{t-1}+W_{ci}c_{t-1}+b_{i}) 
\end{equation}

\begin{equation}
    f_{t} = \sigma_{f}(W_{xf}x_{t}+W_{hf}h_{t-1}+W_{cf}c_{t-1}+b_{f})
\end{equation}

\begin{equation}
    c_{t}=f_{t}c_{t-1}+i_{t}\sigma_{c}(W_{xc}x_{t}+W_{hc}h_{t-1}+b_{c})
\end{equation}

\begin{equation}
    o_{t}=\sigma_{o}(W_{xo}x_{t}+W_{ho}h_{t-1}+W_{co}c_{t}+b_{o})
\end{equation}

\begin{equation}
    h_{t}=o_{t}\sigma_{h}(c_{t})
\end{equation}

\noindent where $i$, $o$, $f$, and $c$ are, respectively, the input gate output, the output gate output, the forget gate output, and the unit activation vector. $h_{t}$ is the hidden value of the unit (i.e. the memory state) and has the same size of the previous vectors. The non-linear functions of the input, forget, and output gates are represented by $\sigma_{i}$, $\sigma_{f}$, and $\sigma_{o}$, respectively. The weight matrices of the unit state (\textit{c}), input (\textit{i}), output (\textit{o}), and forgot (\textit{f}) gates are represented by $W_{xi}$, $W_{hi}$, $W_{ci}$, $W_{xf}$, $W_{hf}$, $W_{cf}$, $W_{xc}$, $W_{hc}$, $W_{xo}$, $W_{ho}$, and $W_{co}$, where $x$ is the input and $h$ the hidden value of LSTM unit. Finally, $b_{i}$, $b_{f}$, $b_{c}$, and $b_{o}$ are, respectively, the bias of input gate, forget gate, cell, and output gate \cite{ordonez2016deep}.
While LSTM addresses gradient problems, critics have noted that the LSTM architecture is \emph{ad hoc}, has a substantial number of components whose purpose is not immediately apparent, and that it is characterized by long training times \cite{jozefowicz2015empirical,lin2016abnormal}. 

Given its ability to model time series data and predictive capacity, we use LSTM in this study. 

\subsubsection{GRU}
\label{subsubsec:bkg-gru}
GRU is a variation of LSTM which only uses two gates, an update gate and a reset gate. Indeed the update gate in a GRU replaces the input and forget gates used in LSTM and decides what input data will be kept \cite{chung2014empirical}. Furthermore and unlike LSTM, GRU exposes its memory content at each step balancing between the previous and new memory content \cite{chung2015gated}. 

The GRU activation, $h_{t}^{j}$, is represented in Equation \ref{eq:gru-activation}. Considering the input data as a time series, at the timestamp $t$, $h_{t}^{j}$ is the linear interpolation between the previous unit data ($h_{t-1}^{j}$) and the current data ($\tilde{h}_{t}^{j}$).

\begin{equation}
    \label{eq:gru-activation}
    h_{t}^{j}=(1-z_{t}^{j})h_{t-1}^{j}+z_{t}^{j}\tilde{h}_{t}^{j}
\end{equation}

\noindent where $z_{t}^{j}$ is the output gate, and defines what should be forgotten and what should be kept in the GRU unit. The output gate is defined in Equation \ref{eq:update-gate-gru} \cite{chung2015gated}:

\begin{equation}
    \label{eq:update-gate-gru}
    z_{t}^{j}=\sigma(W_{z}x_{t}+U_{z}h_{t-1})
\end{equation}

\noindent where the current and previous weight matrices are $W_{z}$ and $U_{z}$, respectively. In a simplified way, this procedure taking a linear sum between the previous hidden states $h_{t-1}$ and the current input $x_{t}$ and applies the sigmoid function ($\sigma$).

The new memory unit is calculated as described in Equation \ref{eq:new-memory-gru}:

\begin{equation}
    \label{eq:new-memory-gru}
    \tilde{h}_{t}=than(Wx_{t}+r_{t}\odot Uh_{t-1})
\end{equation}

\noindent where $\odot$ is the element-wise multiplication, $r_{t}$ refers to reset gate, and its output can be calculated as defined in Equation \ref{eq:reset-gru}:

\begin{equation}
    \label{eq:reset-gru}
    r_{t}=\sigma(W_{r}x_{t}+U_{r}h_{t-1})
\end{equation}

Research suggests that GRUs are easier to generalize with faster training times while achieving comparable performance outcomes \cite{jozefowicz2015empirical,chung2015gated,kaiser2015neural,arkhipenko2016comparison}. As such, we also propose a GRU for comparison against an LSTM in this study.


\subsection{Clustering}
\label{subsec:clustering}
Clustering is an unsupervised ML technique that breaks down a data set into multiple groups (clusters) of observation with similar characteristics \cite{arora2016analysis}. A number of different clustering techniques exist \cite{rai2010survey} but K-Means clustering is of the most-widely used \cite{arora2016analysis}. The K-Means algorithm is a partitioning method which groups unlabeled data into a predefined $k$ number of clusters based on the Euclidean distance between different data vectors \cite{arora2016analysis}.

The K-Means algorithm starts by randomly selecting a $k$ number of centroids; it then calculates the distance between a data vector and each centroid and assigns the vector to the cluster with the closest centroid as outlined in Equation \ref{eqn:kmeans_assign_step} \cite{arora2016analysis}.

\begin{equation}
S_i^{(t)} = \big \{ x_p : \big \| x_p - \mu^{(t)}_i \big \|^2 \le \big \| x_p - \mu^{(t)}_j \big \|^2 \ \forall j, 1 \le j \le k \big\}
\label{eqn:kmeans_assign_step}
\end{equation}

\noindent where $S$ is the distance between the element and the centroid; $i$ is the cluster number;  $t$ is the number of iterations; $x_p$ is point value; $\mu^{(t)}$ is the value of centroid; and $j$ is the  measure of dissimilarity.

Every time a new data vector is assigned to a cluster, the K-Means algorithm calculates a new centroid based on the average of distances of all data points in each cluster as presented in Equation \ref{eqn:kmeans_update_step}. This iterative process is completed when all data vectors have been allocated to different clusters and the ultimate centroids are identified.

\begin{equation}
\mu^{(t+1)}_i = \frac{1}{|S^{(t)}_i|} \sum_{x_j \in S^{(t)}_i} x_j
\label{eqn:kmeans_update_step}
\end{equation}

We are seeking to predict mobile Internet traffic across a city comprising multiple cells. Traffic behavior may vary across different cells and the behavior of one cell is not necessarily similar to its neighboring cell. As such, we propose a methodology based on cell clustering to predict the Internet traffic using K-means clustering.

\section{Related Works}
\label{sec:related-works}
Predicting traffic for the next day, hour, or even the next minute can be used to optimize the available system resources, for example by reducing the energy consumption, applying opportunistic scheduling, or preventing problems in the infrastructure \cite{li2017learning}. 
Zhang et al. \cite{zhang2018citywide} proposed a CNN model that was able to capture the spatial dependency and two temporal dependencies, closeness and period. Prediction matched the ground truth trend well, and the peaks of both in and out traffic was effectively captured and predicted compared to three existing algorithms - Historical Average value, Auto-Regressive Integrated Moving Average (ARIMA) and LSTM. Huang et al. \cite{huang2017study} propose a multi-task learning architecture with three different types of DL models including LSTM, three-dimensional CNN, and a combination of CNN and RNN (CNN-RNN) to model spatial and temporal aspects of the traffic. They also compare the performance against ARIMA and non-DL methods. The CNN-RNN model was found to be reliable for all tasks with 70 to 80\% forecasting accuracy. Chen et al. \cite{chen2018deep} propose a two-phase framework to dynamically find optimal Remote Radio Head (RRH) clustering and Baseband Unit (BBU) mapping schemes under different contexts. A multivariate LSTM was used to learn the temporal dependency and spatial correlation among base station traffic patterns, and make accurate traffic forecasts for future time periods. The prediction output was used to create RRH clusters and map them to BBU pools to maximize the average BBU capacity utility and minimize the overall deployment cost. Results suggested that the proposed method increased the average capacity utility and reduced the overall deployment cost outperforms baseline methods i.e. Distance-Constrained Complementarity-Aware (DCCA) ARIMA, DCCA Windowed Artificial Neural Networks (DCCA-WANN), and multivariate distance-constrained LSTM. Wang et al. \cite{wang2017spatiotemporal} propose a hybrid DL model for spatio-temporal prediction comprising of an AE-based model for spatial modeling, and LSTM for temporal modeling. Results suggest that the proposed hybrid model significantly improves prediction accuracy compared to ARIMA and Support Vector Regression (SVR). In Alawe et al. \cite{alawe2018improving}, an LSTM and a deep neural network model were compared for predicting traffic load on a 5G network to inform a scalability mechanism. Performance was evaluated using simulation and suggests that the forecast-based scalability mechanism outperformed threshold-based solutions. 

Unlike the other Telecom Italia studies cited above, Zhang and Patras \cite{zhang2018long} focus on reliable \emph{long-term} mobile data traffic forecasting. They propose an ensemble system that leverages convolutional LSTM and 3D-ConvNets structures to model long-term trends and short-term variations of the mobile traffic volume, respectively. Results suggest that the proposed system provided highly accurate long term (10-hour long) traffic predictions, while operating with short observation intervals (2 hours), irrespective of the time of day.  The ensemble system outperformed baseline methods including Holt Winters (HW), ARIMA, MLP, and Support Vector Machine (SVM). 

Wang et al. \cite{wang2017spatiotemporal} propose a novel decomposition of in-cell and inter-cell urban data traffic, and apply a graph-based neural network (GNN) using LSTM to accurately predict mobile traffic. They compare their proposed GNN with a number of baseline methods including NAIVE, ARIMA, LSTM, HW and variations of their GNN. Results suggest that the proposed DL variant of their GNN consistently and significantly outperformed all the baselines in both MAE and Mean Absolute Relative Error (MARE). Feng et al. \cite{feng2018deeptp} propose an LSTM-based end-to-end model (DeepTP) to forecast traffic demands from spatial-dependent and long-period cellular traffic. DeepTP outperforms ARIMA, SVR and GRU (although to a lesser extent) based on MRSE. While outperforming other models, DeepTP was much slower than other models including GRU. Qiu et al. \cite{qiu2018spatio} also use LSTM combined with unified multi-task learning frameworks to explore spatio-temporal correlations among base stations to improve traffic prediction. They evaluate their proposal against Online SVR, Non-parametric Regression, the Adaptive Kalman filters, and an AE approach. The proposed LSTM-approach outperforms other methods using MSE as a performance metric.


Even though a number of studies have already tried to address the challenges posed by mobile traffic prediction, our work proposes a different methodology. In fact, while other studies have clustered cells based on their geographical location, in this study we group them based on traffic similarity in order to mitigate traffic variation and therefore improve traffic prediction. In addition, we test and compare different DL approaches in order to identify the most effective model for traffic prediction.


\section{Material and Methods}
\label{sec:methodology}
\subsection{Data Set}
\label{subsubsec:data-set}
\begin{figure}[h!]
\centering
\includegraphics[width=0.6\columnwidth]{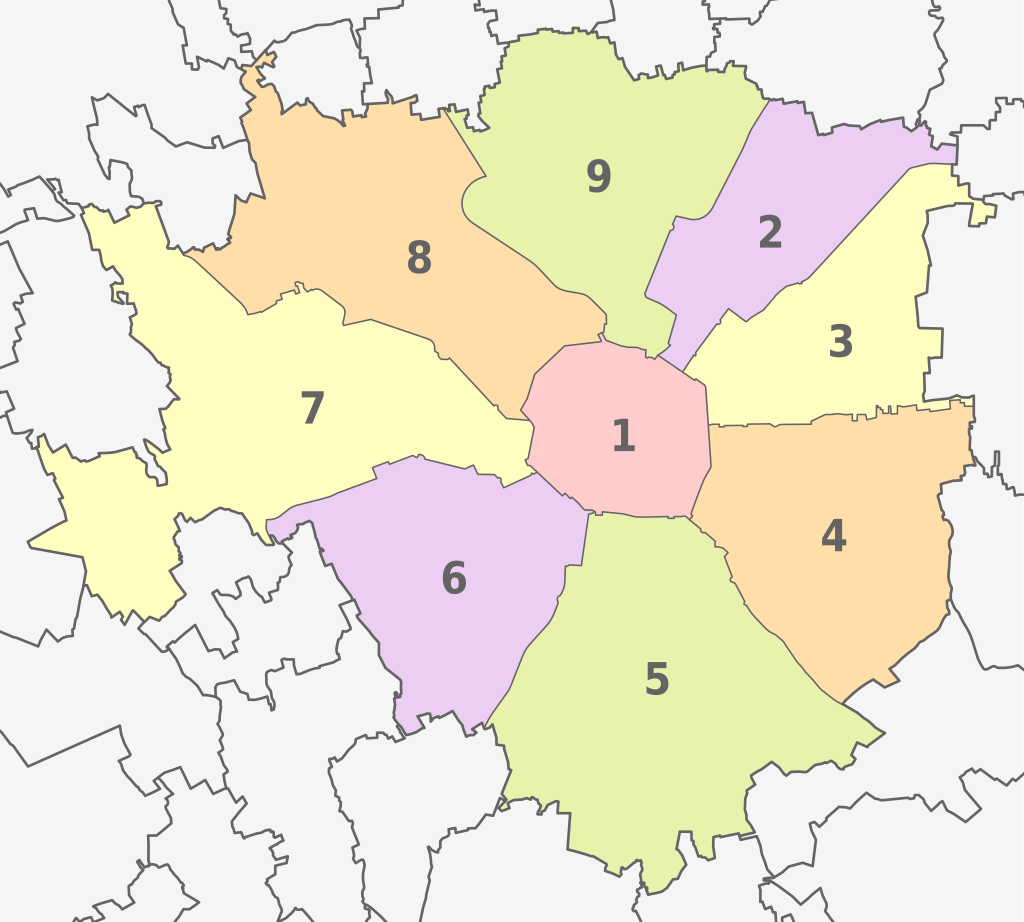}
\caption{The Milan Metropolitan Area\\Source: \url{https://www.wikiwand.com/en/Municipalities\_of\_Milan}}
\label{fig:milan}
\end{figure}

The metropolitan area of Milan is located in northern Italy and consists of nine different municipalities (see Figure \ref{fig:milan}). Milan is the largest metropolitan area in Italy and one of the ten most populous in the European Union \cite{eurostat20}.  

In this study, we use the Telecom Italia data set for Milan from the \textit{Big Data Challenge} \cite{barlacchi2015multi}. The data set is organized into 10,000 cells (100 x 100) comprising over 10 million user activity logs, each related to a particular cell. The data set has log data for two months (62 days) from 1 November 2013 to 1 January 2014 \cite{hussain2019mobile}.
Although this data set was collected between 2013 and 2014, it still proves to be quite valuable for researchers exploring mobile traffic prediction and it has been used in a number of recently published articles (see, for example, \cite{mededovic2019node,zhang2020citywide,amin2020hotspots}). The Telecom data set adopted in this study in fact is one of the few telecommunication data sets that are publicly available in contrast to the large number of data sets that are typically accessible to a restricted number of researchers under non-disclosure agreements (NDAs), or by third parties that have a contractual relationship with telecommunication providers.

The log activity is structured as Call Detail Records (CDRs) on the following activities: (i) incoming and outgoing voice calls, (ii) short message service (SMS) messages, and (iii) Internet activity. A CDR is generated every time a user starts or finishes a voice call, sends or receives an SMS, and starts or terminates an Internet session (the data is recorded if the connection takes more than 15 minutes, or more than 5 MB is transferred during the session). In this paper, we specifically focus on predicting Internet traffic. 

As the data set has periods with no Internet traffic (e.g., a few minutes at night where there are no records), we aggregate all CDRs for Internet traffic into 30-minute periods. Consequently, we have 48 records per day related to Internet traffic. We use a sliding window strategy with a window size of four time periods thus we are use the previous two hours to predict the Internet activity of the subsequent 30 minutes.

To create the training and testing data sets, the original data set is divided in two parts: the first 80\% of the time series for training and the last 20\% for testing. We also normalized these data sets to the [0,1] range to facilitate the training of DL models, since their parameters are very small, close to zero.

\subsection{Clustering the Cells}
\label{subsubsec:clustering-methodology}
The data set is composed of traffic data of different cells for Milan. As discussed in \ref{subsec:clustering}, we propose a methodology based on cell clustering using Internet activity as a statistical metric to propose DL models to predict Internet traffic. We calculate the total number of Internet activities considering six periods in each day, as described in Table \ref{tab:periods-day}\footnote{In Northern Italy, unlike other countries where lunch is 1230-1400, the working day often includes a break from 1200-1330 or 1430-1600. For the purposes of this study, we have aggregated this as a four hour block. Researchers may need to modify this for other countries.}. 

\begin{table}[h]
\caption{Periods of the day}
\label{tab:periods-day}
\centering
\begin{tabular}{cc}
\hline
\textbf{Period} & \textbf{Time (in hour)} \\ \hline
Late Night      & 00:00 - 04:00  \\
Early Morning   & 04:00 - 08:00   \\
Morning         & 08:00 - 12:00  \\
Afternoon       & 12:00 - 16:00  \\
Evening         & 16:00 - 20:00  \\
Night           & 20:00 - 00:00  \\ \hline
\end{tabular}
\end{table}

Each cell can be represented by a vector containing six values based on the average Internet activity for each period of the day. Based on these values, we create clusters of cells using the K-Means algorithm \cite{arora2016analysis} which has been widely used in a number of different research domains such as document classification, recommendation systems based on user interests,  classification based on user purchase behavior etc. The K-Means algorithm in fact has many advantages compared to other clustering techniques such as ease of implementation and fast convergence even in the context of big data \cite{yuan2019research}.

\begin{figure}[h!]
\centering
\includegraphics[width=0.8\columnwidth]{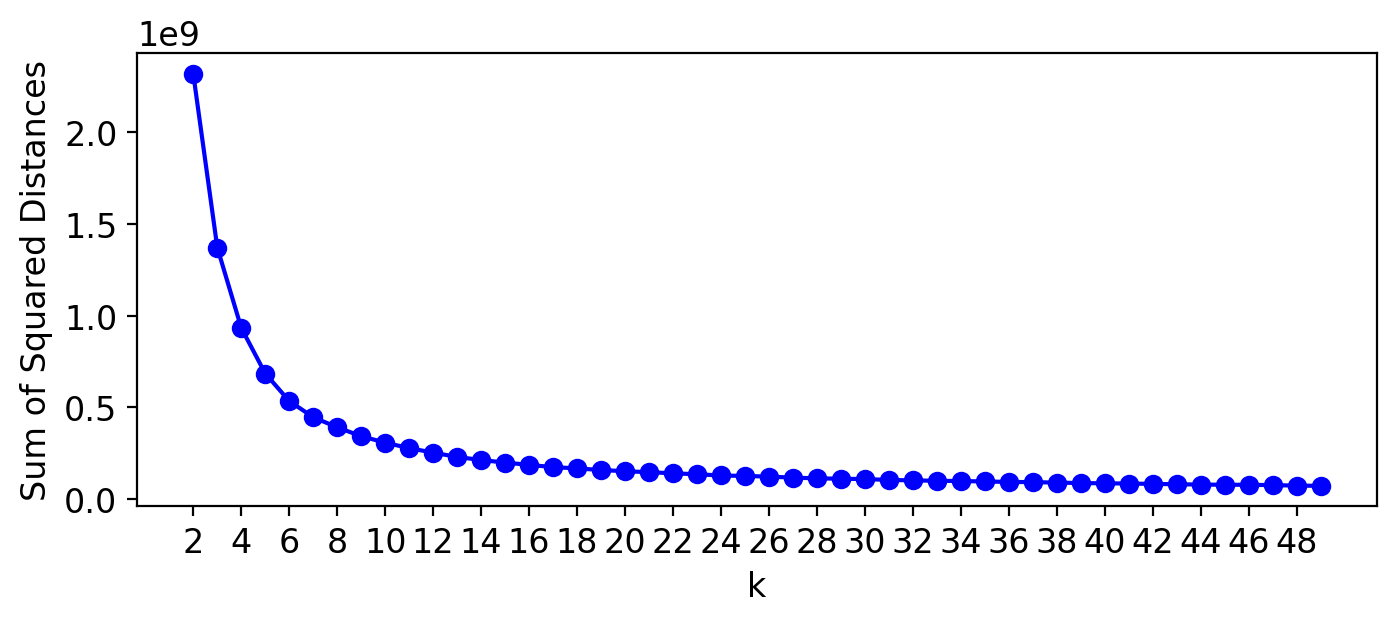}
\caption{Elbow method results}
\label{fig:elbow-method-results}
\end{figure}

To automatically estimate the optimal number of clusters ($k$) of cells, we applied the Elbow method. This method varies the number of clusters within a range to find the optimal $k$ based on the sum of square error \cite{syakur2018integration}. We varied $k$ from one to 50 as shown in the Figure \ref{fig:elbow-method-results}. As the number of cluster increases, the sum of squared distance tends towards zero with the elbow of curve being the optimal value. In our case, we select $k=12$ since it is at the end of the elbow and the beginning of the stabilization of the sum of the squared distances. Based on the results presented in Figure \ref{fig:elbow-method-results}, using more than 12 clusters would only increase the complexity of the algorithm with no significant gains in terms of performance.

\begin{figure}[h!]
\centering
\includegraphics[width=0.6\columnwidth]{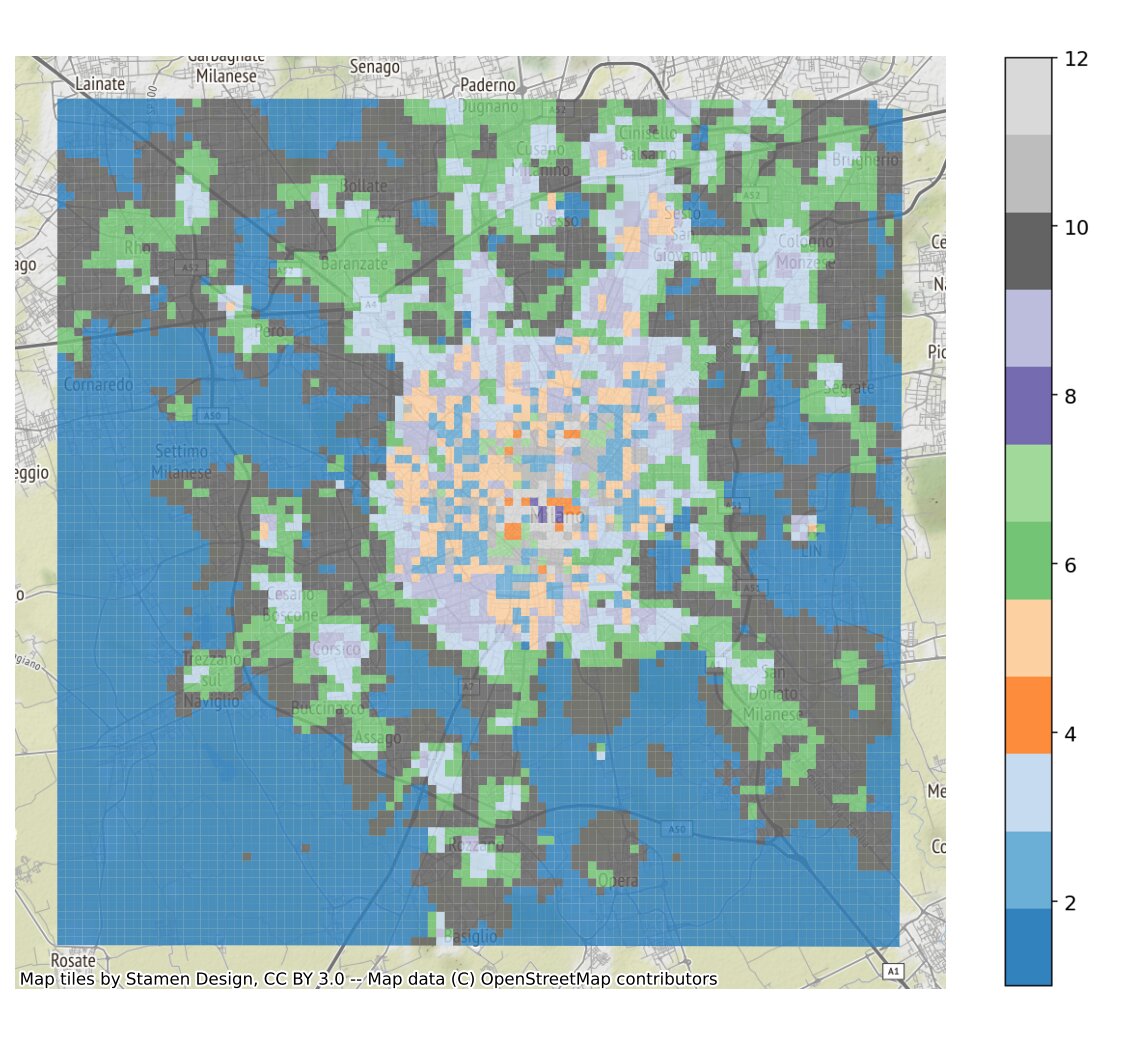}
\caption{Overlay of the 12 clusters on a map of the metropolitan area of Milan}
\label{fig:cluster}
\end{figure}

The 12 clusters have a similar Internet activity pattern across different time periods within each day, regardless of the cell location. Figure \ref{fig:cluster} shows the 12 clusters overlaid on a map of the Milan Metropolitan Area.     

To some extent, Cluster 1 represents the external areas of Milan; Cluster 10 represents the border of the municipalities (excluding the municipality 1, at the center), while Cluster 6 is at the center of such municipalities. Other Clusters (2, 3, 4, 5, 7, 8, 9, 11 and 12) are regions within the city.

We calculate the mean Internet traffic for each of the 12 clusters using the 30-minute aggregated traffic of all cells included in each cluster divided by the number of cells of the cluster. Thus, for each cluster, we have a time series related to their mean cell's traffic. This time series is used to train and test the proposed DL models.

\subsection{Metric}
To assess the performance of the models, we used the RMSE metric as described in Equation \ref{eq:rmse-metric}:

\begin{equation}
    \label{eq:rmse-metric}
    RMSE=\sqrt{\frac{1}{N}\sum_{i=1}^{N}(f_i-y_i)^2}
\end{equation}

\noindent where $N$ is the number of points from the traffic series, $f_i$ is the model prediction at timestamp $i$, and $y_i$ is the the real value at timestamp $i$ \cite{wang2018analysis}. 


\subsection{DL Model Configuration}
\label{subsubsec:dl-models-configuration}

In this paper, we propose two different RNNs that are widely used in the DL literature for regression problems, LSTM and GRU. To find the best configuration of the models, we apply a technique called Grid Search. This technique performs an exhaustive search in a subset of the previously defined parameters and provides the near optimal parameter combination within the given range \cite{syarif2016svm}. 

To apply the Grid Search, we vary the number of hidden layers and their units for both LSTM and GRU (see the parameters and levels in Table \ref{tab:parameters_grid}). 

\begin{table}[h]
\caption{Parameters and levels of Grid Search}
\label{tab:parameters_grid}
\centering
\begin{tabular}{cc}
\hline
\textbf{Parameters} & \textbf{Levels} \\ \hline
Number of layers      & 1 to 4, step 1  \\
Number of units   & 50 to 150, step 50   \\ \hline
\end{tabular}
\end{table}

The first layer of the model is a fixed recurrent layer (the same as the hidden layers) where the number of units equals the input data length. The last layer is a fully connected layer with one neuron that gives the prediction value. Table \ref{tab:parameters} shows the fixed parameters (empirically chosen) to train the DL models. 

\begin{table}[h]
\caption{Parameters used to train the DL models}
\label{tab:parameters}
\centering
\begin{tabular}{@{}ll@{}}
\toprule
\textbf{Parameter}                      & \textbf{Value} \\ \midrule
Activation function of recurrent layers & sigmoid        \\
Activation function of last layer       & hard sigmoid   \\
Number of epochs                        & 50             \\
Optimizer                               & ADAM \cite{kingma2014adam}           \\
Batch size                              & 32             \\
Loss function                           & mean squared error \\
Number of runs                          & 30             \\ \bottomrule
\end{tabular}
\end{table}

Due to the random characteristics that exist in training, such as initialization of weights, selection of batches etc., we perform the experiments 30 times and calculate the average RMSE.

\section{Results}
\label{sec:results}
\begin{figure*}[]
\centering
\subfigure[]{
\includegraphics[width=0.32\columnwidth]{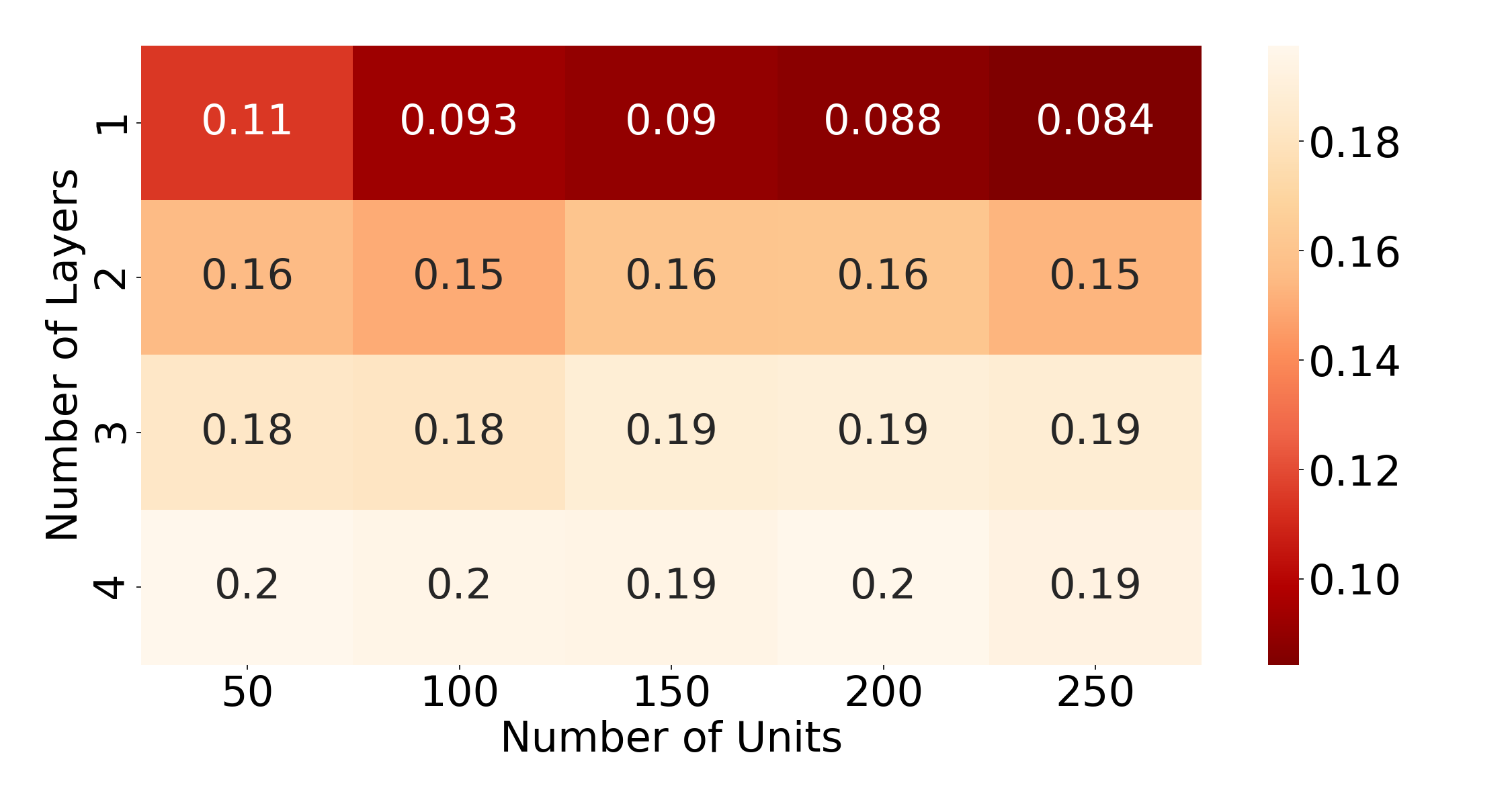}}
\subfigure[]{
\includegraphics[width=0.32\columnwidth]{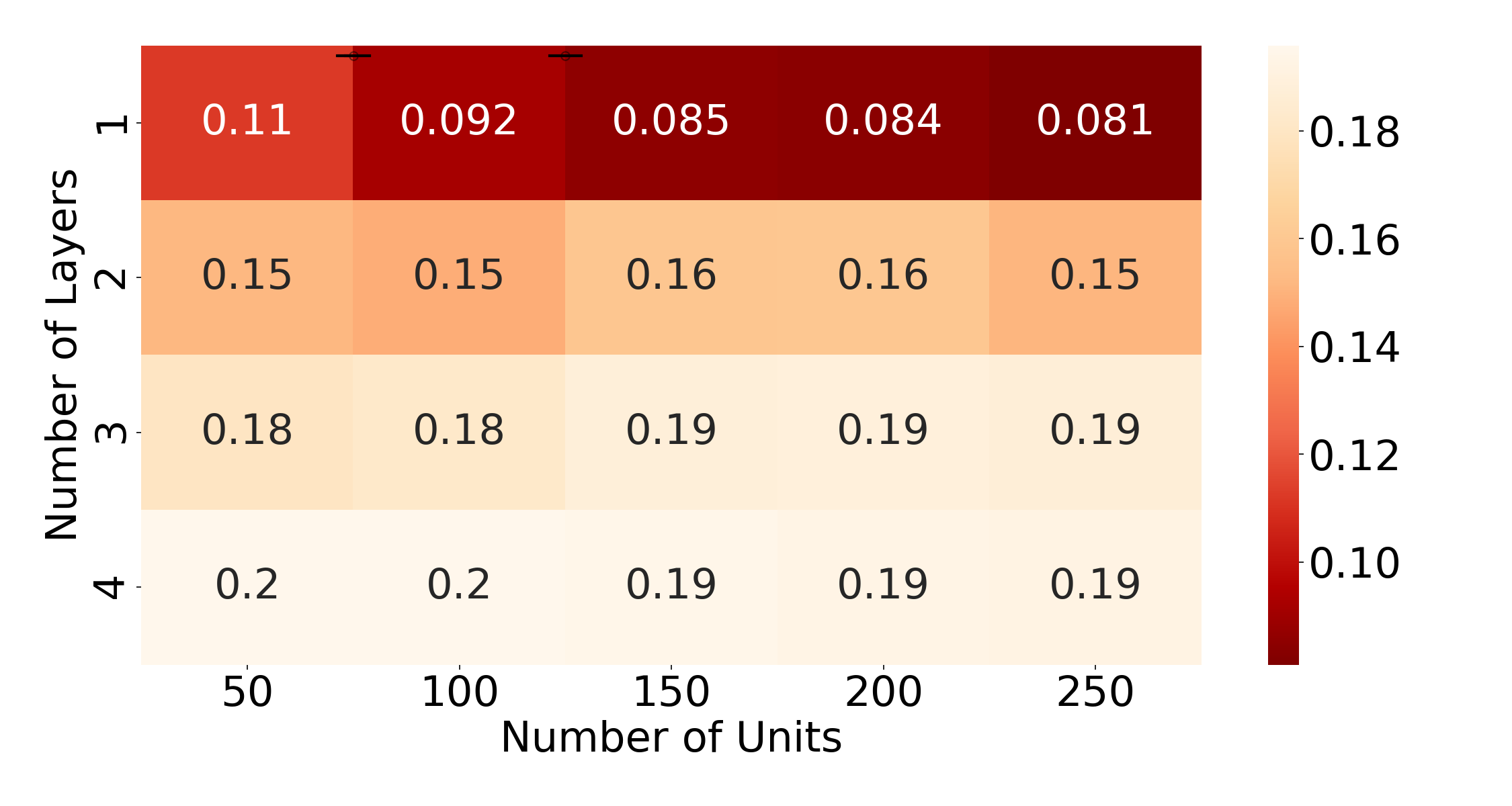}}
\subfigure[]{
\includegraphics[width=0.32\columnwidth]{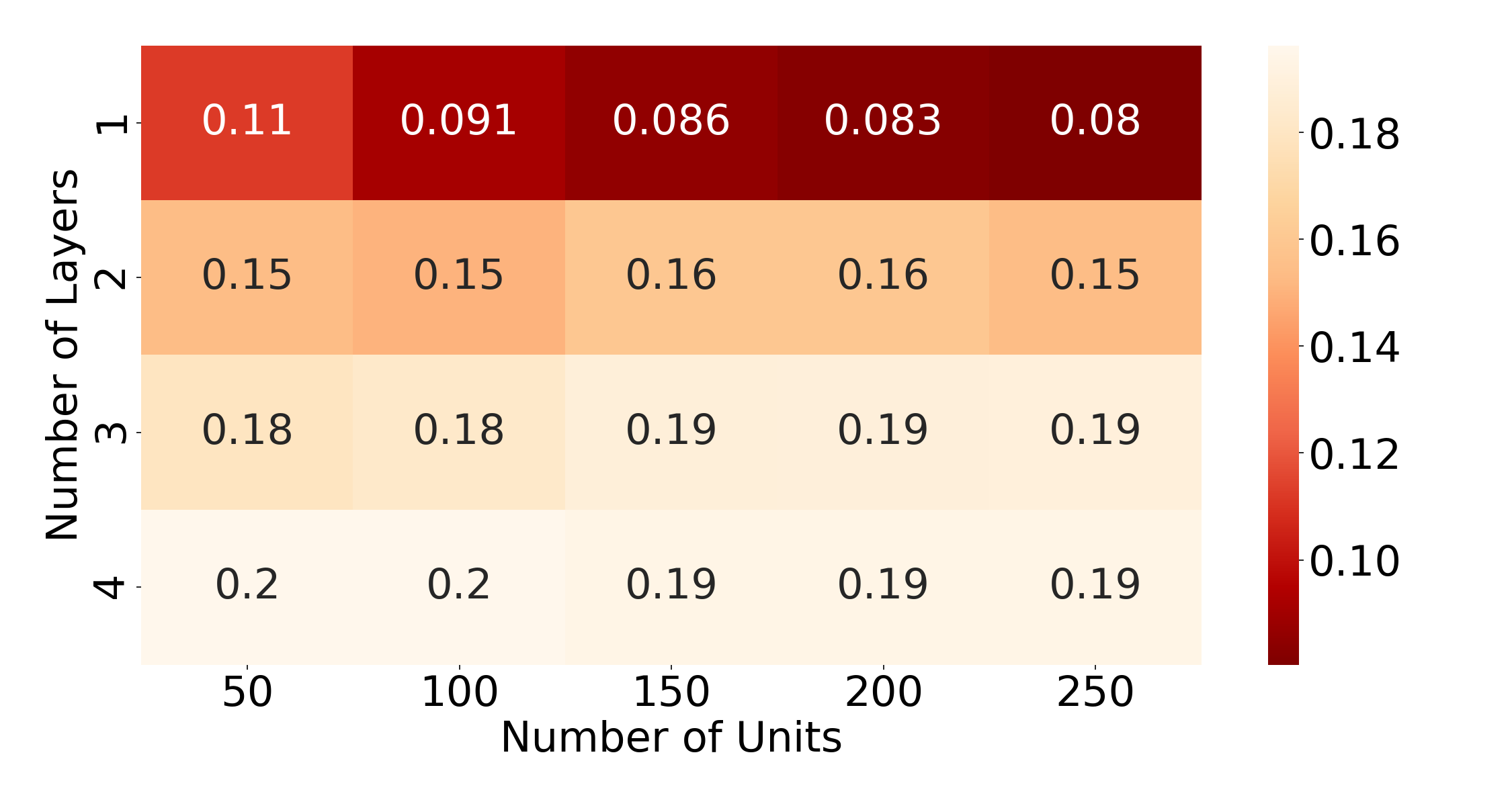}}

\subfigure[]{
\includegraphics[width=0.32\columnwidth]{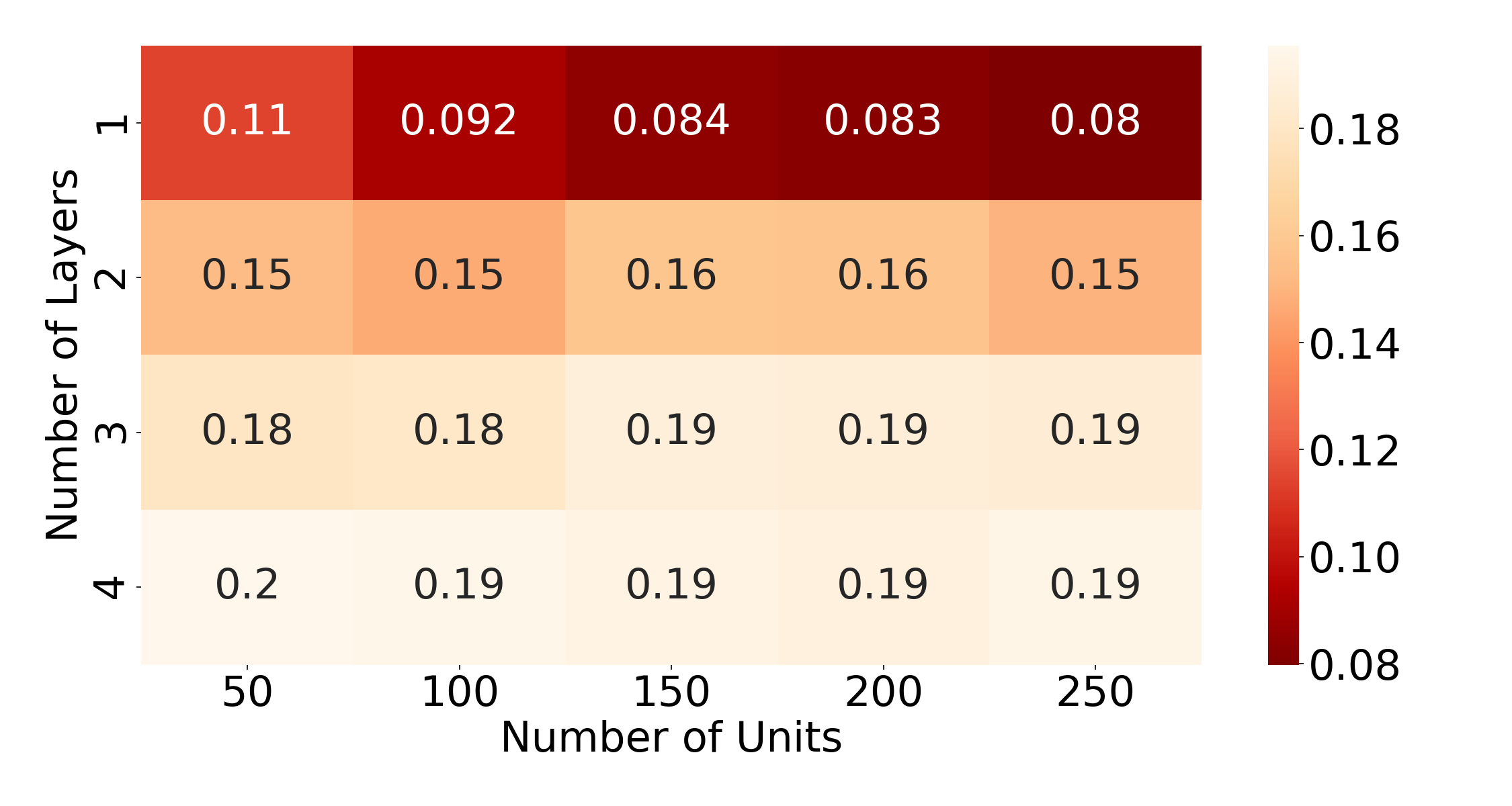}}
\subfigure[]{
\includegraphics[width=0.32\columnwidth]{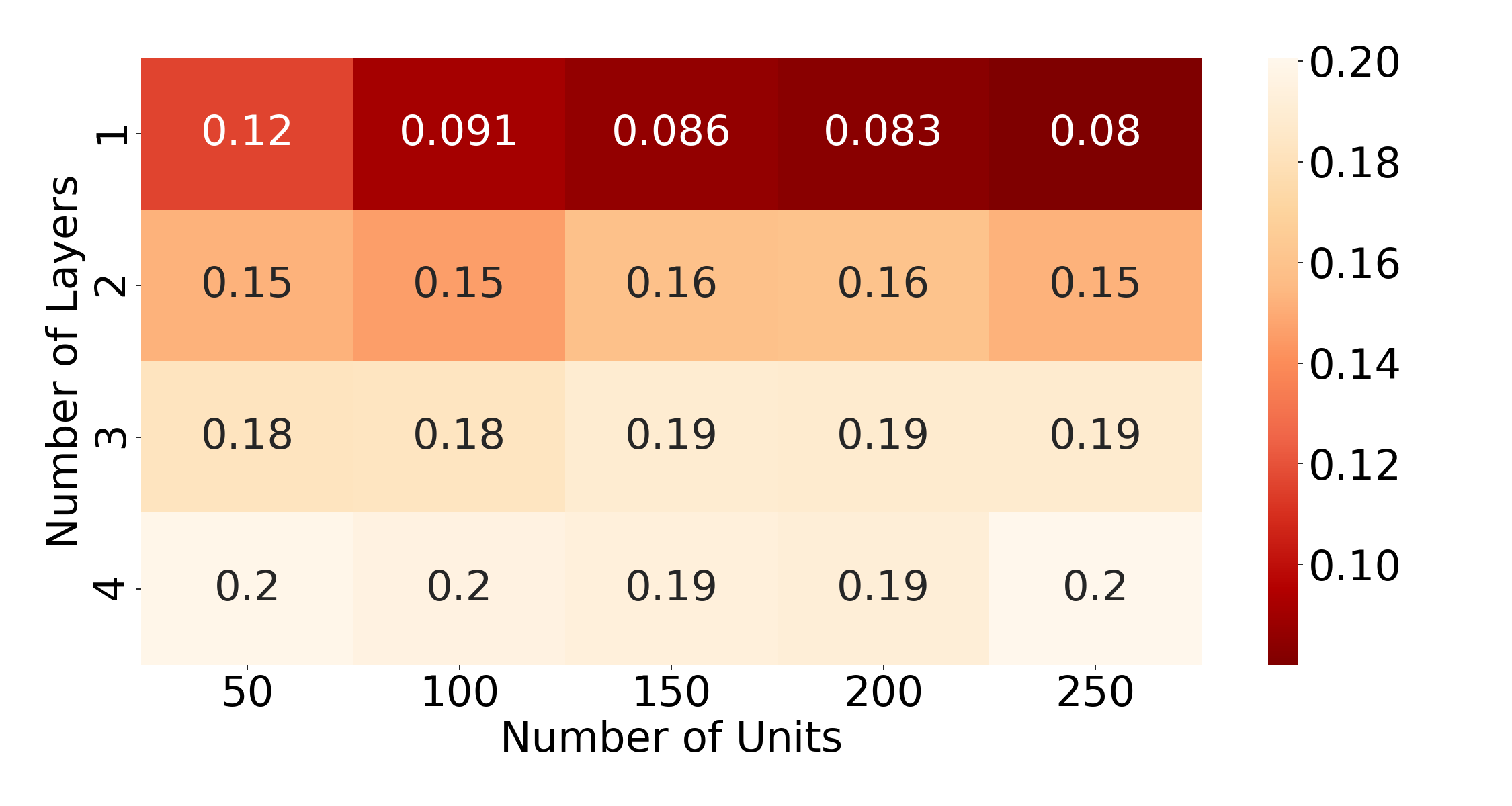}}
\subfigure[]{
\includegraphics[width=0.32\columnwidth]{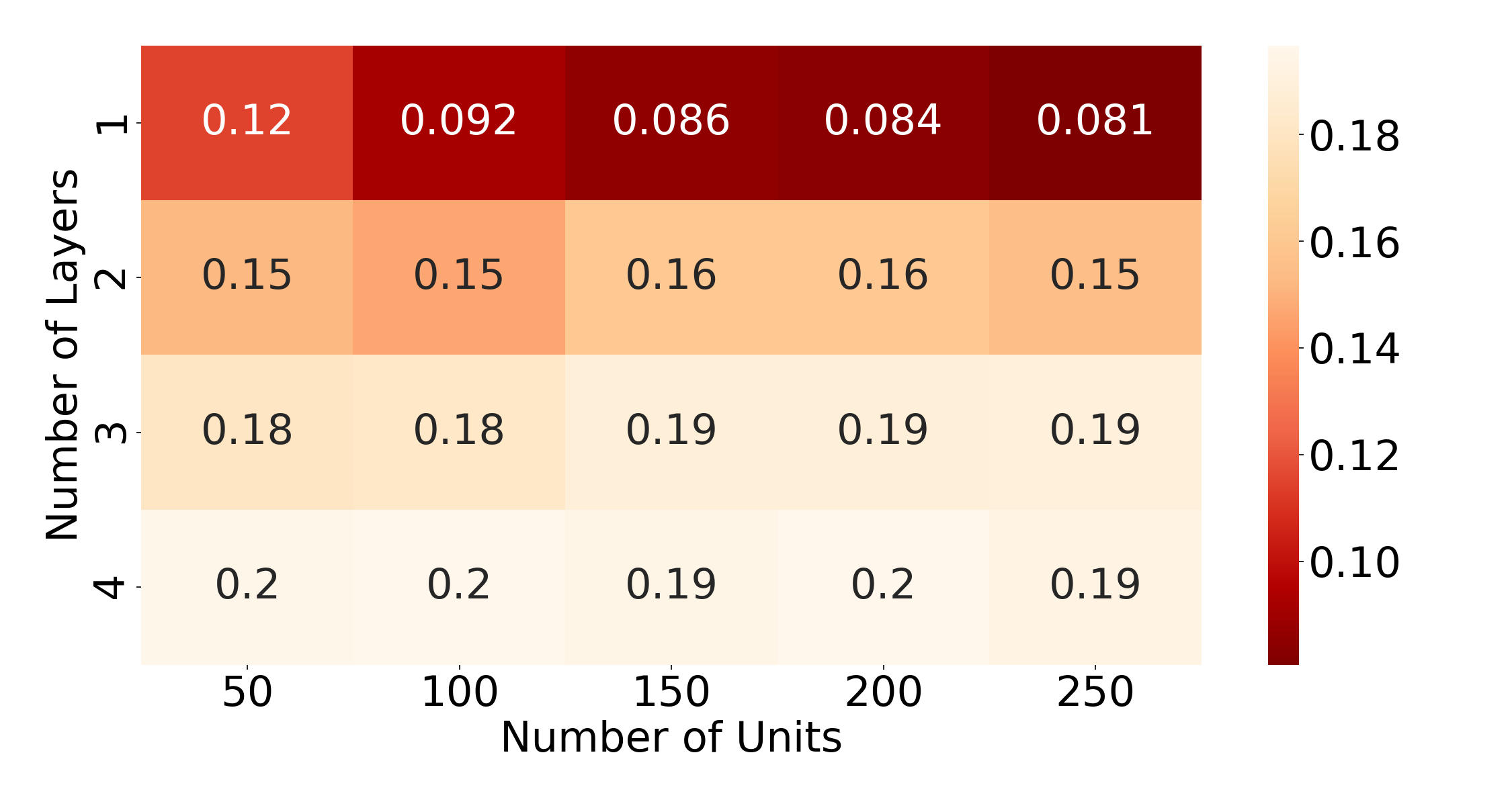}}

\subfigure[]{
\includegraphics[width=0.32\columnwidth]{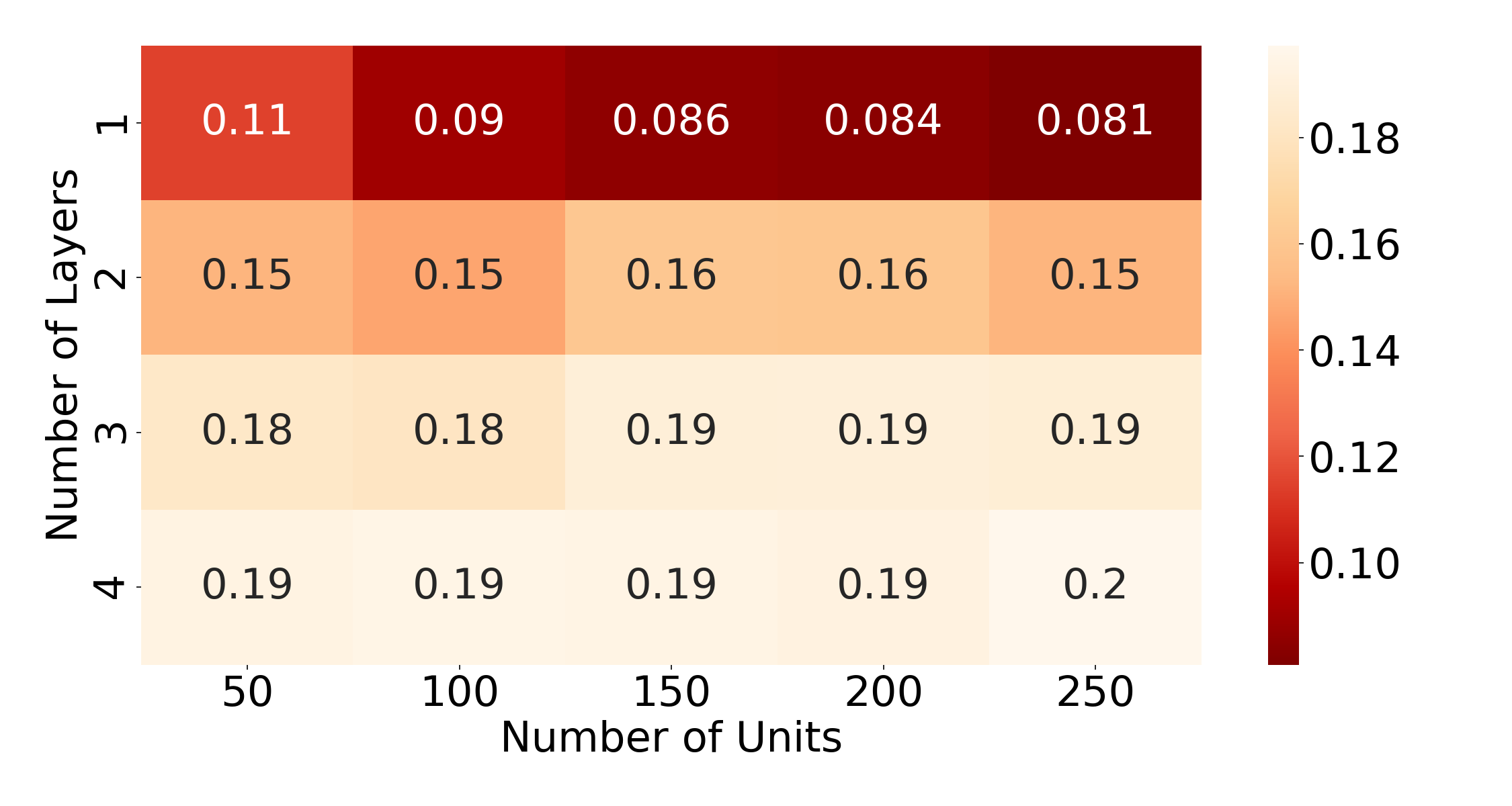}}
\subfigure[]{
\includegraphics[width=0.32\columnwidth]{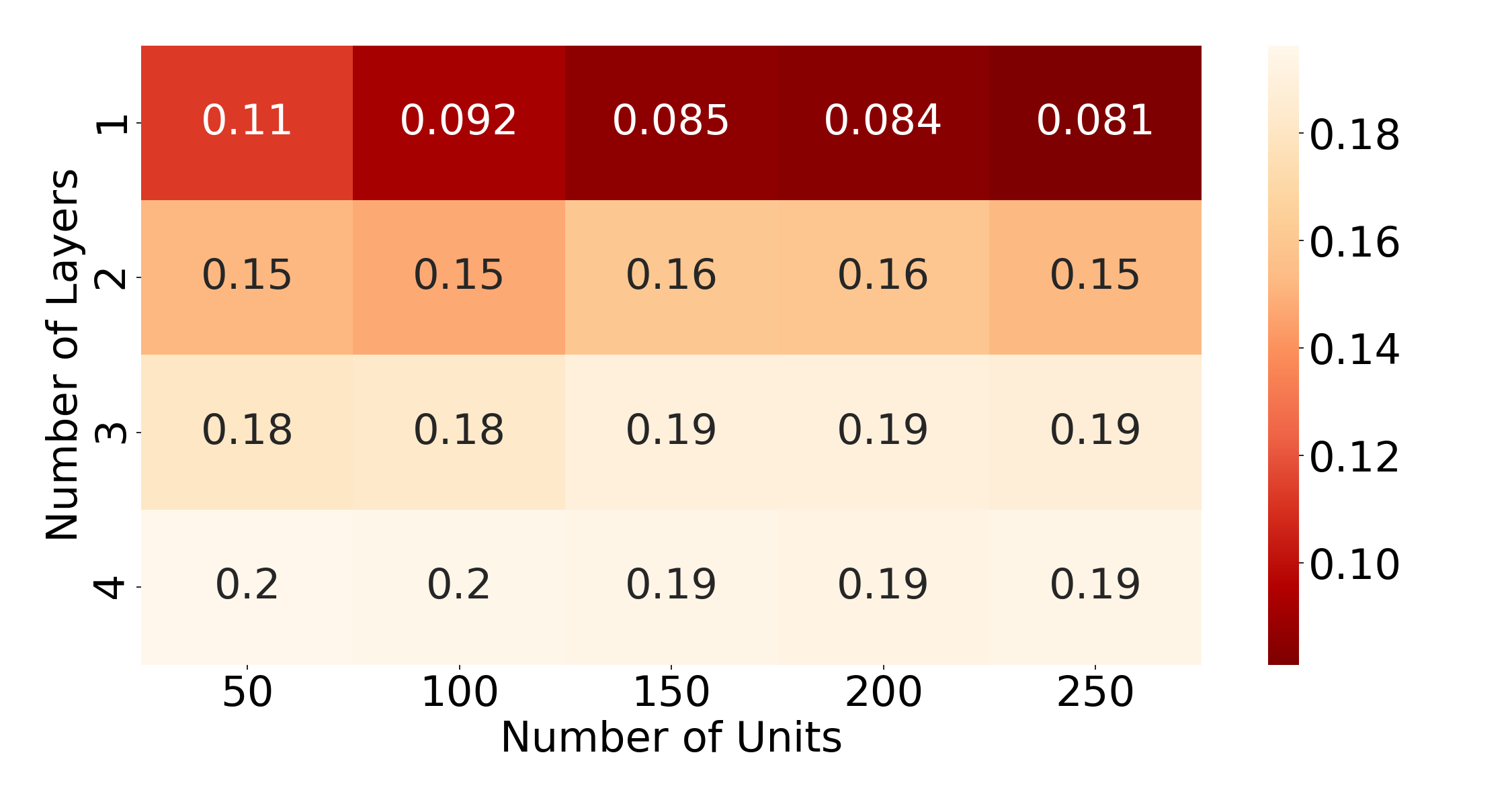}}
\subfigure[]{
\includegraphics[width=0.32\columnwidth]{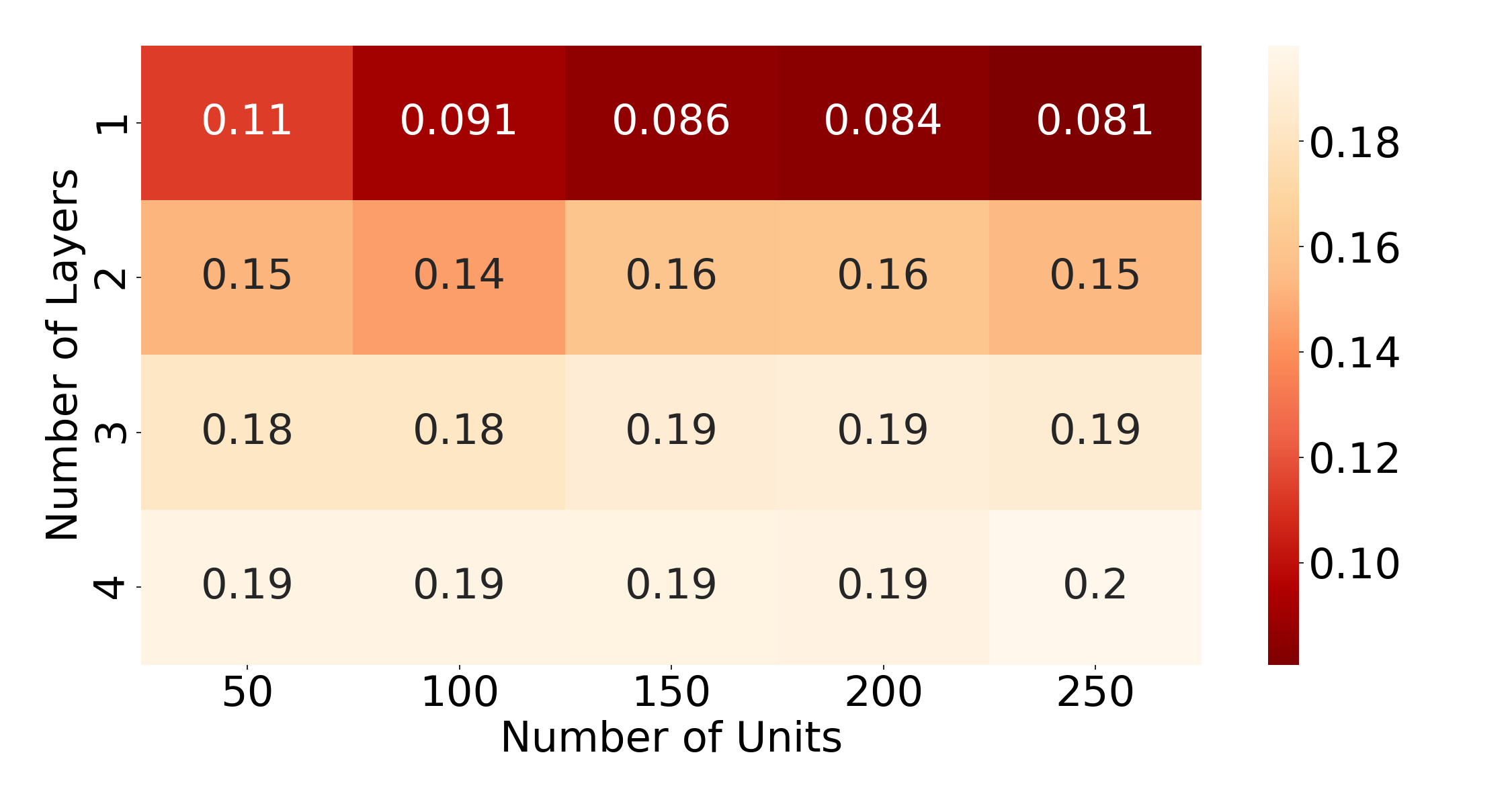}}

\subfigure[]{
\includegraphics[width=0.32\columnwidth]{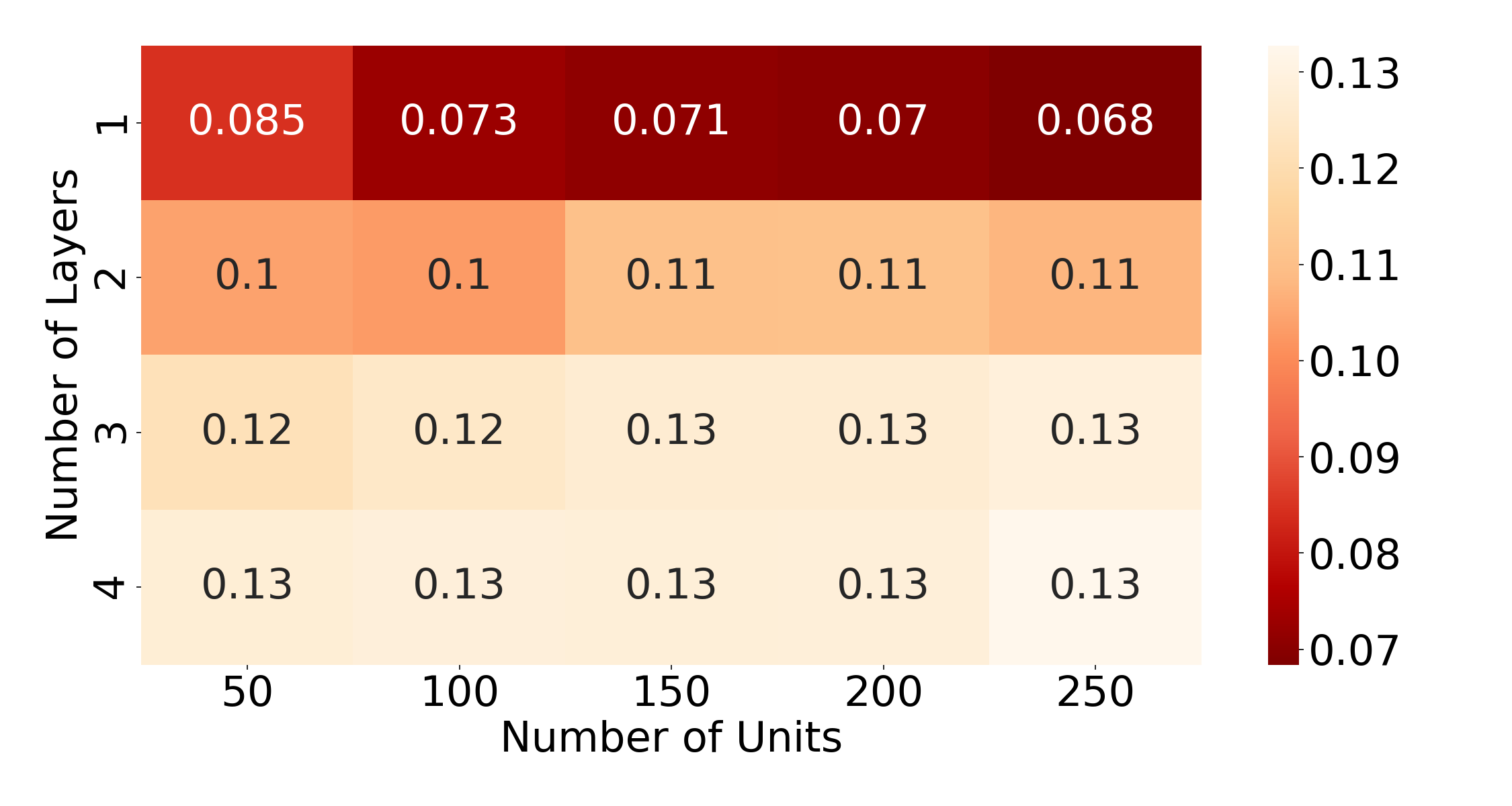}}
\subfigure[]{
\includegraphics[width=0.32\columnwidth]{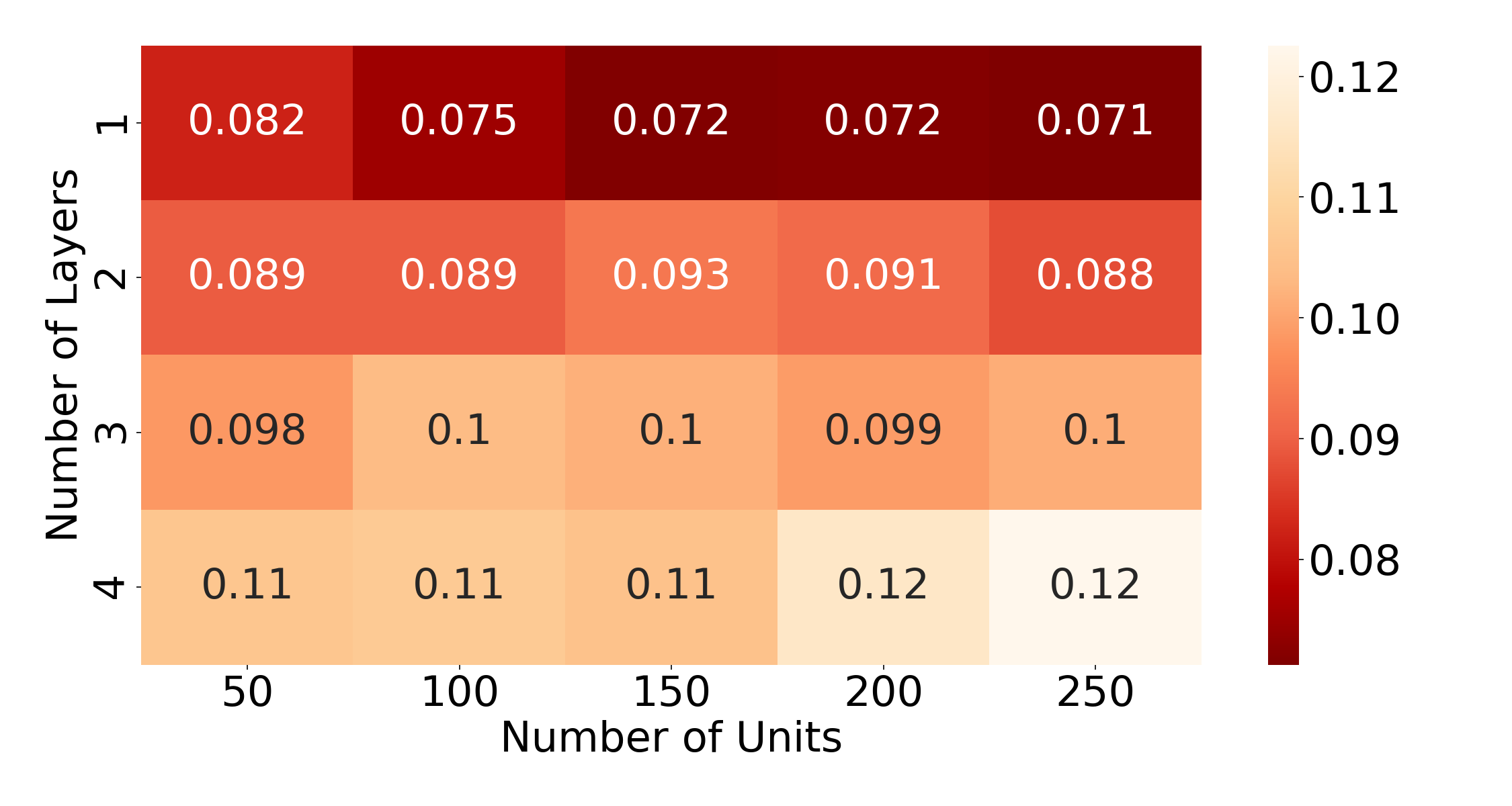}}
\subfigure[]{
\includegraphics[width=0.32\columnwidth]{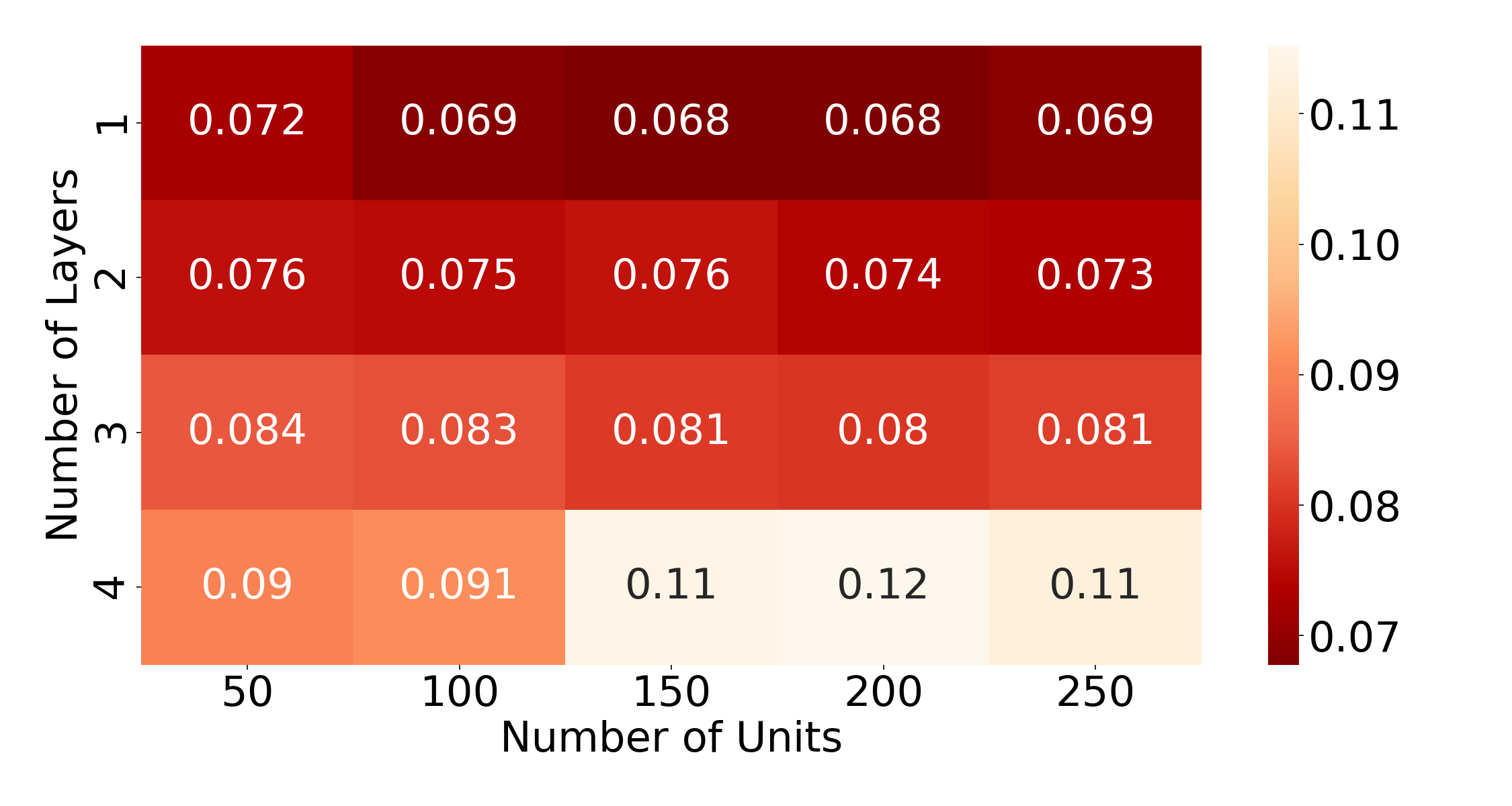}}

\caption{Mean RMSE of LSTM model for (a) cluster 1, (b) cluster 2, (c) cluster 3, (d) cluster 4, (e) cluster 5, (f) cluster 6, (g) cluster 7, (h) cluster 8, (i) cluster 9, (j) cluster 10, (k) cluster 11, and (l) cluster 12.}
\label{fig:lstm-grid-search}
\end{figure*}

\begin{figure*}[]
\centering
\subfigure[]{
\includegraphics[width=0.32\columnwidth]{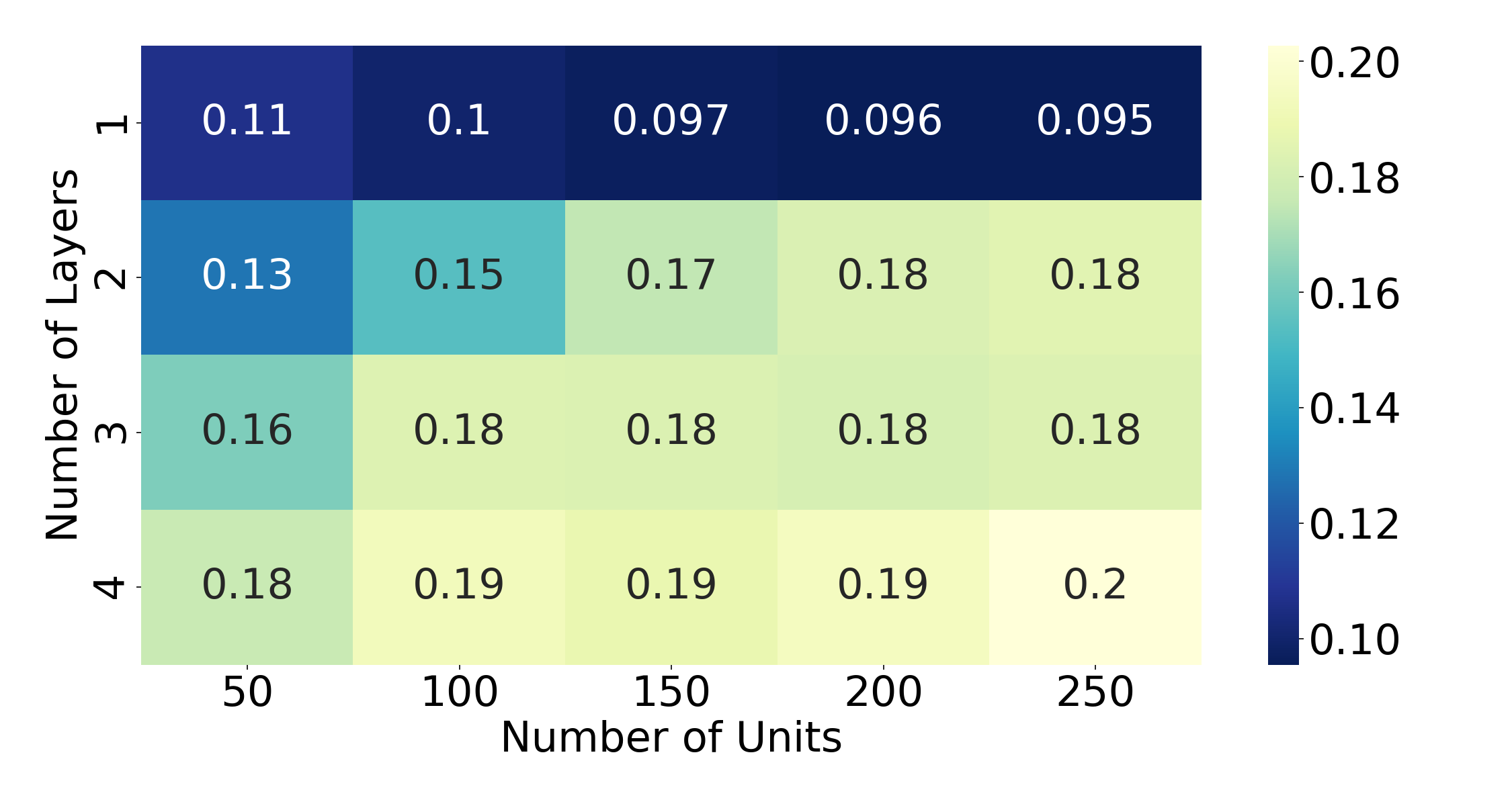}}
\subfigure[]{
\includegraphics[width=0.32\columnwidth]{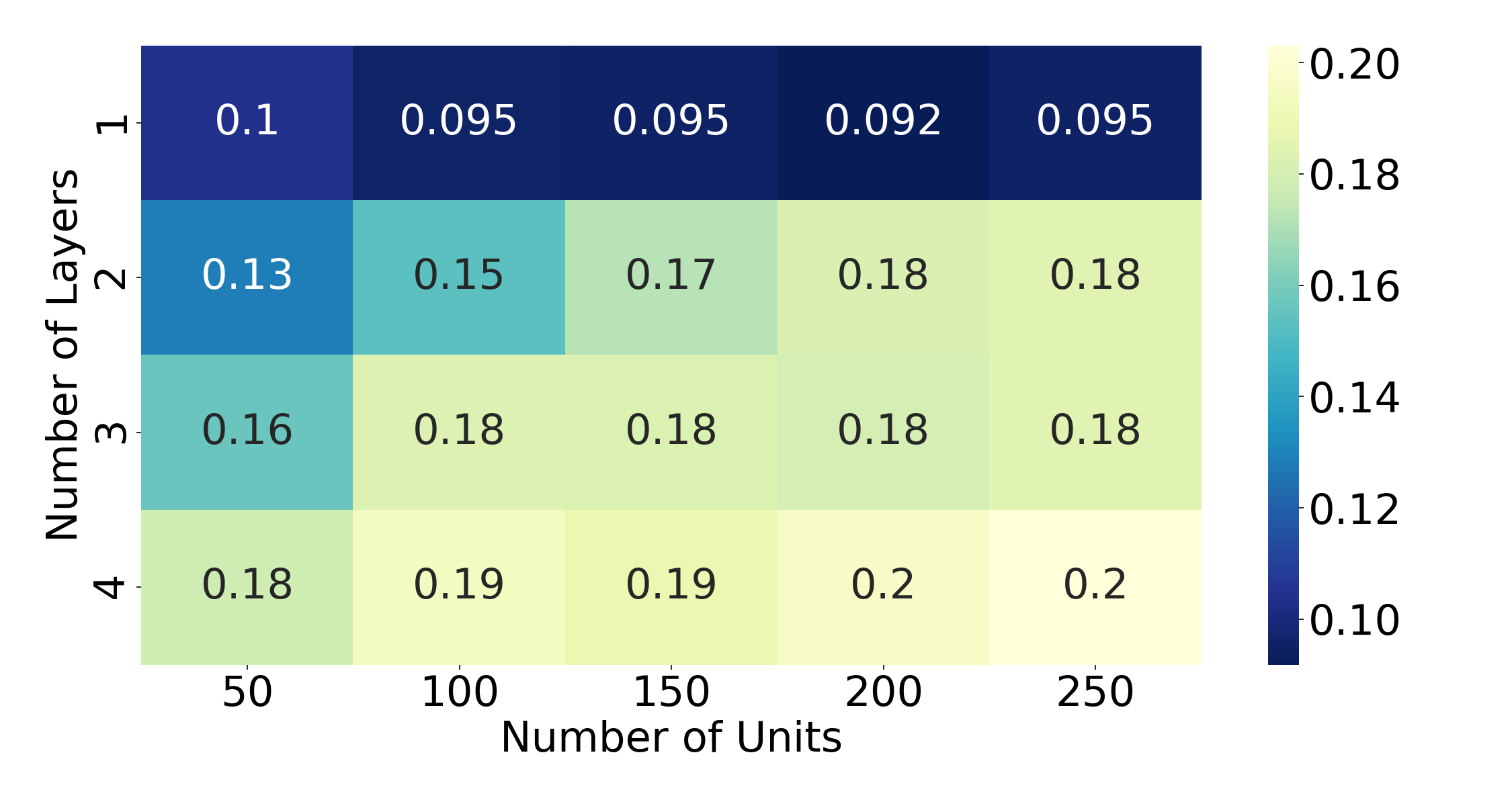}}
\subfigure[]{
\includegraphics[width=0.32\columnwidth]{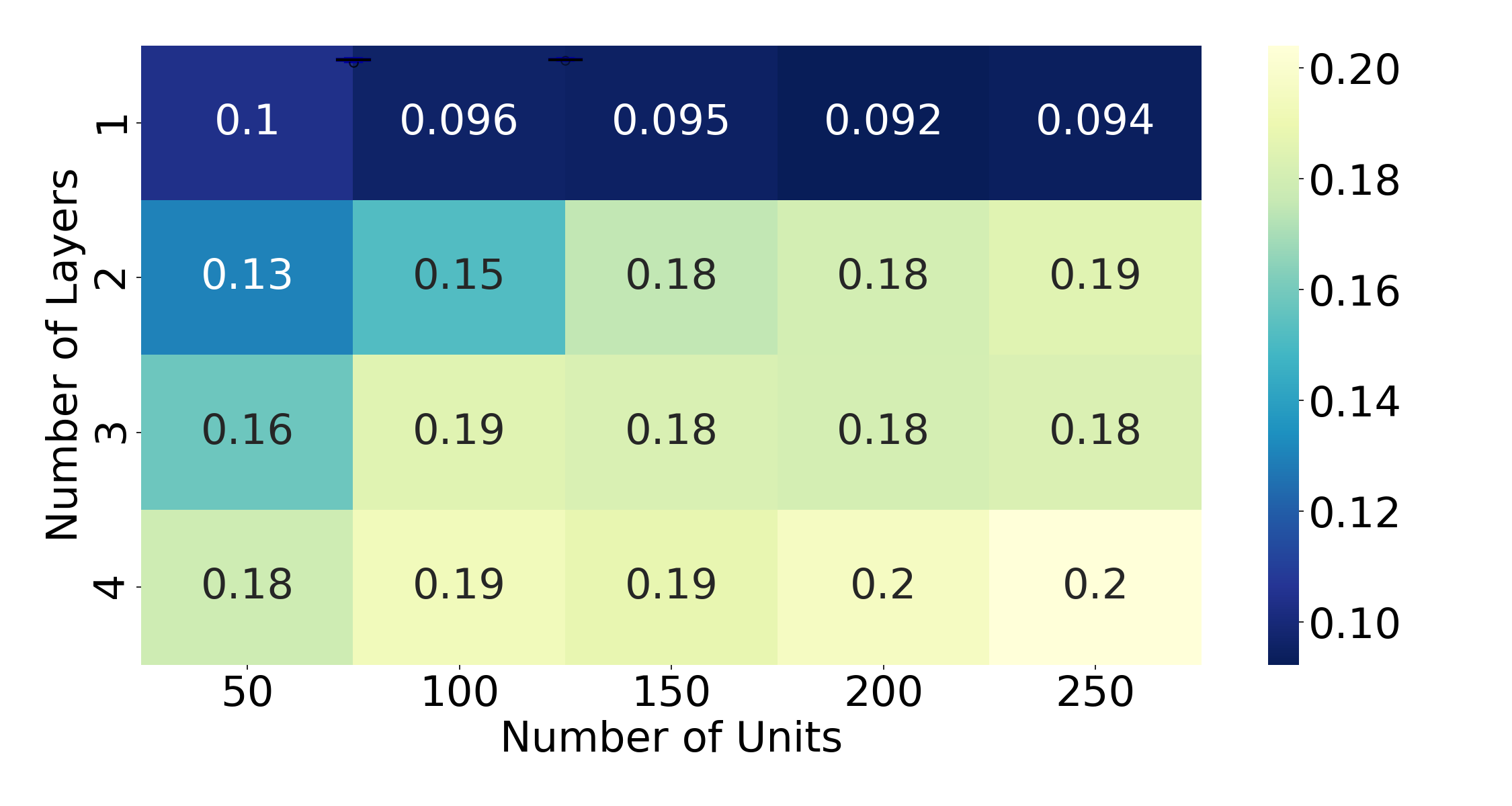}}

\subfigure[]{
\includegraphics[width=0.32\columnwidth]{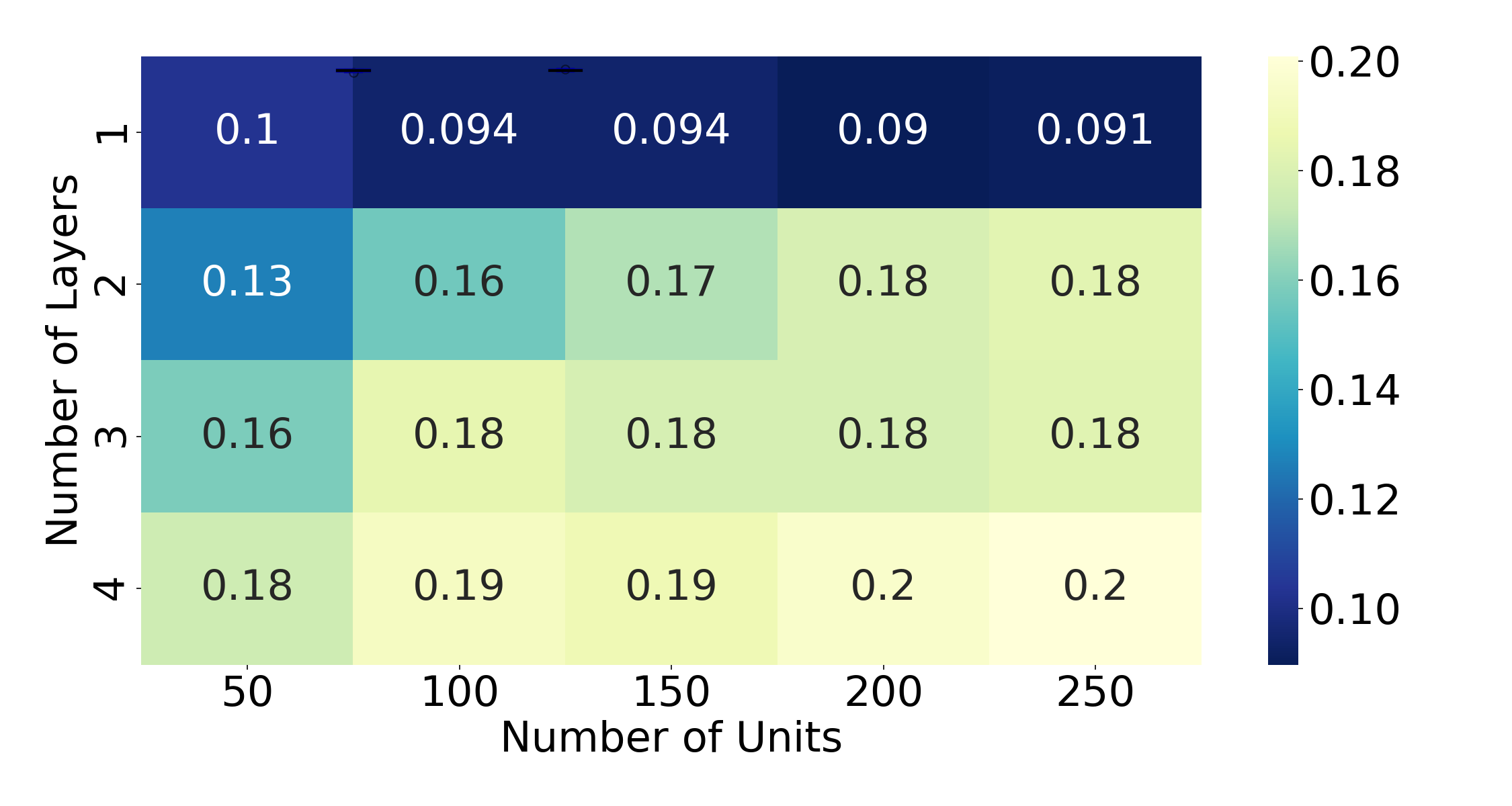}}
\subfigure[]{
\includegraphics[width=0.32\columnwidth]{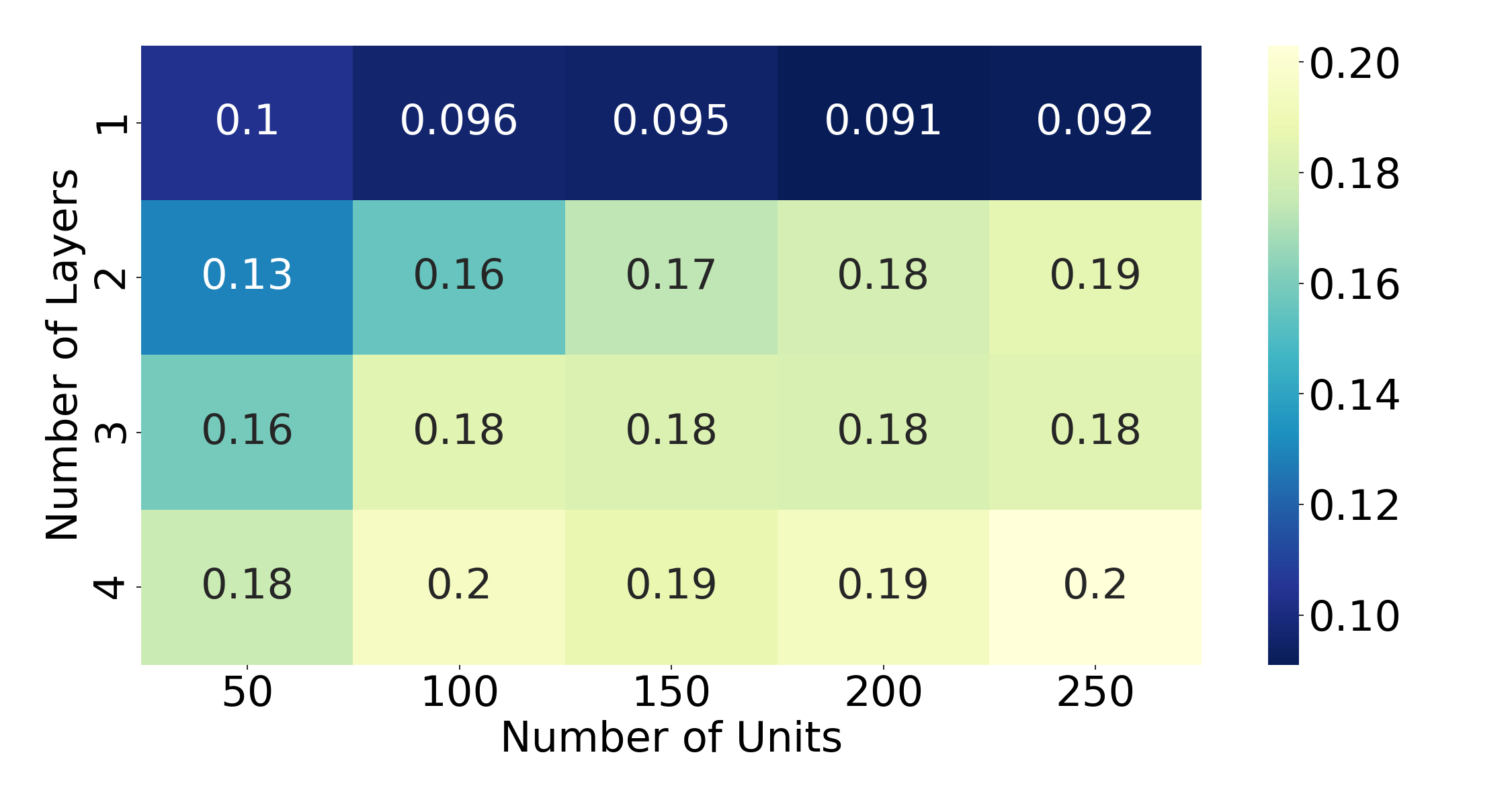}}
\subfigure[]{
\includegraphics[width=0.32\columnwidth]{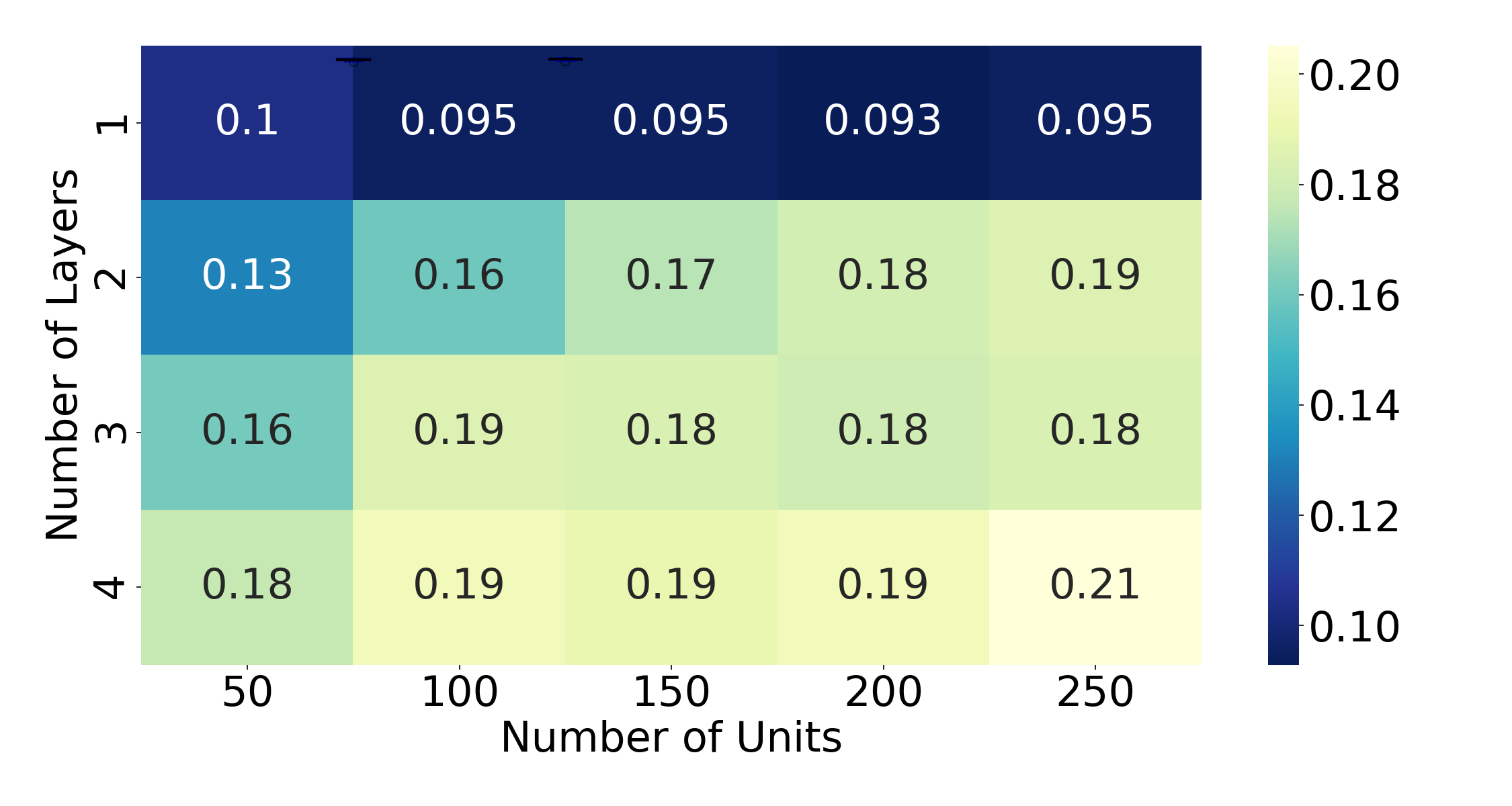}}

\subfigure[]{
\includegraphics[width=0.32\columnwidth]{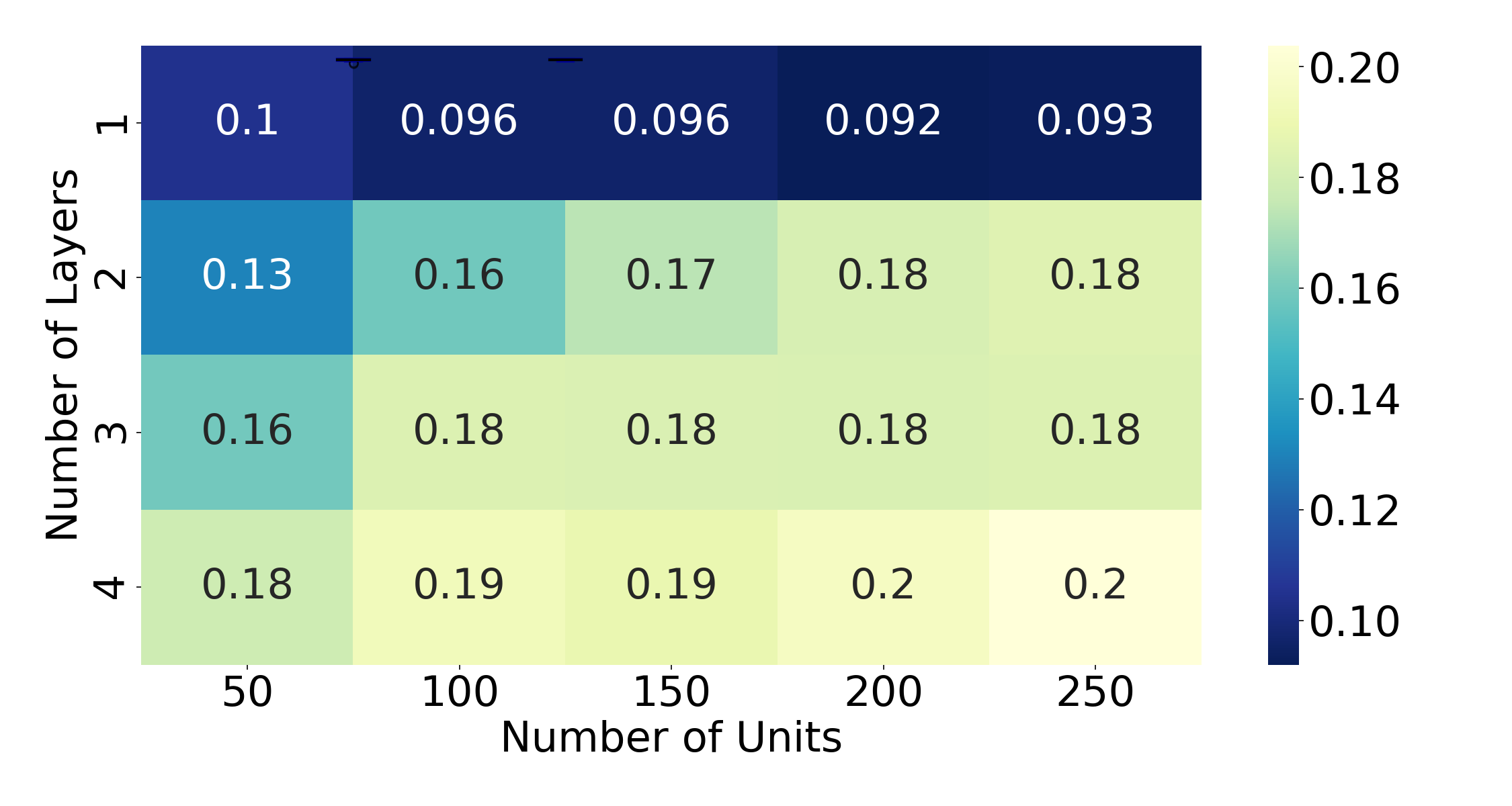}}
\subfigure[]{
\includegraphics[width=0.32\columnwidth]{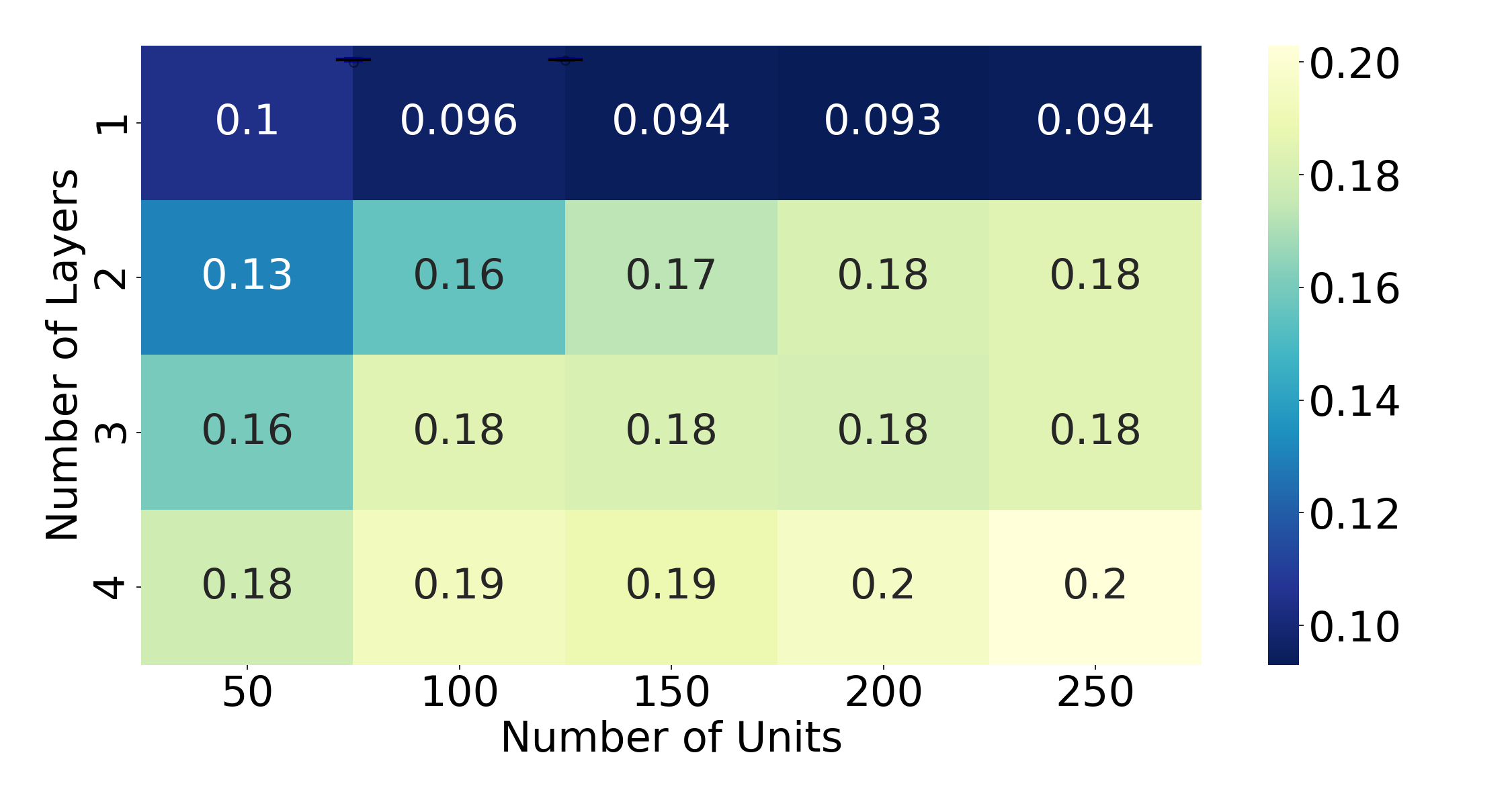}}
\subfigure[]{
\includegraphics[width=0.32\columnwidth]{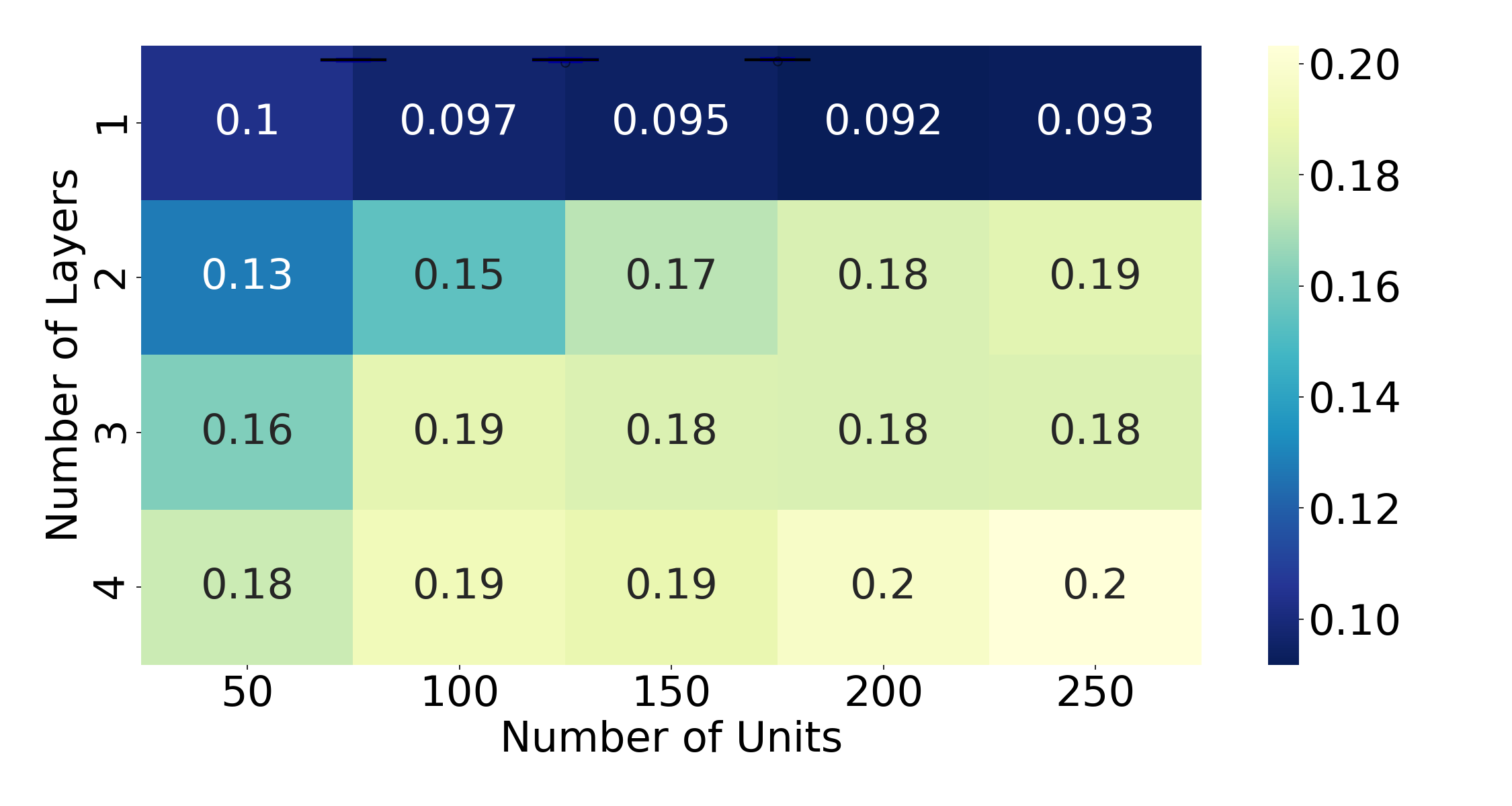}}

\subfigure[]{
\includegraphics[width=0.32\columnwidth]{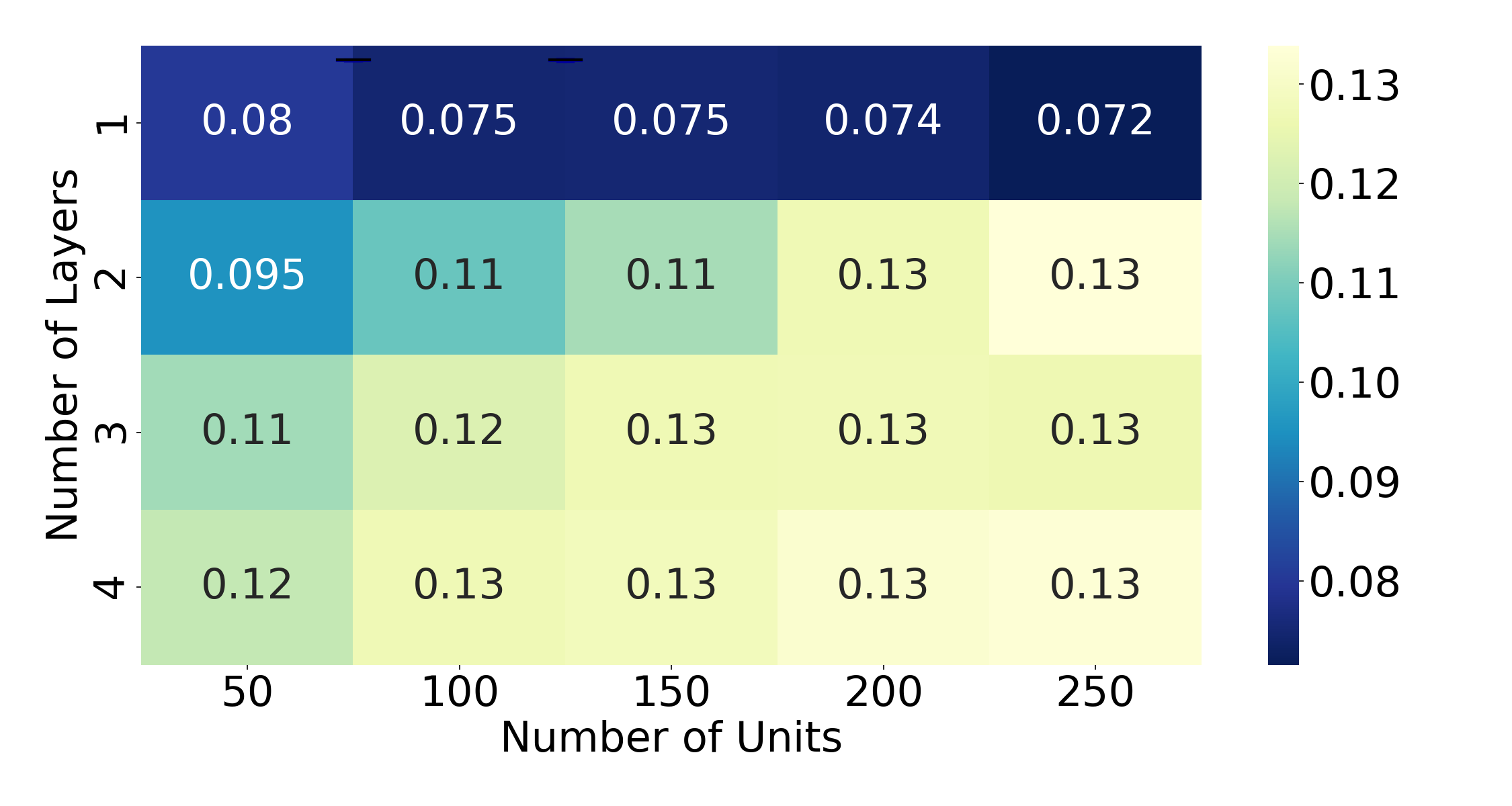}}
\subfigure[]{
\includegraphics[width=0.32\columnwidth]{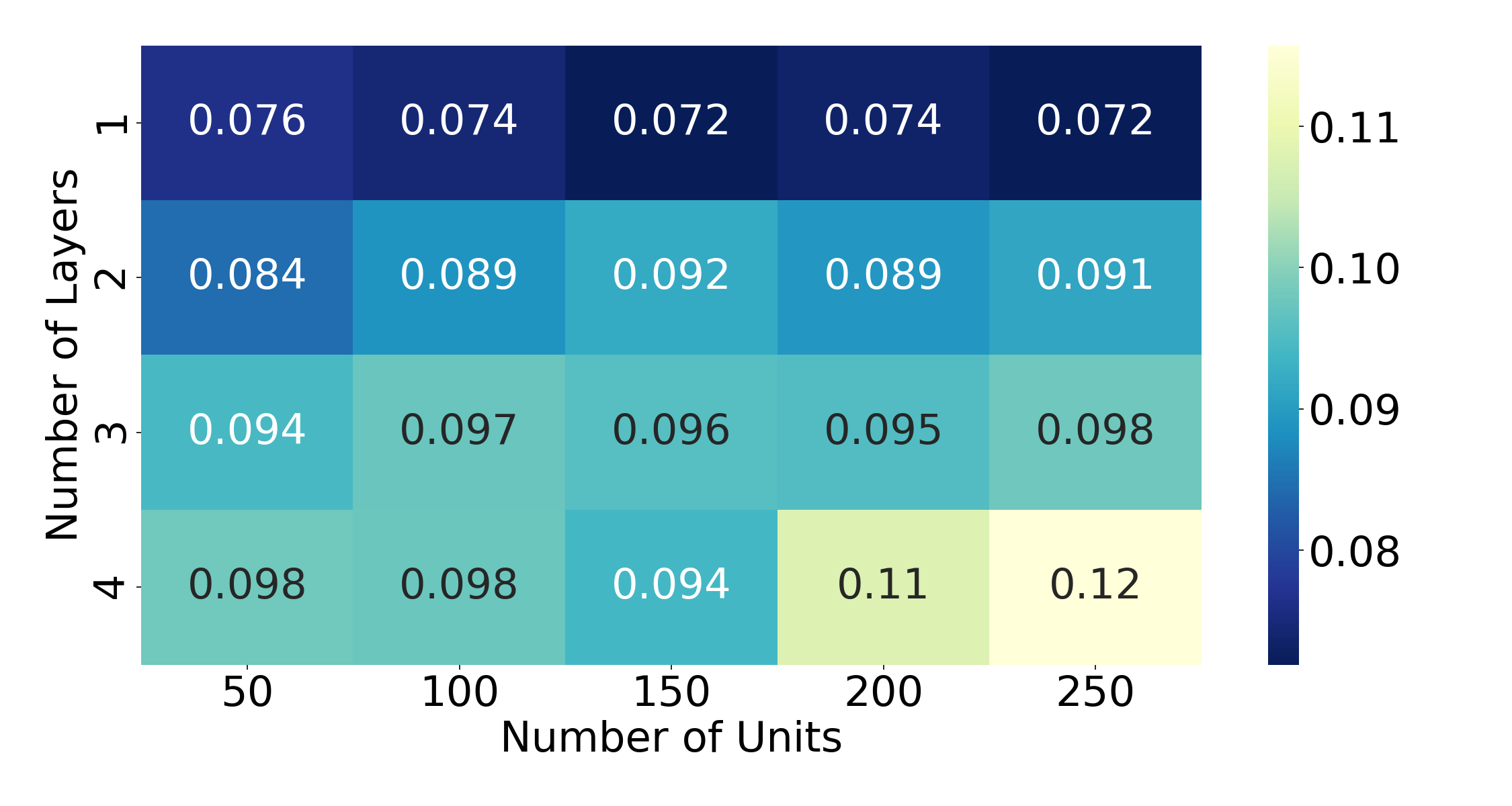}}
\subfigure[]{
\includegraphics[width=0.32\columnwidth]{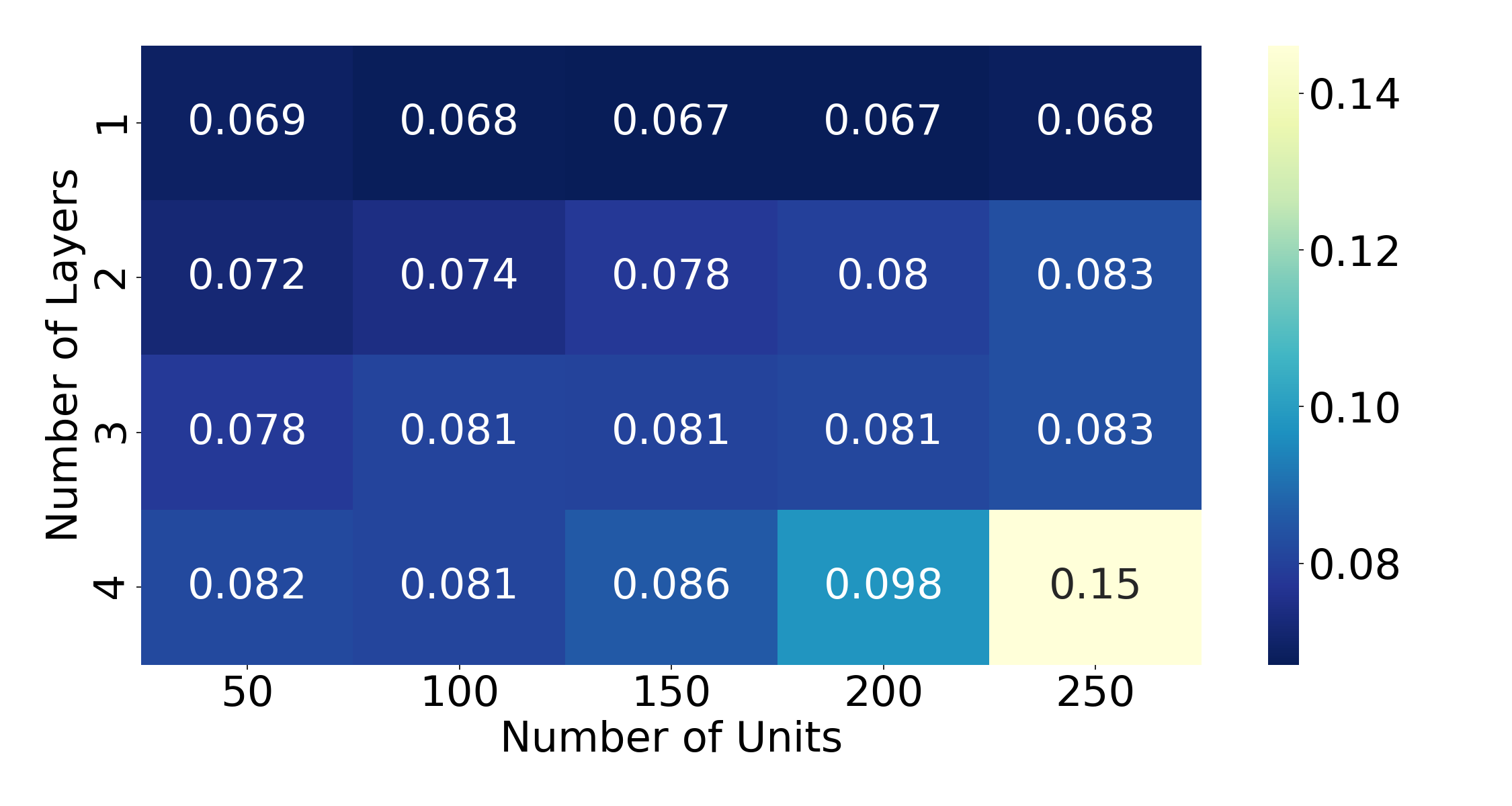}}

\caption{Average RMSE of GRU model for (a) cluster 1, (b) cluster 2, (c) cluster 3, (d) cluster 4, (e) cluster 5, (f) cluster 6, (g) cluster 7, (h) cluster 8, (i) cluster 9, (j) cluster 10, (k) cluster 11, and (l) cluster 12.}
\label{fig:gru-grid-search}
\end{figure*}

Figure \ref{fig:lstm-grid-search} and \ref{fig:gru-grid-search} present the Grid Search results for each cluster for each of the LSTM and GRU models, respectively.


For the LSTM models (Figure \ref{fig:lstm-grid-search}),  for the majority of the clusters, the best configuration had one layer and 250 units. The only exception is Cluster 12, where the best overall average RMSE was achieved using a configuration with one layer and 150 units (LSTM-12-1L-150U) and one layer with 200 units (LSTM-12-1L-200U). Cluster 1 achieved the worth average RMSE result (0.084) while Cluster 12 achieved the best average RMSE result (0.068).

For GRU (Figure \ref{fig:gru-grid-search}), the configurations with one layer presented the best average RMSE, however the performance of different clusters varied based on the number of units. From Cluster 2 to Cluster 9 (GRU-2-1L-200U, GRU-3-1L-200U, GRU-4-1L-200U, GRU-5-1L-200U, GRU-6-1L-200U, GRU-7-1L-200U, GRU-8-1L-200U, andd GRU-9-1L-200U), the configuration with the lowest average RMSE was one layer with 200 units. For Clusters 1 and 10, the configuration that provided the best average RMSE was one layer with 250 units (GRU-1-1L-250U and GRU-10-1L-250U). In Clusters 11 and 12, two configurations had the best average RMSE, all with only one layer. For Cluster 11, the best configurations had 150 units (GRU-11-1L-150U) and 250 (GRU-11-1L-250U) units while for Cluster 12 the best performing configuration had 150 units (GRU-12-1L-150U) and 200 units (GRU-12-1L-250U). Similar to LSTM, the clusters with the worst and best average RMSE were Cluster 1 (0.0091) and Cluster 12 (0.0045), respectively.

The complexity of the model is directly related to the number of layers and units i.e. the more layers and units, the more complex the model becomes. Model complexity has to be adjusted according to the data. Figures \ref{fig:lstm-grid-search} and \ref{fig:gru-grid-search} illustrate that very complex models (those with many layers and units) resulted in poorer performance, due to model overfitting. Simpler models with less complexity also performed poorly, most likely due to model underfitting. 

In general, the DL models with one hidden layer obtained better average RMSE results than models with more layers; while those with 150 units or more resulted in better average RMSE results. Fine-tuning the models by increasing the number of units rather than layers resulted in better performance. Adding hidden layers resulted in performance degradation.

\subsection{Statistical analysis}
Some configurations achieved by the Grid Search obtained very similar average RMSE. To explore this further, we used Kruskal-Wallis non-parametric analysis to compare independent samples to check whether they are similar or not, based on the mean ranks of these samples \cite{elliott2011sas}.

\begin{figure}[h!]
\centering
\includegraphics[width=0.6\columnwidth]{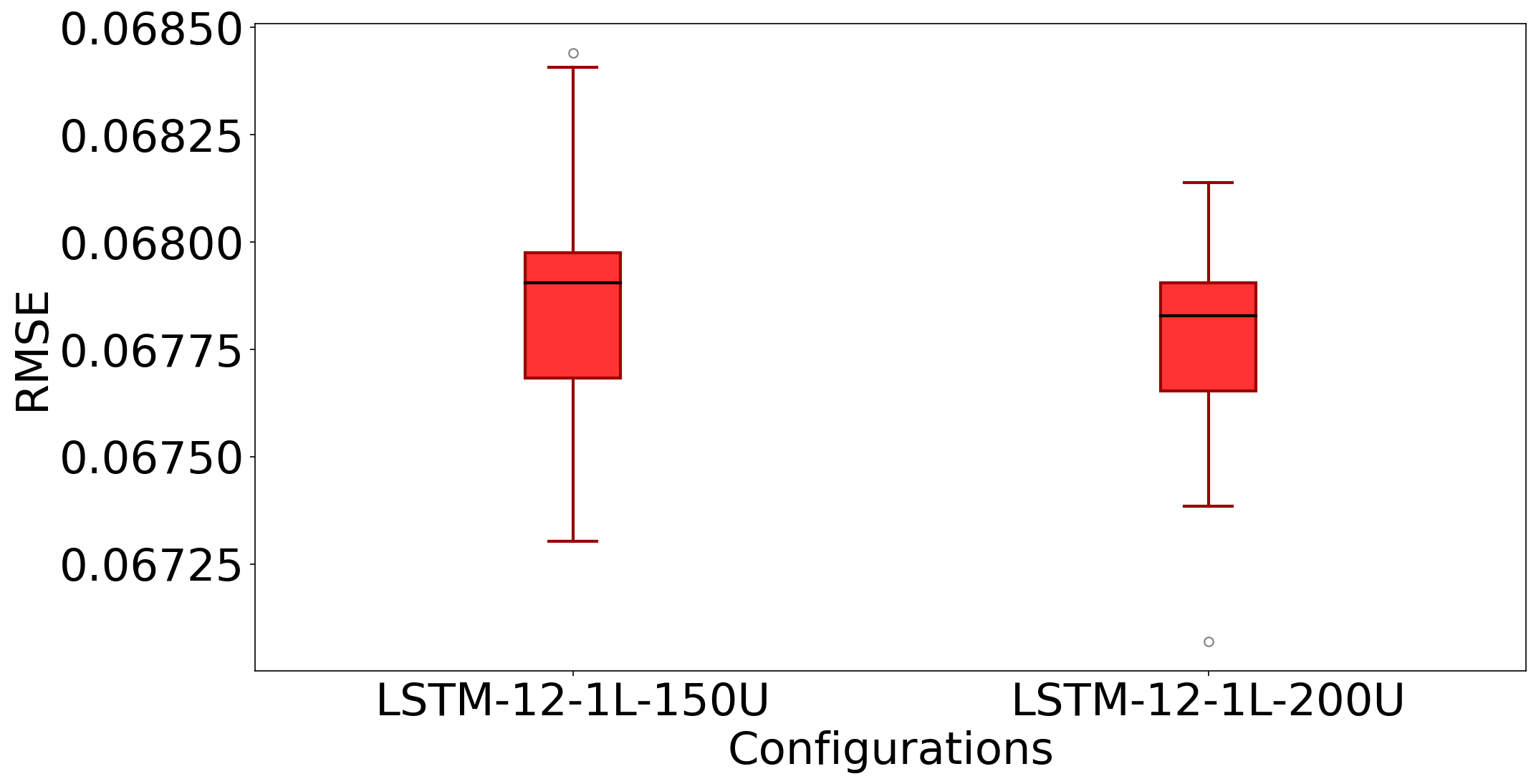}
\caption{Box plot of the RMSE of the LSTM best configurations of cluster 11}
\label{fig:boxplot_cluster11_lstm}
\end{figure}

In our LSTM results, Cluster 12 (LSTM-12-1L-150U and LSTM-12-1L-200U) have two configurations with the same average RMSE. Figure \ref{fig:boxplot_cluster11_lstm} presents the box plot of the RMSE of these configurations of Cluster 12.

While the best configurations of Cluster 12 have the same average RMSE (0.068), they have different RMSE distributions. We can note that the LSTM-12-1L-200U has a lower dispersion and a lower median than LSTM-12-1L-150U. LSTM-12-1L-200U has an outlier below the minimum RMSE value, and LSTM-12-1L-150U has an outlier above of the maximum RMSE value. This analysis suggests that LSTM-12-1L-200U is the best configuration since it has the lowest dispersion and the lowest median.

\begin{figure}[t]
\centering
\subfigure[]{
\includegraphics[width=0.32\columnwidth]{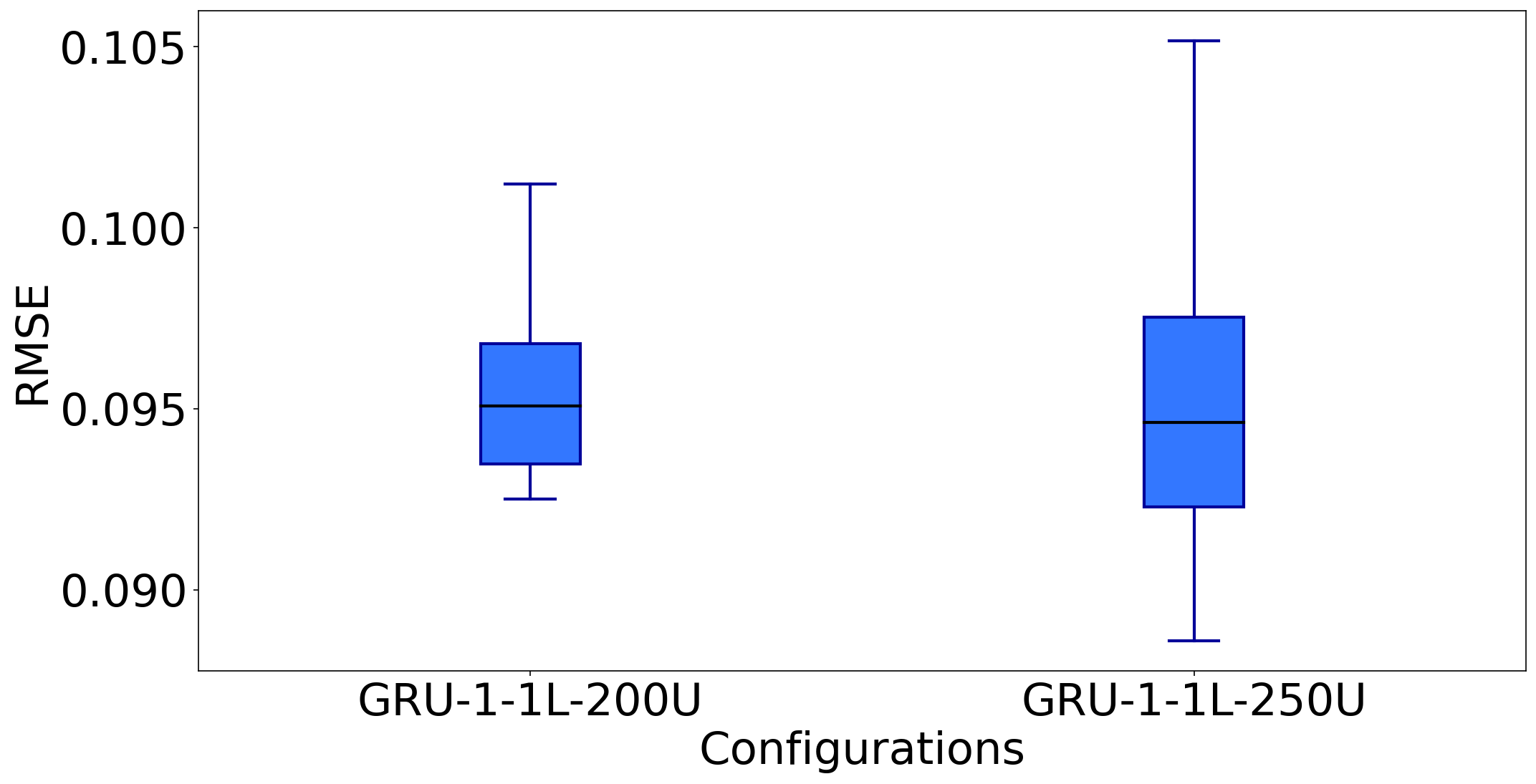}}
\subfigure[]{
\includegraphics[width=0.32\columnwidth]{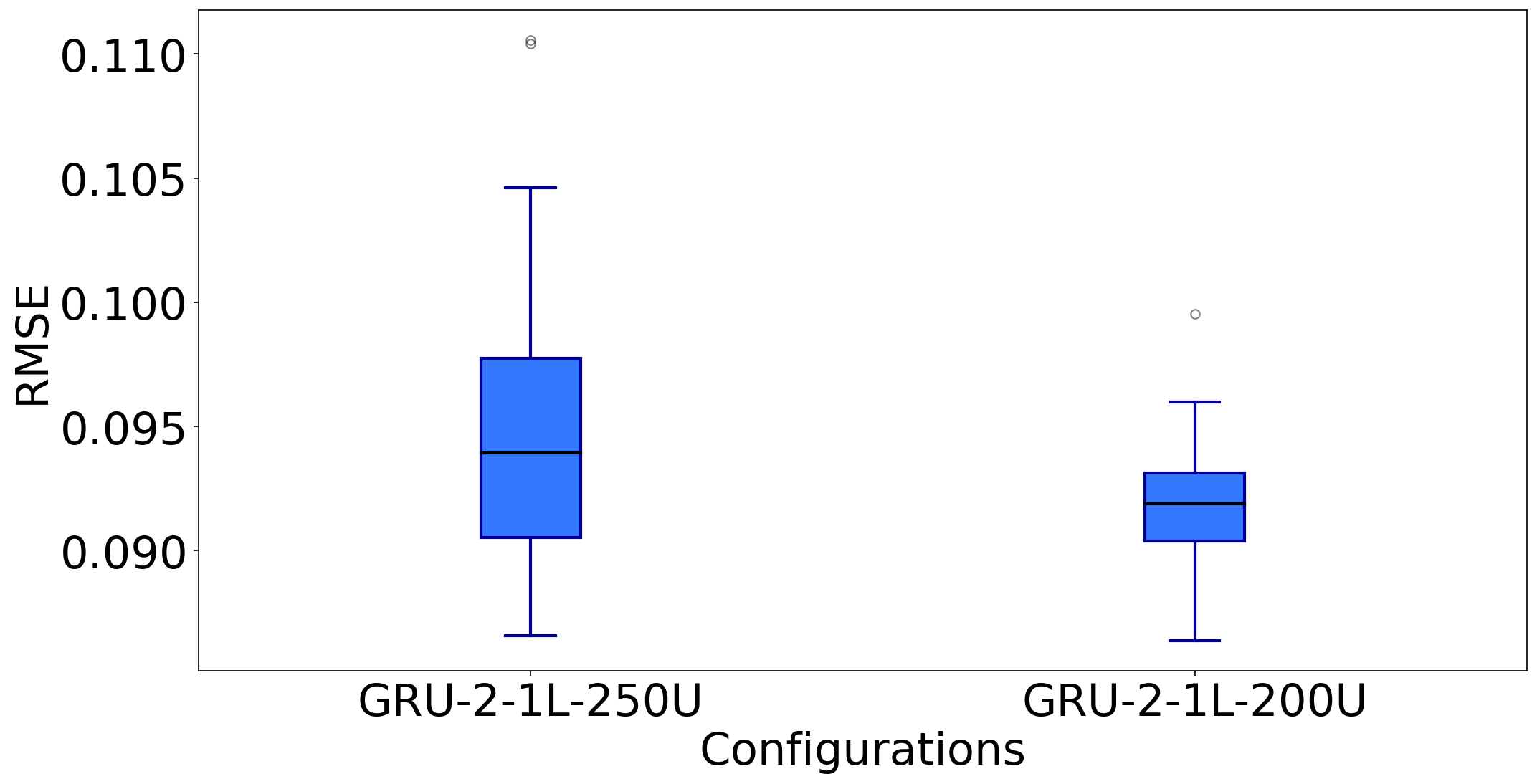}}
\subfigure[]{
\includegraphics[width=0.32\columnwidth]{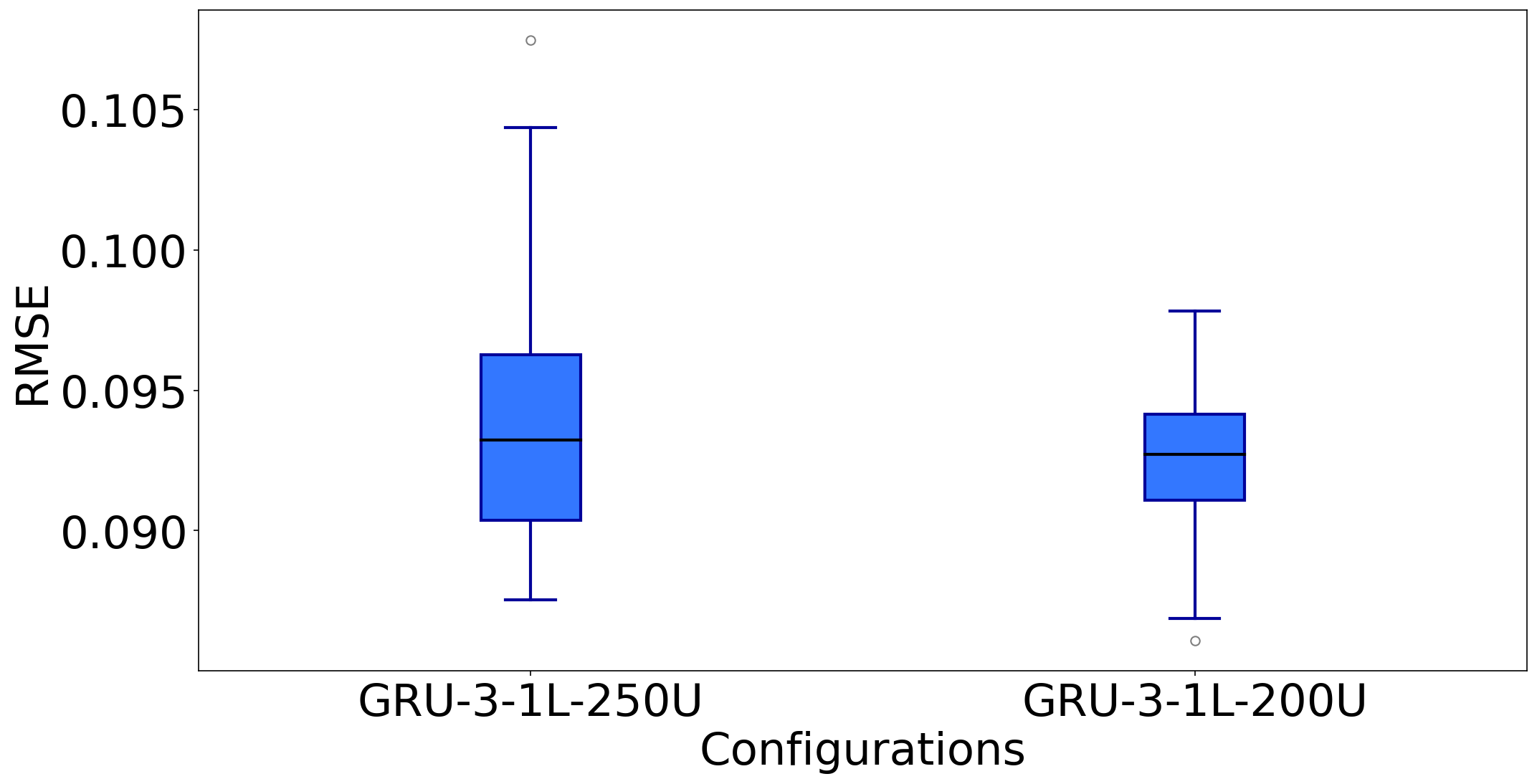}}
\subfigure[]{
\includegraphics[width=0.32\columnwidth]{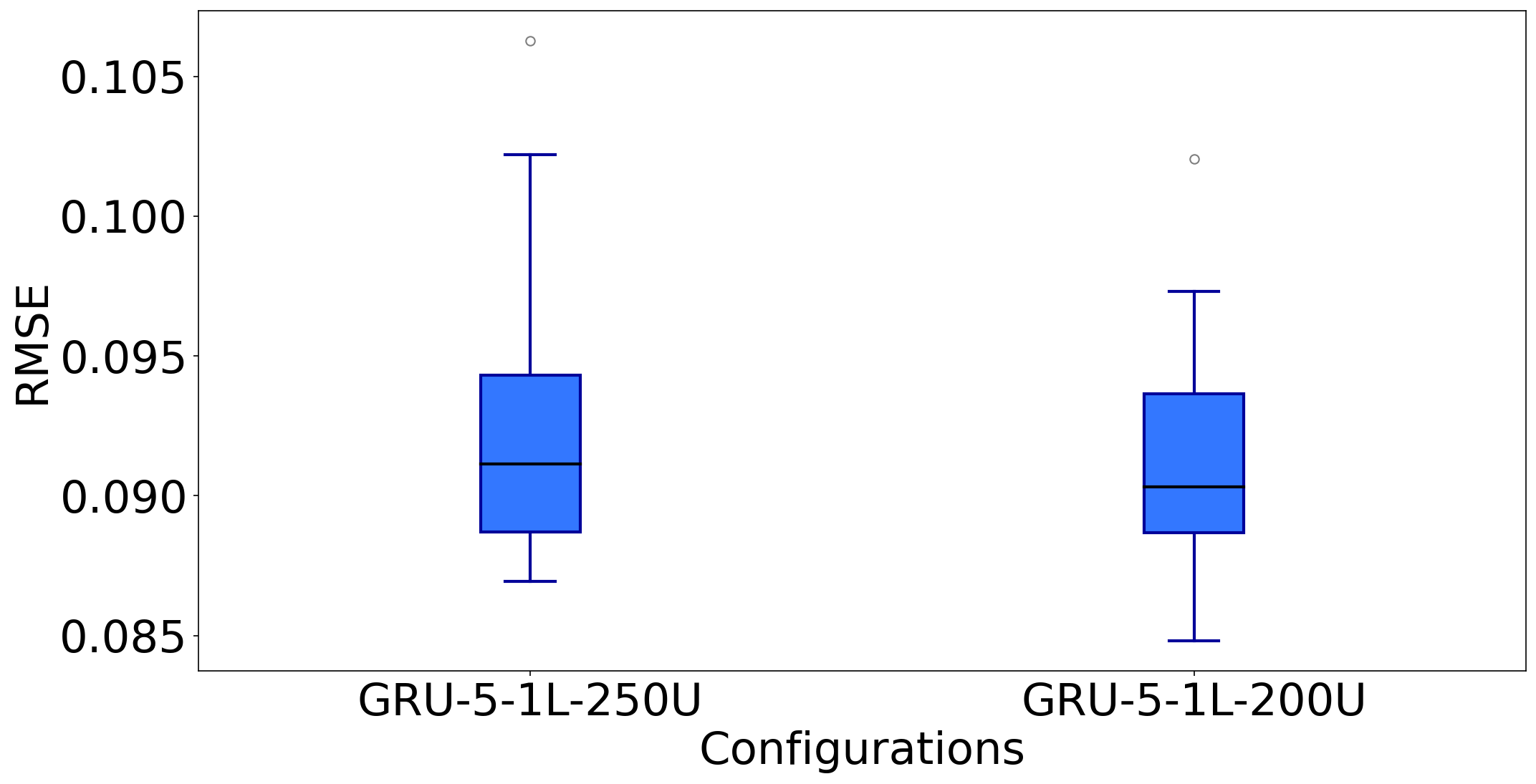}}
\subfigure[]{
\includegraphics[width=0.32\columnwidth]{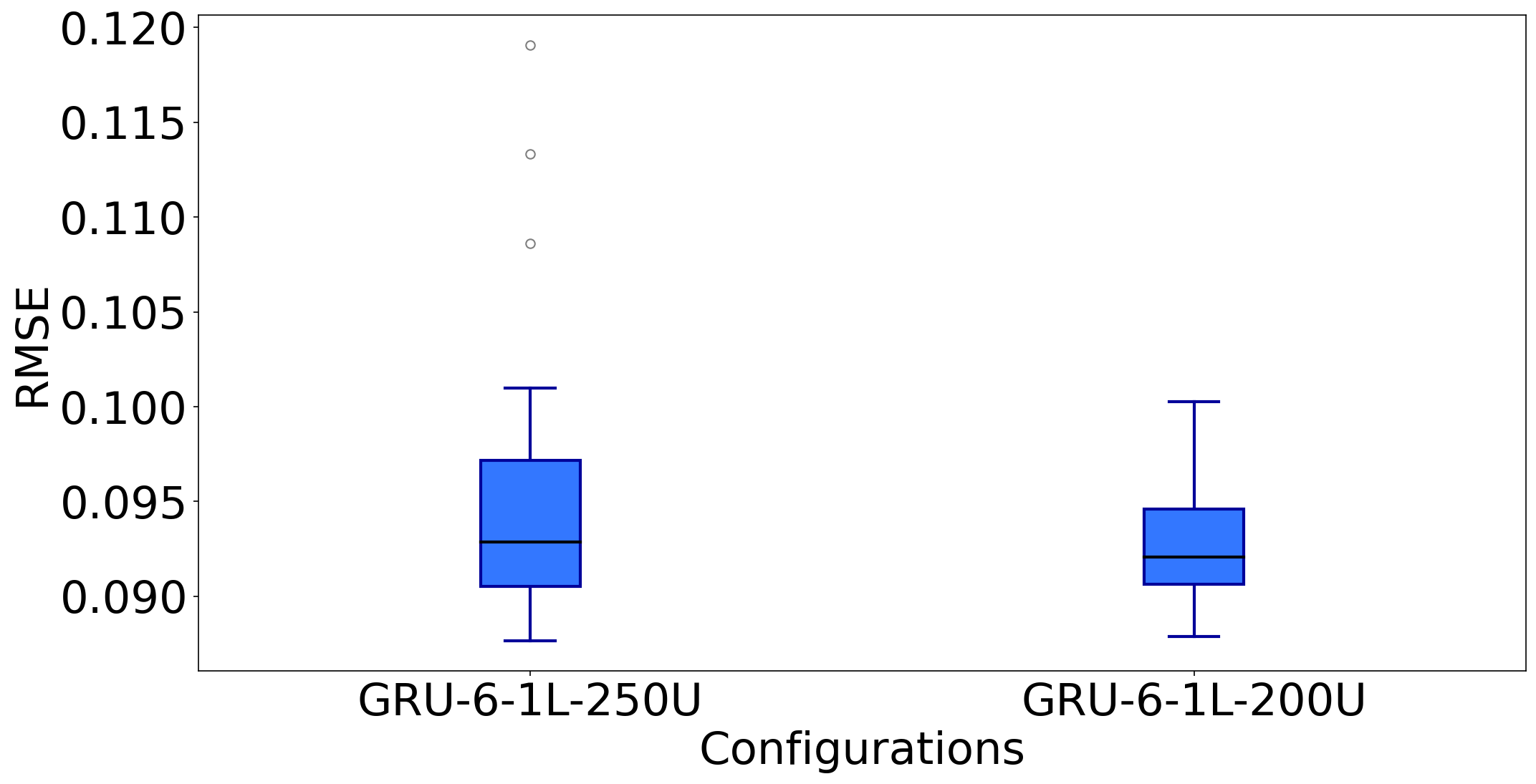}}
\subfigure[]{
\includegraphics[width=0.32\columnwidth]{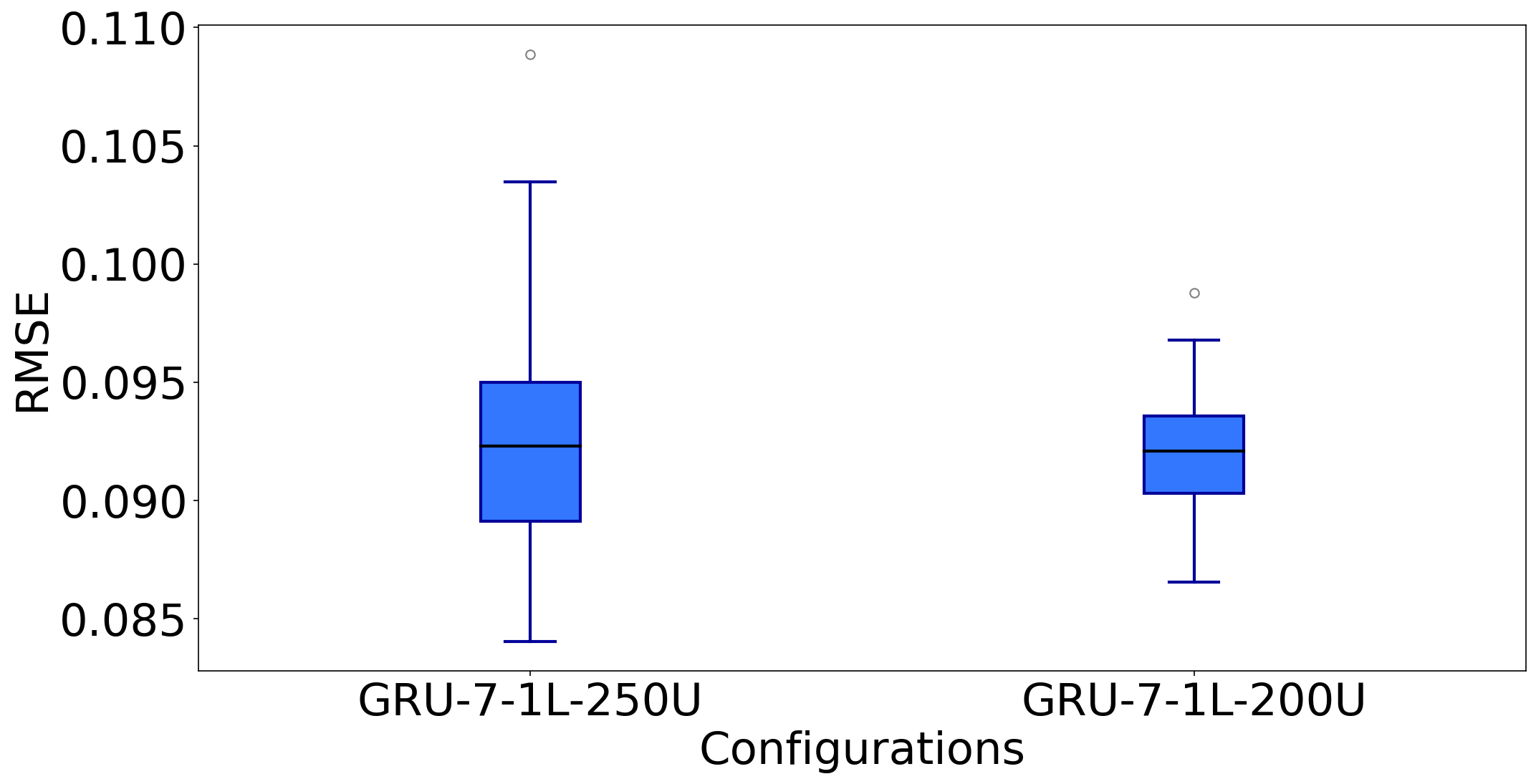}}
\subfigure[]{
\includegraphics[width=0.32\columnwidth]{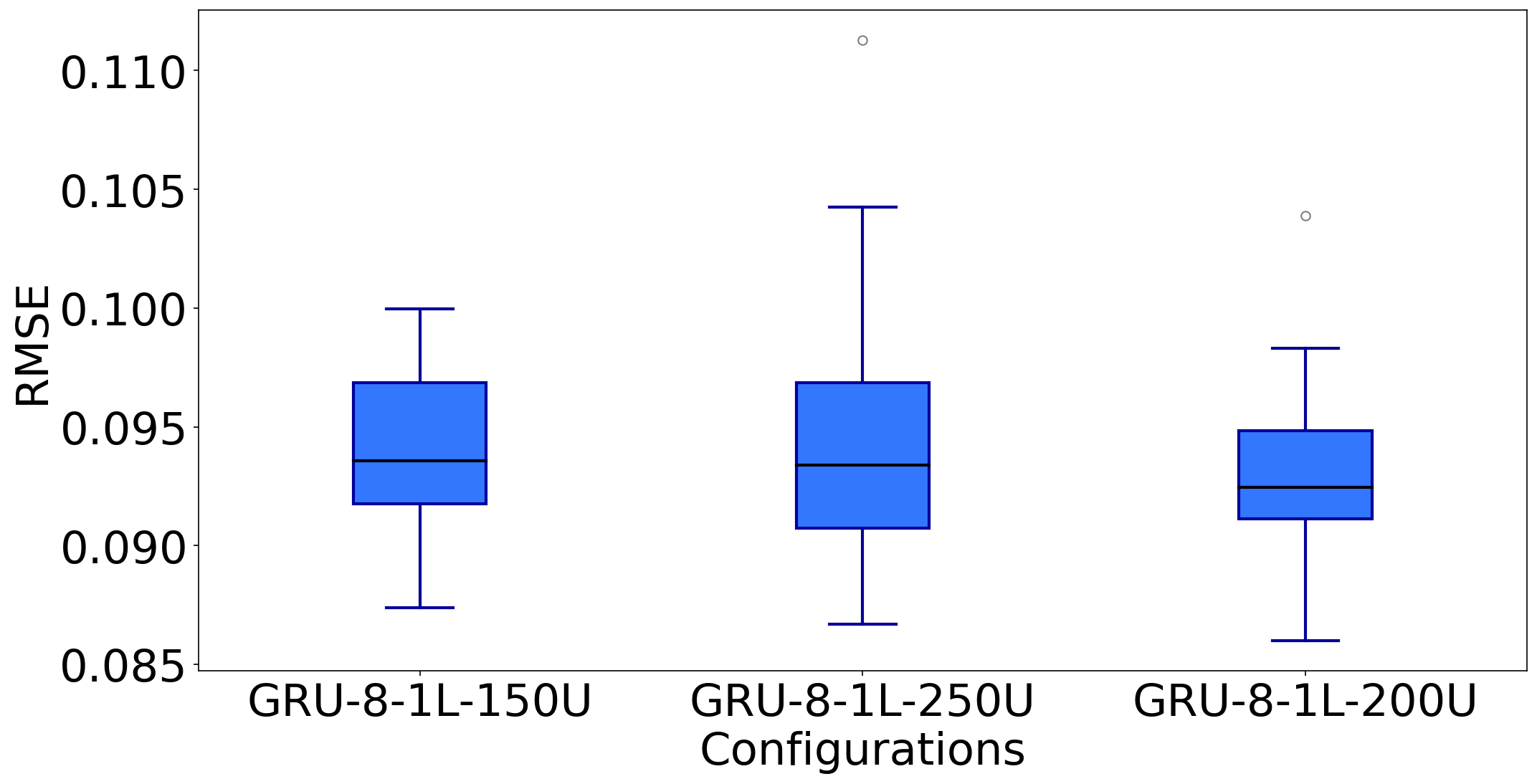}}
\subfigure[]{
\includegraphics[width=0.32\columnwidth]{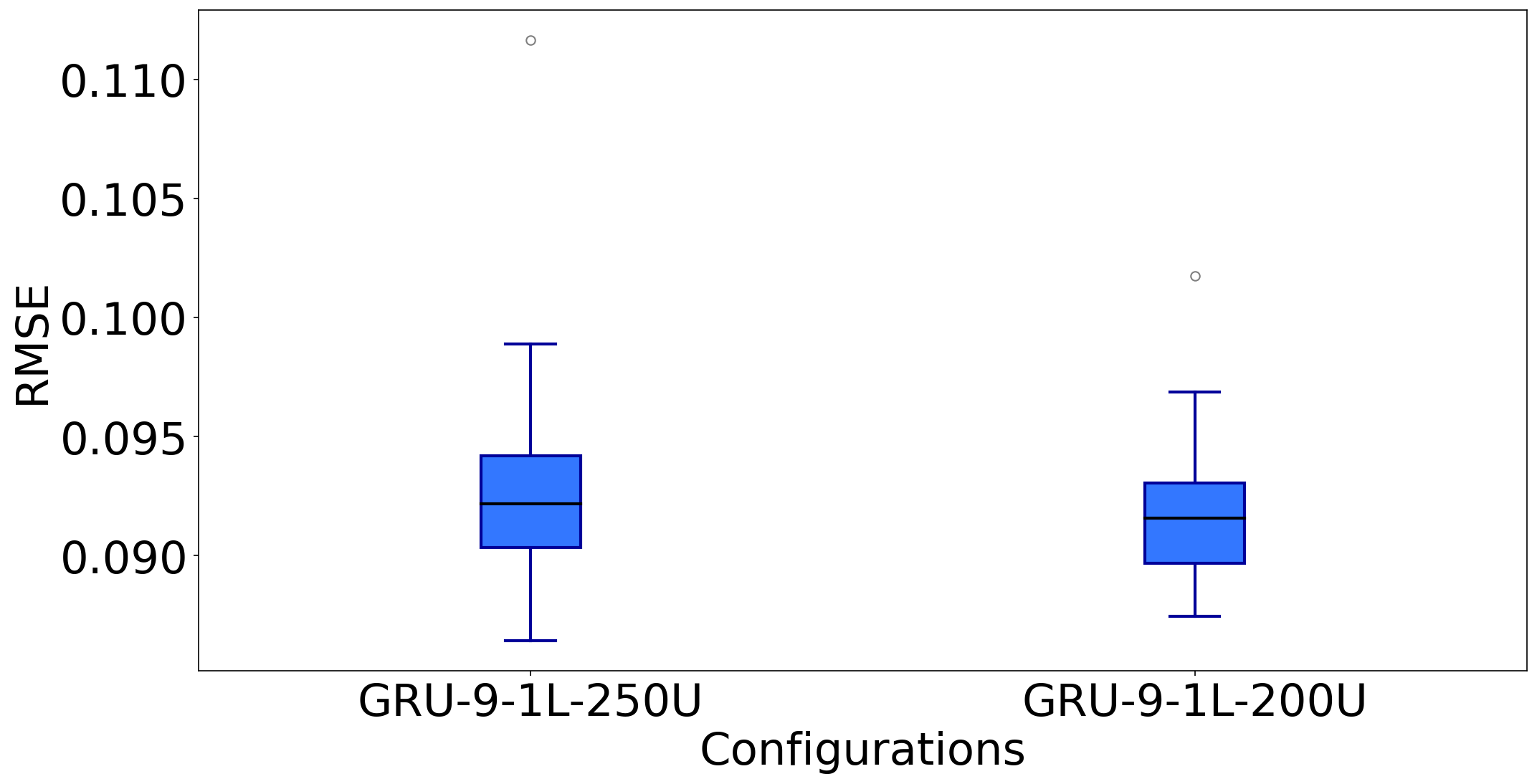}}
\caption{Boxplot of the RMSE of the best configurations of (a) cluster 1, (b) cluster 2, (c) cluster 3, (d) cluster 5, (e) cluster 6, (f) cluster 7, (g) cluster 8, and (h) cluster 9.}
\label{fig:boxplot_gru}
\end{figure}

For the GRU models, Clusters 1-3 and 5-9 each had at least one statistically similar best configuration (Figure \ref{fig:boxplot_gru}). For Clusters 1-3 and 5-7, the configurations with 250 units (GRU-1-1L-250, GRU-2-1L-250, GRU-3-1L-250, GRU-5-1L-250, GRU-6-1L-250 and GRU-7-1L-250) presented a higher dispersion than the configurations with 200 units (GRU-1-1L-200, GRU-2-1L-200, GRU-3-1L-200, GRU-5-1L-200, GRU-6-1L-200 and GRU-7-1L-200); GRU-1-1L-200 presented a higher median than GRU-1-1L-250. For Cluster 8, three models are statistically similar, those with one layer and 150, 200, and 250 units. Again, the configuration with 200 units presented lower dispersion and a lower median than the other configurations. Finally, both configurations of Cluster 9 (GRU-9-1L-200 and GRU-1-1L-250) had very similar distributions, with similar dispersion and a similar median. 

In general, for these clusters with statistically similar configurations, the configuration with one layer and 200 units presented a lower dispersion and lower median; this is considered the best performing configuration for the GRU models.

\subsection{LSTM vs. GRU comparison}
\label{subsec:lst-gru-comparison}

\begin{figure*}[]
\centering
\subfigure[]{
\includegraphics[width=0.32\columnwidth]{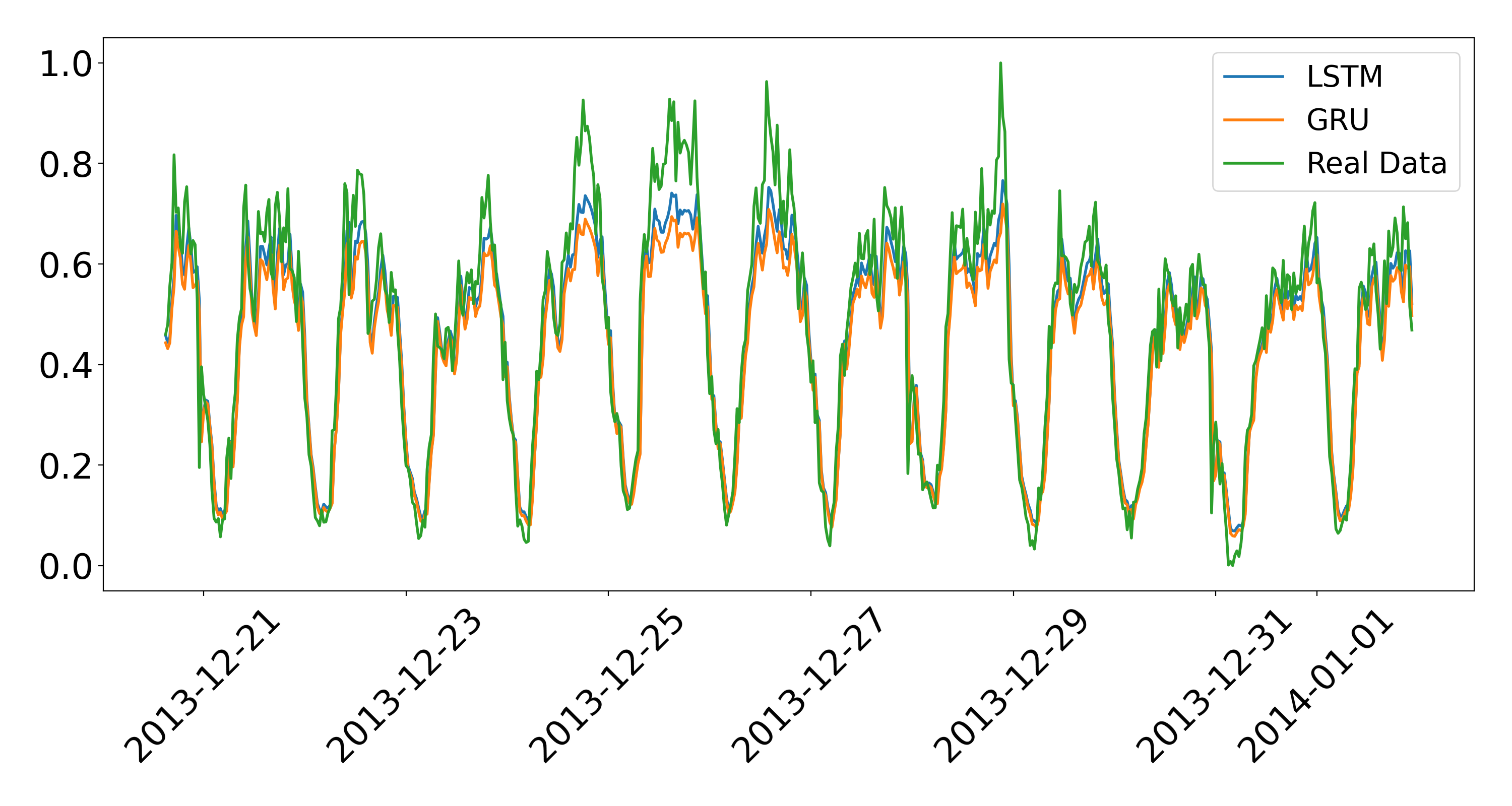}}
\subfigure[]{
\includegraphics[width=0.32\columnwidth]{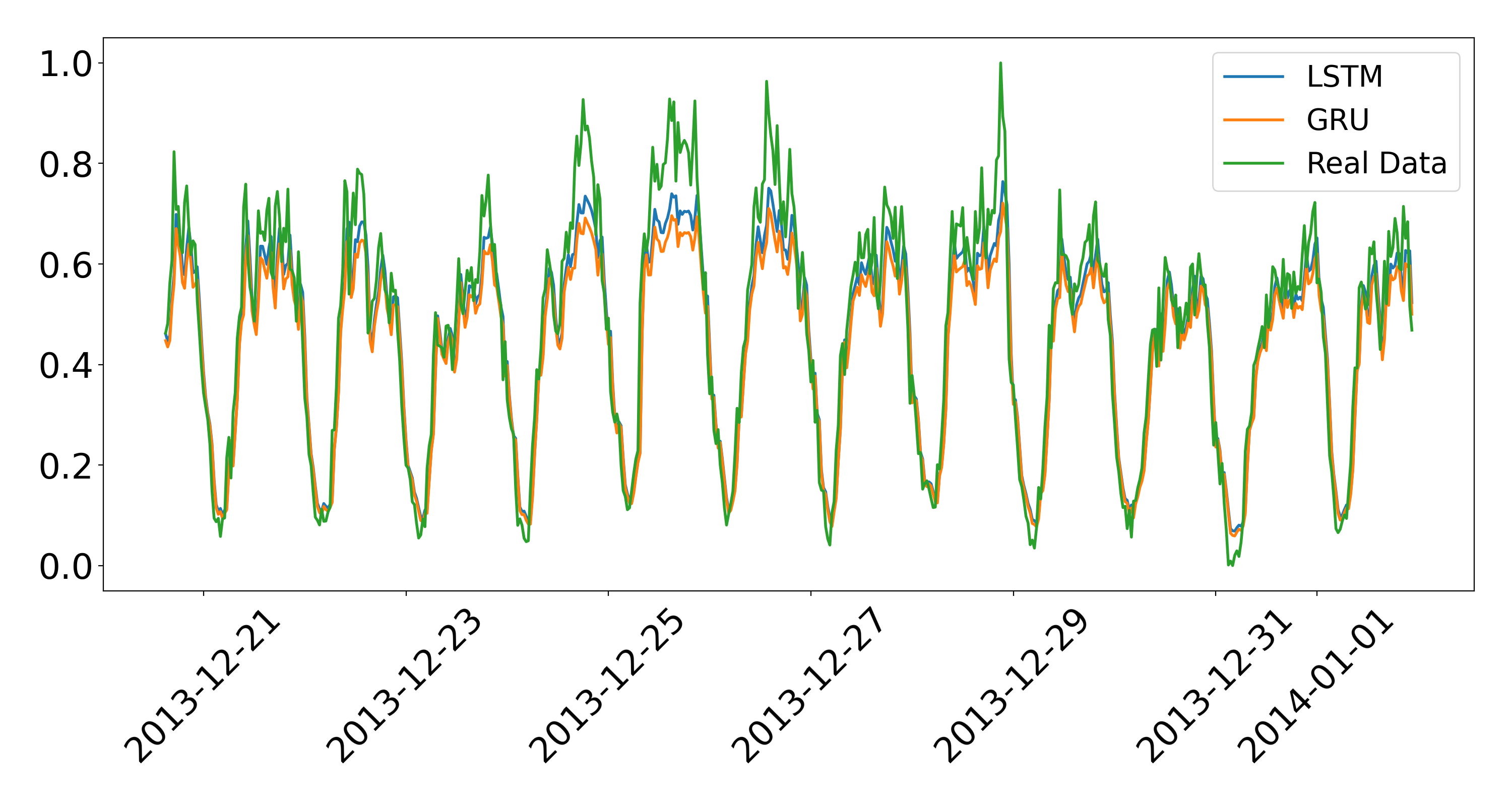}}
\subfigure[]{
\includegraphics[width=0.32\columnwidth]{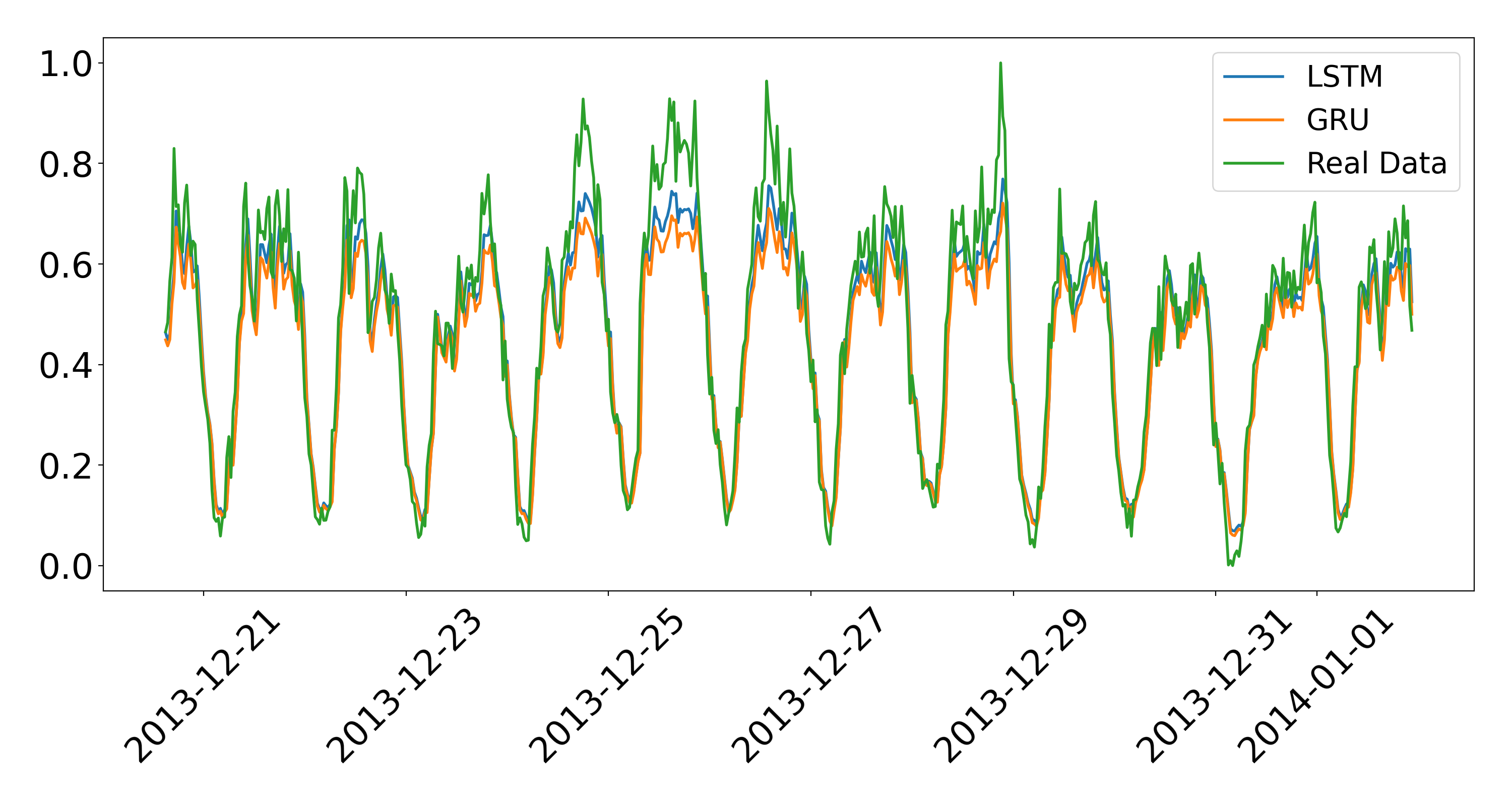}}

\subfigure[]{
\includegraphics[width=0.32\columnwidth]{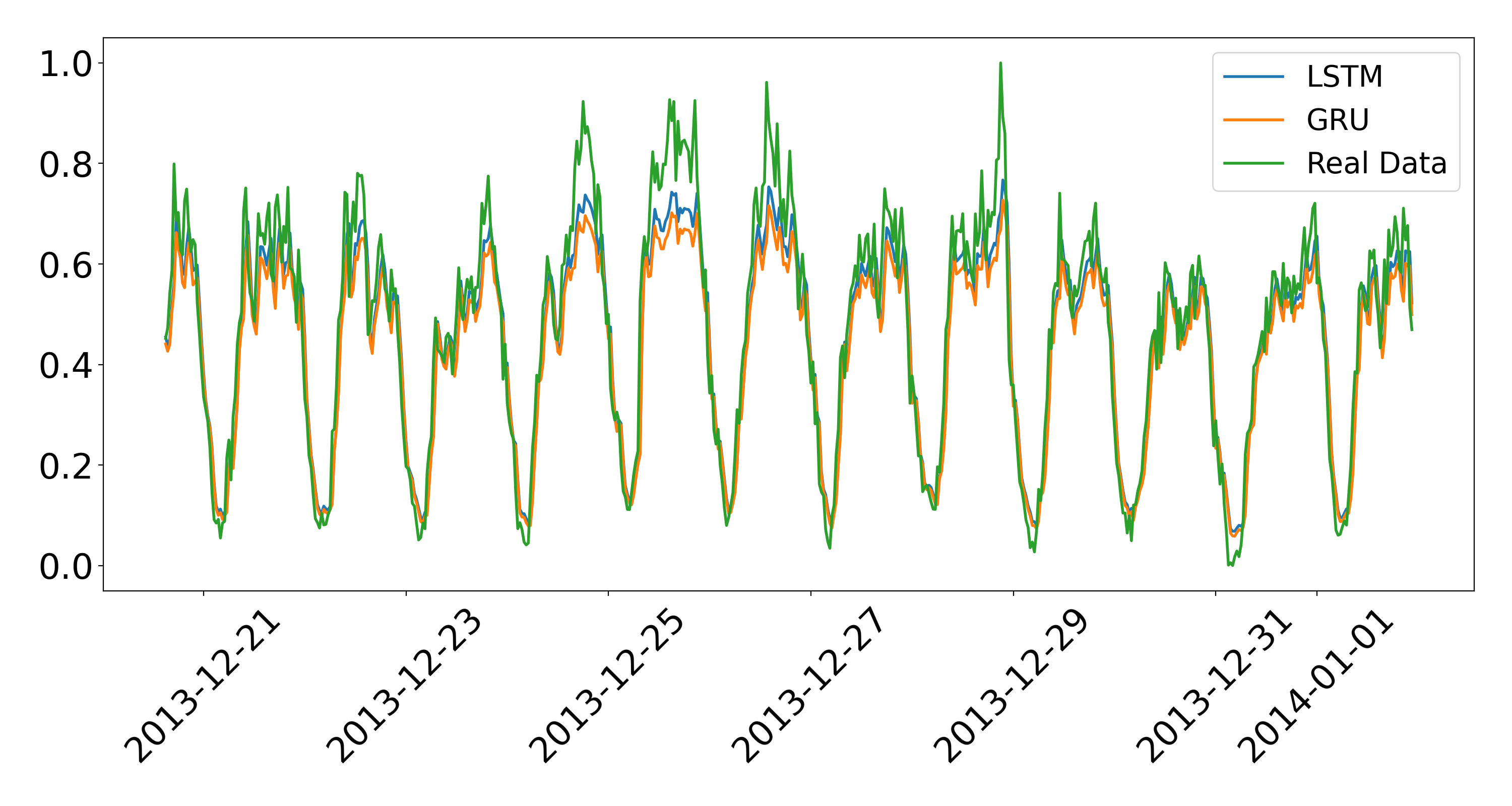}}
\subfigure[]{
\includegraphics[width=0.32\columnwidth]{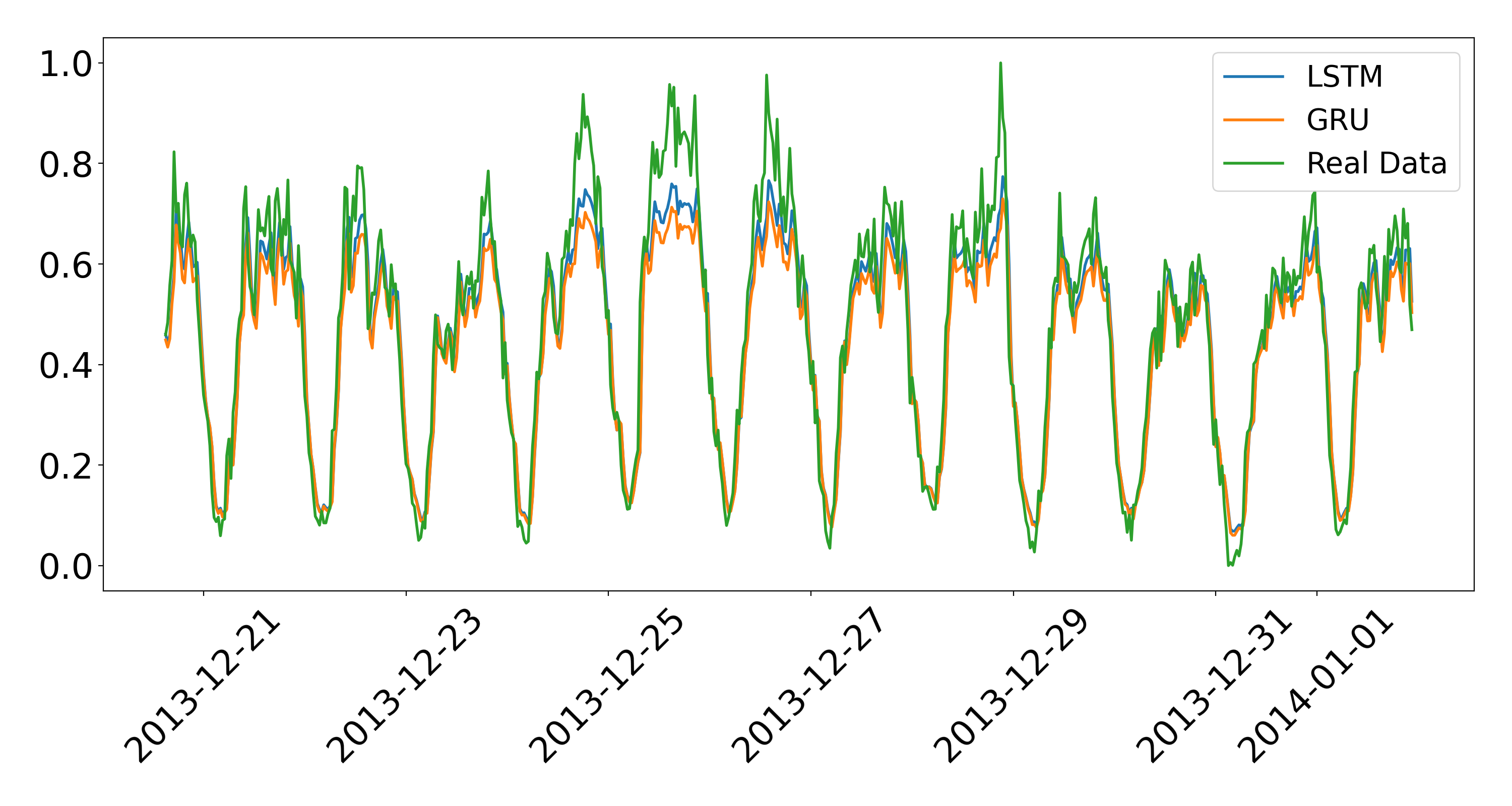}}
\subfigure[]{
\includegraphics[width=0.32\columnwidth]{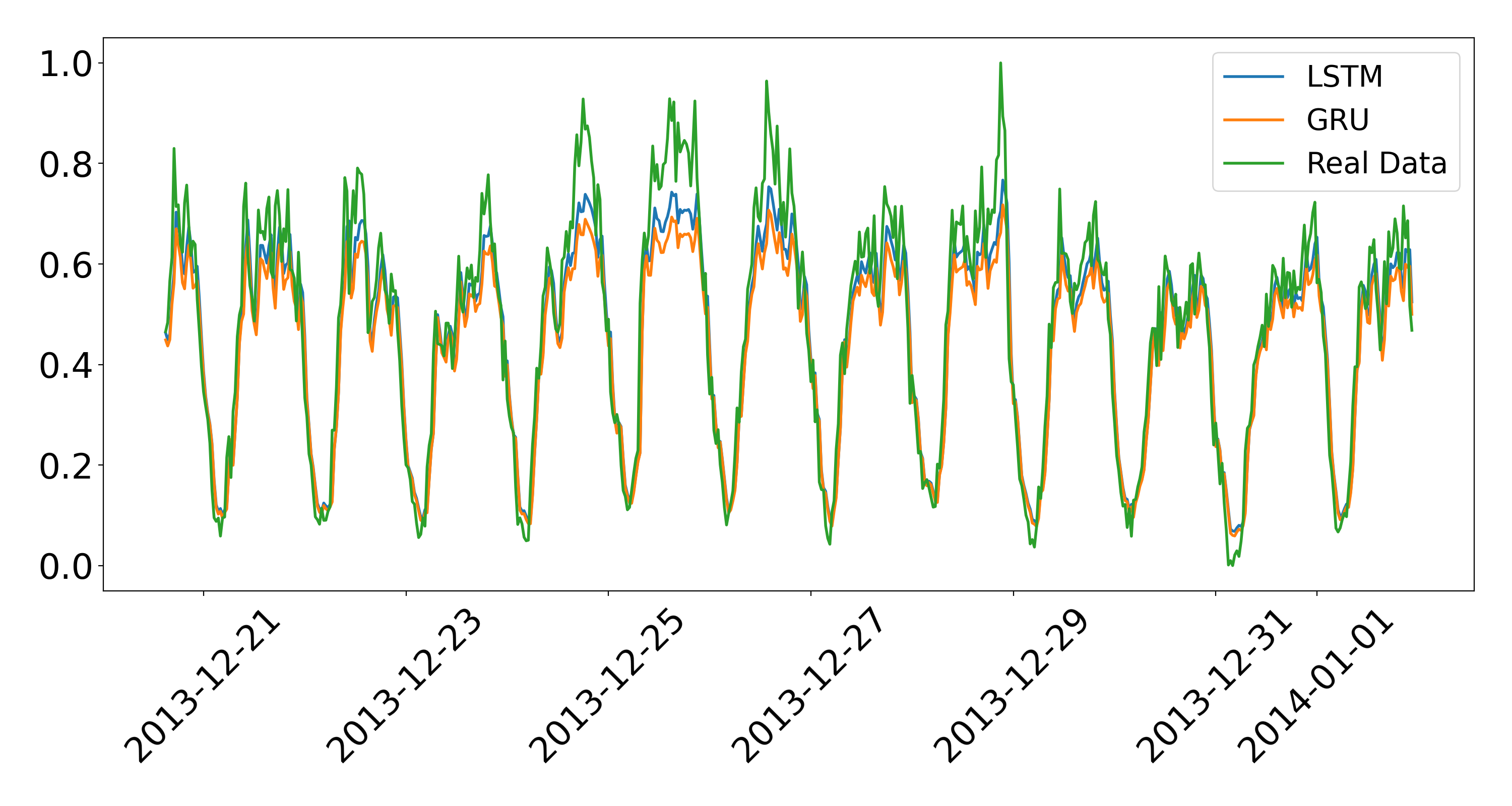}}

\subfigure[]{
\includegraphics[width=0.32\columnwidth]{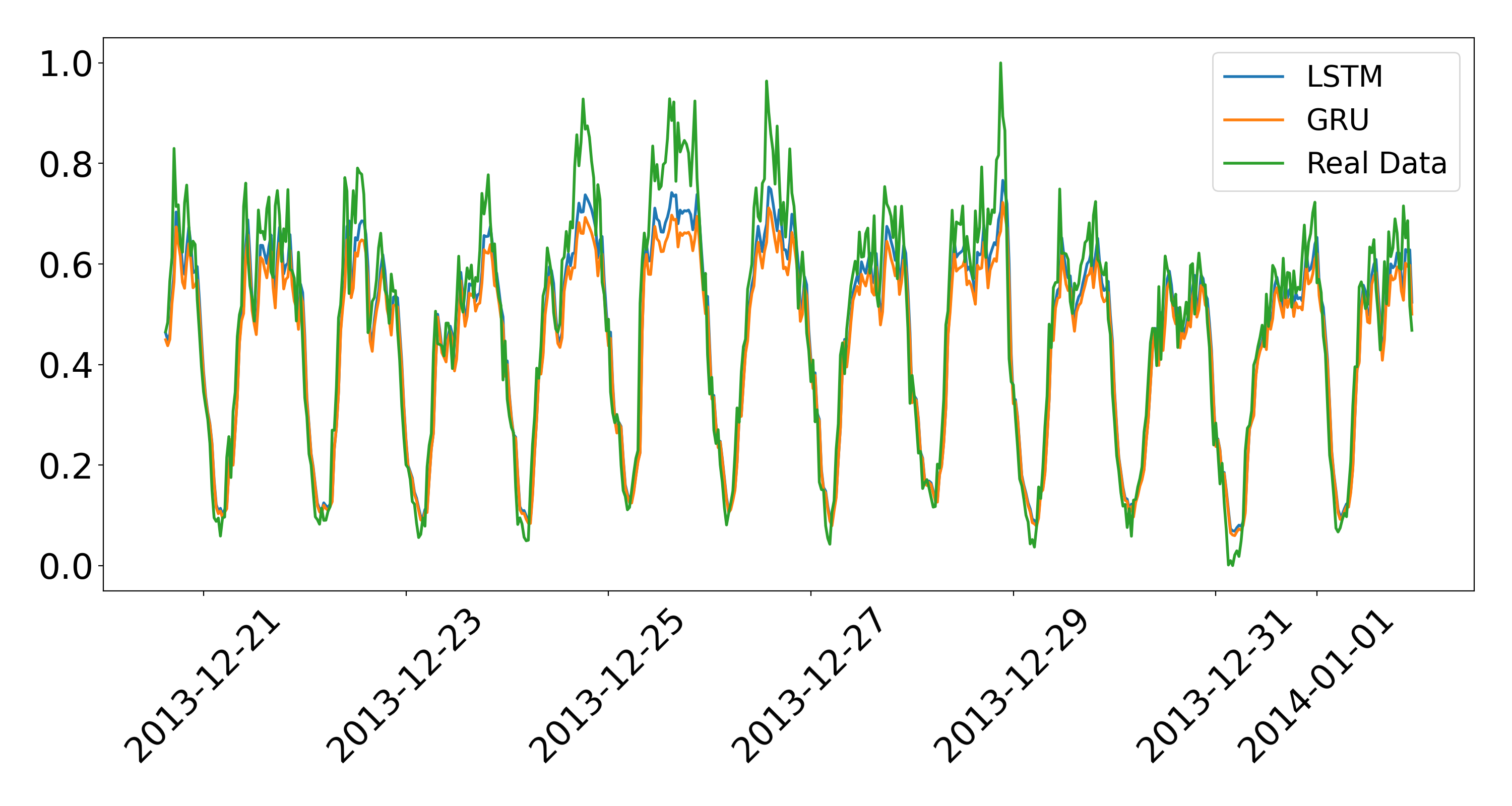}}
\subfigure[]{
\includegraphics[width=0.32\columnwidth]{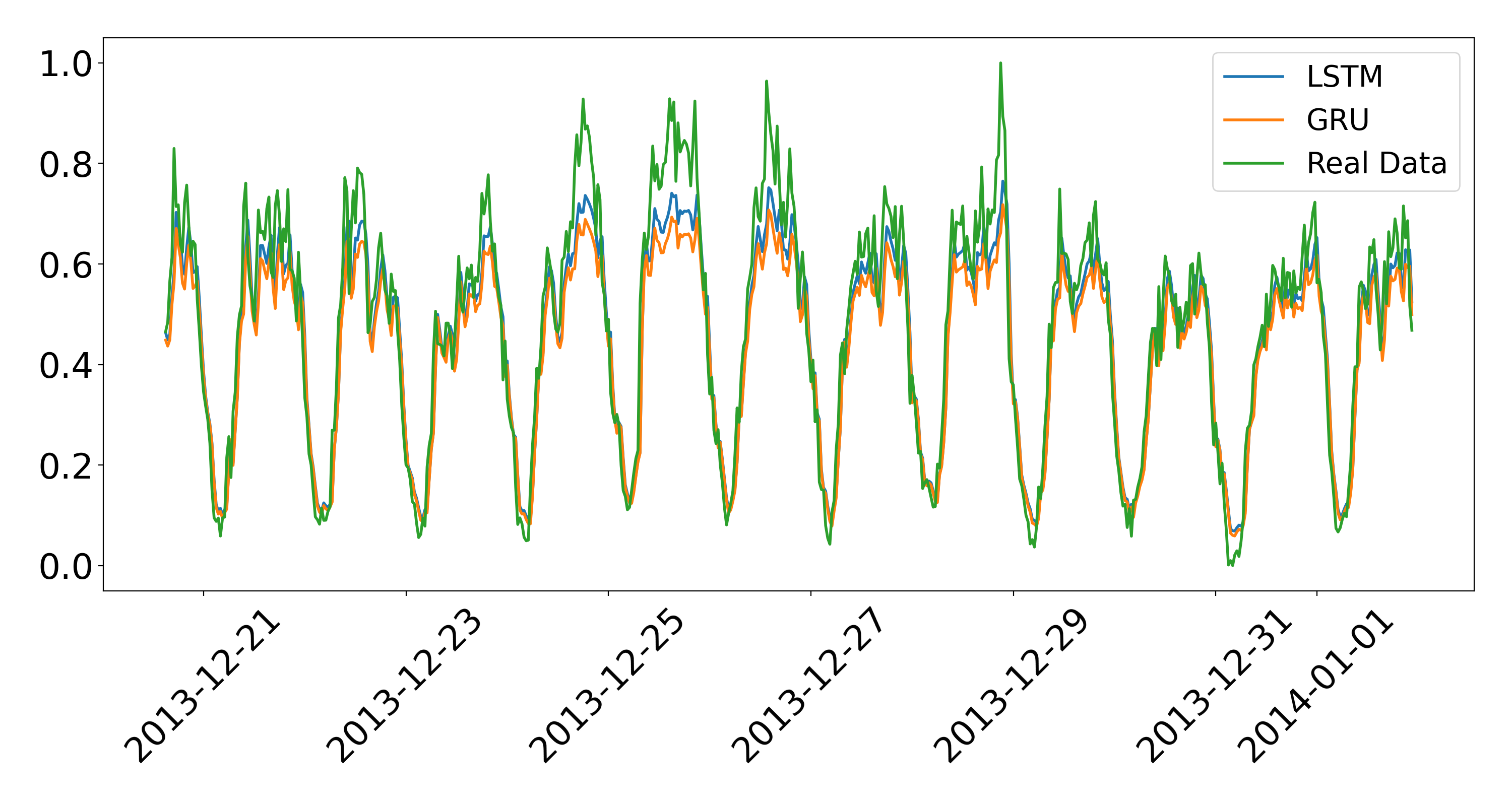}}
\subfigure[]{
\includegraphics[width=0.32\columnwidth]{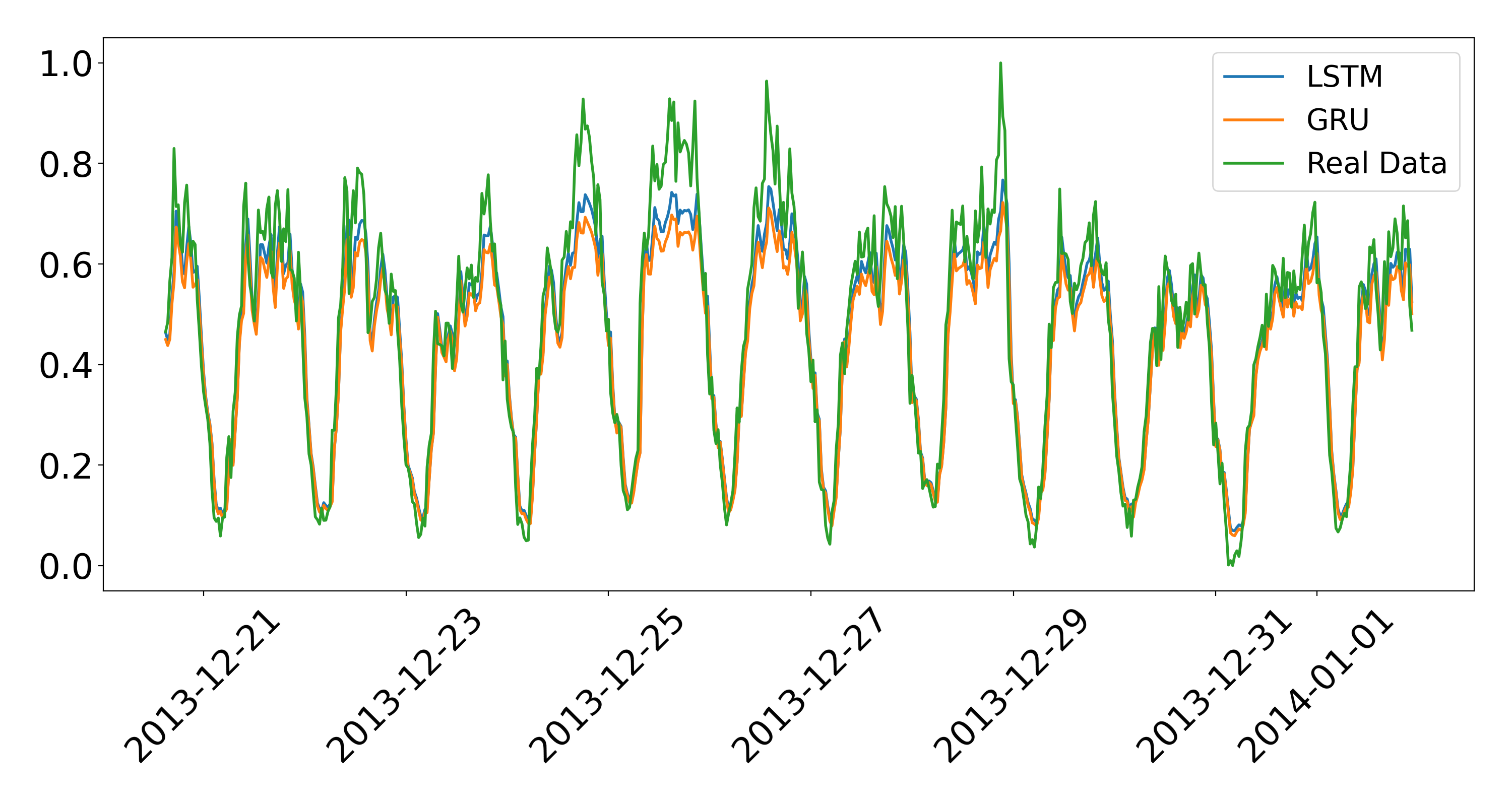}}

\subfigure[]{
\includegraphics[width=0.32\columnwidth]{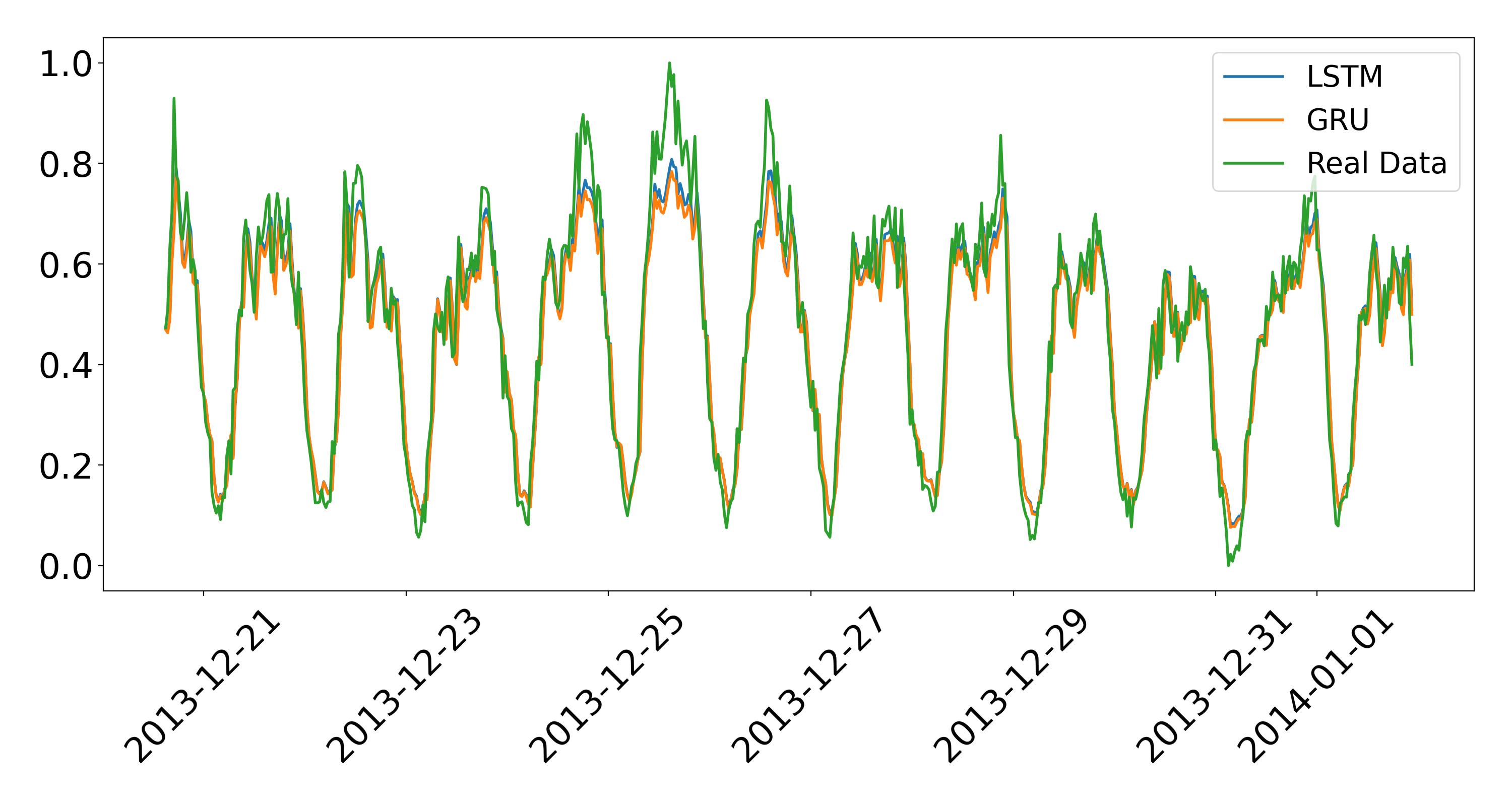}}
\subfigure[]{
\includegraphics[width=0.32\columnwidth]{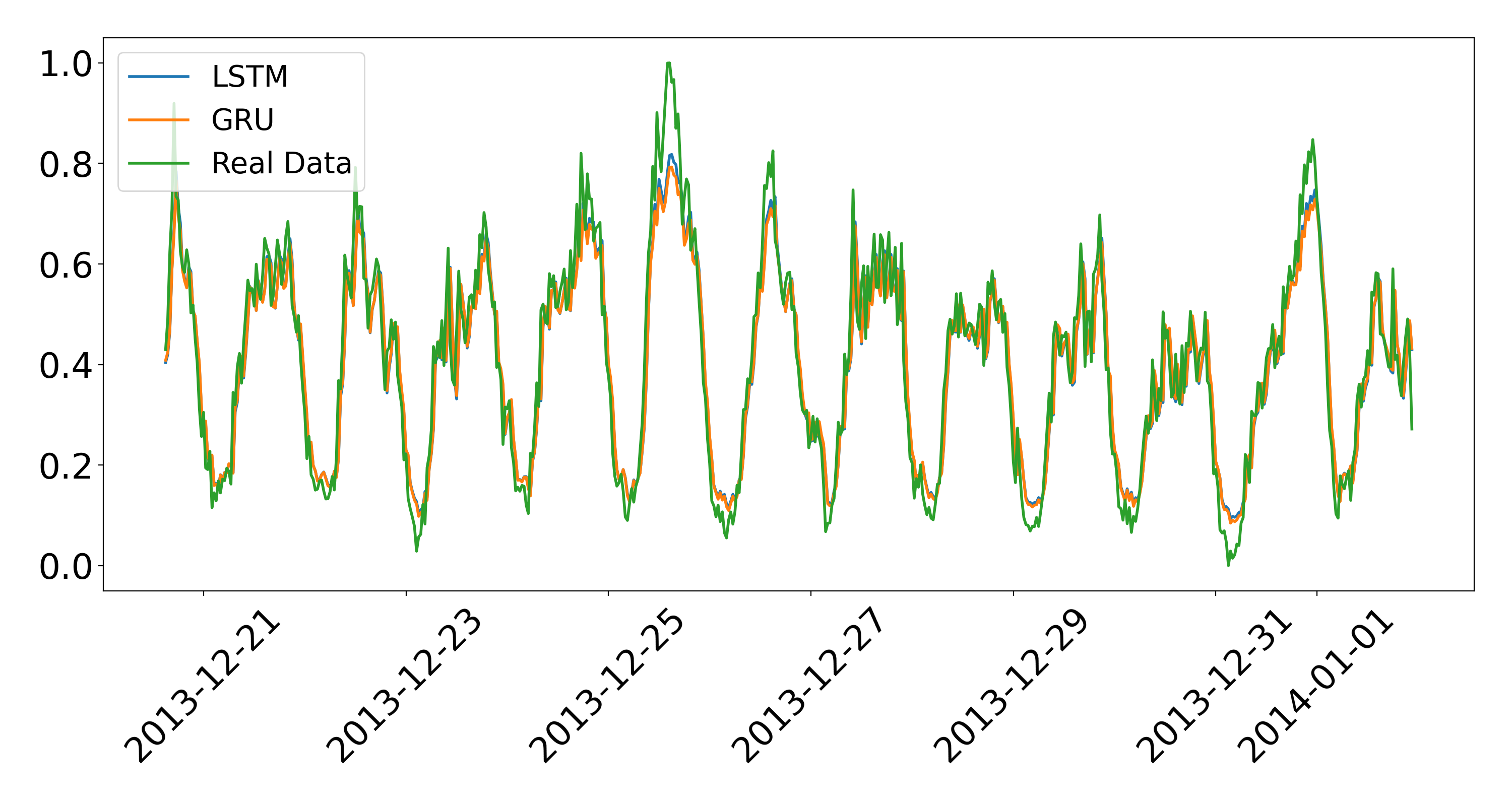}}
\subfigure[]{
\includegraphics[width=0.32\columnwidth]{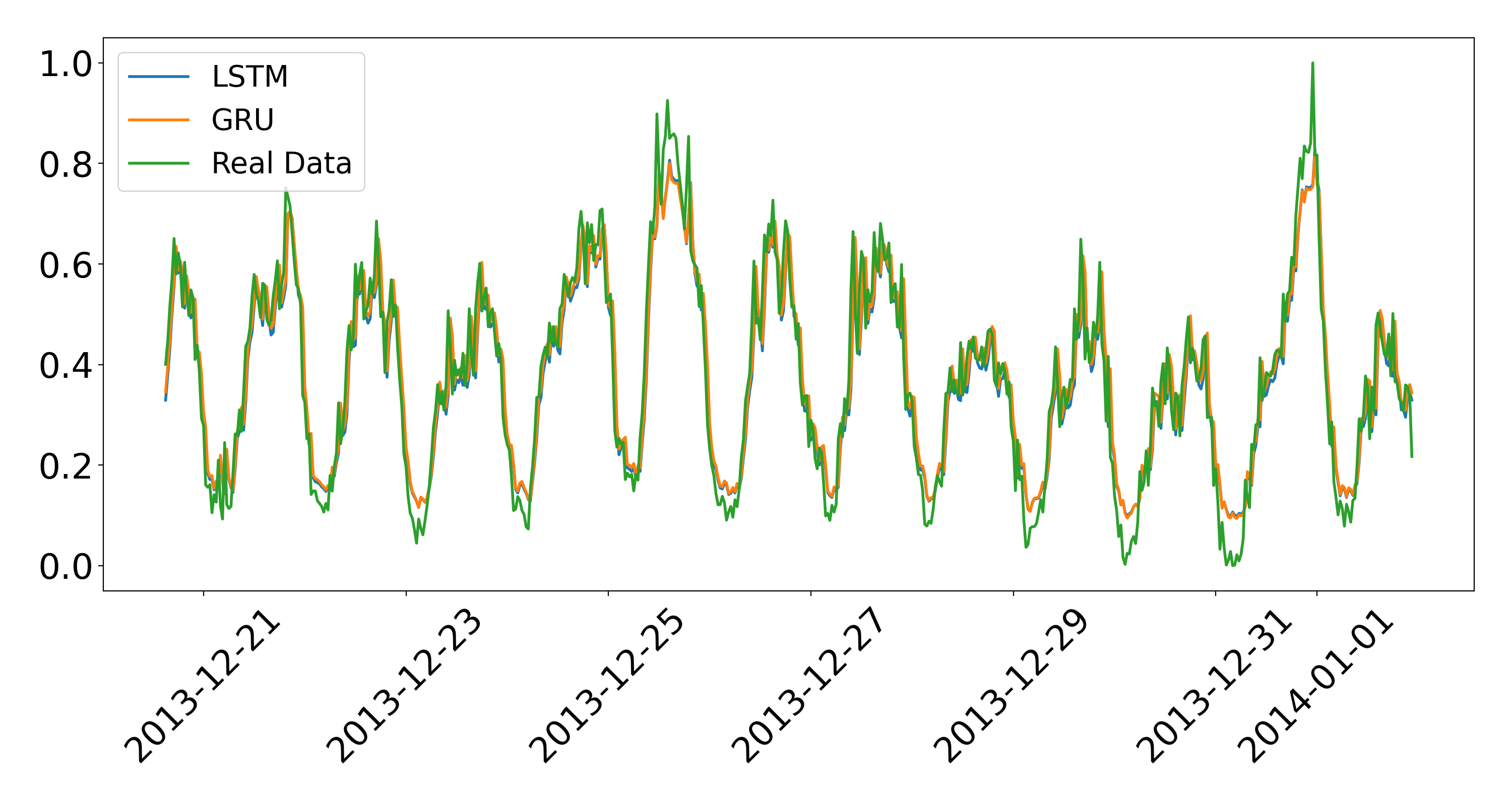}}

\caption{Comparison against ground truth Internet activity and the predictions of LSTM and GRU models for (a) cluster 1, (b) cluster 2, (c) cluster 3, (d) cluster 4, (e) cluster 5, (f) cluster 6, (g) cluster 7, (h) cluster 8, (i) cluster 9, (j) cluster 10, (k) cluster 11, and (l) cluster 12.}
\label{fig:predictions-real-data}
\end{figure*}

We also compared the best configurations for LSTM and GRU models for each cluster using the Kruskal-Wallis test. For all clusters, the RMSE distributions of the LSTM and GRU models are statistically different. The LSTM models obtained better average RMSE results except for Cluster 12, where the GRU slightly outperformed the LSTM (See Table \ref{tab:comparison-lstm-gru-performance}). Cluster 1 presented the worst average RMSE for both LSTM (0.084) and GRU (0.095) while Cluster 12 had the best average RMSE, 0.068 and 0.067 for LSTM and GRU, respectively. Figure \ref{fig:predictions-real-data} compares the actual Internet activity (green line) and the predictions of LSTM model (blue line) and GRU model (orange line). 

One can see that both models learned the pattern of Internet activity data, capturing the data seasonality. For Cluster 1-9, we can see that the models' predictions are slightly lower than the ground truth data in some periods, with the GRU prediction values lower than the LSTM predictions, consistent with the GRU models' higher average RMSE results. For Clusters 10-12, the predictions of both models are closer to the ground truth data in comparison to the other clusters, with the LSTM model closer to the ground truth data in more periods than the GRU models. It is important to notice that Cluster 10 is situated at the outskirt of the metropolitan area of Milan while Clusters 11 and 12 are closer to the center of the city.

The periods where the predictions of the DL models (for all clusters) were more distant from the ground truth data are between in Christmas period (24-26 December). During this period, the predictions of both models were much lower than the ground truth data. This can be explained easily by the seasonal, although predictable, traffic at that time. This could be addressed by augmenting the overall DL scheme with historical statistics that summarize prior knowledge of predictable long-term trends as per \cite{zhang2018long}.
 
Based on our statistical tests, the LSTM models outperformed the GRU models in all but one cluster. In that case, Cluster 12, the superior performance of the GRU model was relatively small. As shown in Figure \ref{fig:predictions-real-data}, both RNNs captured the data pattern, however, the values predicted by the LSTM models were closer to the ground truth Internet activity data. This is not entirely surprising. As discussed in Section \ref{subsubsec:bkg-gru}, typically LSTM is expected to outperform its less complex variant, GRU. However, the GRU took less time to train and thus where a model needs to be retrained for multi-step traffic prediction, the time saving may outweigh the differences in accuracy.

\begin{table}[h!]
\caption{Best configurations for LSTM and GRU models for each cluster}
\label{tab:comparison-lstm-gru-performance}
\resizebox{\columnwidth}{!}{
\begin{tabular}{@{}cccc@{}}
\toprule
\multicolumn{2}{c}{\textbf{LSTM}}           & \multicolumn{2}{c}{\textbf{GRU}}           \\ \midrule
\textbf{Configuration model}  & \textbf{Average RMSE} & \textbf{Configuration model} & \textbf{Average RMSE} \\ \midrule
LSTM-1-1L-250U       &0.084       & GRU-1-1L-250U       &0.095       \\
LSTM-2-1L-250U       &0.081        & GRU-2-1L-200U      &0.092       \\
LSTM-3-1L-250U       &0.080        & GRU-3-1L-200U      &0.092       \\
LSTM-4-1L-250U       &0.080        & GRU-4-1L-200U      &0.090       \\
LSTM-5-1L-250U       &0.080        & GRU-5-1L-200U      &0.091       \\
LSTM-6-1L-250U       &0.081        & GRU-6-1L-200U      &0.093       \\
LSTM-7-1L-250U       &0.081        & GRU-7-1L-200U       &0.092       \\
LSTM-8-1L-250U       &0.081        & GRU-8-1L-200U       &0.093       \\
LSTM-9-1L-250U       &0.081        & GRU-9-1L-200U       &0.092       \\
LSTM-10-1L-250U      &0.068        & GRU-10-1L-250U      &0.072       \\
LSTM-11-1L-250U & 0.071            & GRU-11-1L-150U/250U &0.072       \\
LSTM-12-1L-150U/200U & 0.068      & GRU-12-1L-150U/200U & 0.067      \\ \bottomrule
\end{tabular}}
\end{table}

As discussed earlier, Chen et al. \cite{chen2018deep} also used the Telecom Italia data set for Milan and an LSTM to predict base station traffic. While they used a clustering strategy at a group base stations level, it was \emph{post hoc} i.e. they forecast the traffic patterns using the LSTM and \emph{then} cluster complementary base stations to BBUs based on the traffic patterns. As such, the results of Chen et al. \cite{chen2018deep} may not be entirely comparable with this study. In fact, the different clustering approaches may result in different traffic patterns and consequently impact model performance. However, some similarities between both studies exist so we calculated the MAE for our LSTM and GRU models to compared results.

MAE was calculated as presented in Equation \ref{eq:mae}:

\begin{equation}
    \label{eq:mae}
    MAE=\frac{1}{N} \sum_{i=1}^{N}|f_i - y_i|
\end{equation}

\noindent where $N$ is the length of time series, $f_i$ is the prediction, and $y_i$ is the actual value at timestamp $i$.

Table \ref{tab:comparison-lstm-gru-mae} presents the MAE of our LSTM and GRU models for the 12 clusters. All MAE values obtained for LSTM models were lower than the best result presented in \cite{chen2018deep}, which was 0.074. For the GRU models, Cluster 1 obtained the same MAE as \cite{chen2018deep} (0.0739) while the models for the other clusters performed better. Cluster 12 achieved the best MAE. The LSTM presented an improvement of 26.22\% and the GRU model presented an improvement of 27.16\% when compared to the reported results in \cite{chen2018deep}.

\begin{table}[h!]
\caption{MAE of best configurations for LSTM and GRU models for each cluster}
\label{tab:comparison-lstm-gru-mae}
\resizebox{\columnwidth}{!}{
\begin{tabular}{@{}cccc@{}}
\toprule
\multicolumn{2}{c}{\textbf{LSTM}}           & \multicolumn{2}{c}{\textbf{GRU}}           \\ \midrule
\textbf{Configuration model}  & \textbf{Average MAE} & \textbf{Configuration model} & \textbf{Average MAE} \\ \midrule
LSTM-1-1L-250U       &0.0653       & GRU-1-1L-200U/250U  &0.0739       \\
LSTM-2-1L-250U       &0.0639       & GRU-2-1L-200U       &0.0716       \\
LSTM-3-1L-250U       &0.0634       & GRU-3-1L-200U       &0.0720       \\
LSTM-4-1L-250U       &0.0629       & GRU-4-1L-200U       &0.0701       \\
LSTM-5-1L-250U       &0.0632       & GRU-5-1L-200U       &0.0709       \\
LSTM-6-1L-250U       &0.0636       & GRU-6-1L-200U       &0.0724        \\
LSTM-7-1L-250U       &0.0638       & GRU-7-1L-200U       &0.0718       \\
LSTM-8-1L-250U       &0.0639       & GRU-8-1L-200U       &0.0725      \\
LSTM-9-1L-250U       &0.0636       & GRU-9-1L-200U       &0.0716       \\
LSTM-10-1L-250U      &0.0544      & GRU-10-1L-250U       &0.0564       \\
LSTM-11-1L-250U &0.0564       & GRU-11-1L-150U/250U &0.0565       \\
LSTM-12-1L-150U/200U &0.0546       & GRU-12-1L-150U/200U &0.0539       \\ 
\bottomrule
\end{tabular}}
\end{table}

\section{Conclusion}
\label{sec:conclusion}
Global total mobile data traffic is projected to grow 4X to reach 160EB per month in 2025; 5G networks will carry nearly half of the world’s mobile data traffic by that time \cite{ericsson2019}. This massive surge in demand for mobile broadband requires solutions that satisfy QoS and QoE requirements with minimum service delay and within budget constraints. Mobile network traffic prediction will be an essential input in to infrastructure planning as well as dynamic and proactive network resource optimization. Extant approaches may have unacceptable accuracy, training times, turnaround times, lack computational complexity and may not therefore account for characteristics such as bursting, non-linear patterns or other important correlations to meet the QoS and QoE requirements of increasingly demanding end users \cite{narejo2018an,tran2019cellular}. These issues can result in mis-timed resource allocation as well as over- and under-utilization. DL has the potential to address these shortcomings.

In this work, we propose and compare the performance of two RNNs, LSTM and GRU, to predict mobile Internet traffic in a large metropolitan area, Milan. We proposed a novel \emph{a priori} clustering methodology to group cells using K-Means clustering and used the Grid Search method to identify the best configurations for each RNN. We compared RNN performance using RMSE and testing against ground truth data for Milan. Both RNNs were effective in modeling Internet activity and seasonality, both within days and across two months however were sub-optimal in predicting anomalies e.g. Christmas. In this case, this could have been addressed by augmenting the training with historic trend data as per \cite{zhang2018long}. We also find variations by in clusters across the city. While the LSTM outperformed the GRU, the GRU had faster training times which may be relevant for multi-step prediction scenarios. We compared our proposed RNN models against the results in Chen \cite{chen2018deep} using MAE. Notwithstanding the validity issues in such a comparison, results suggest our models present significantly better performance. 

In future work, we plan to compare additional deep learning architectures including ensemble approaches and augmenting the model with longer-term historical trend data. Furthermore, we will extend the data set with more heterogeneous data sources including SMS and voice call log data, amongst others, as well as other areas, e.g. Trentino, available in the Telecom Italia data set.   Improved mobile network prediction can be applied to a wide range of network planning and optimization use cases to optimise utilization, reduce cost and meet QoS. In future works, we will explore the efficacy of these models in a variety of use cases particularly where the faster training times of GRUs may provide advantages over LSTM, such as multi-step prediction and faster optimization time scales.


\bibliographystyle{unsrt}  
\bibliography{references}  

\begin{thebibliography}{10}

\bibitem{gsma20}
GSMA.
\newblock The mobile economy 2020.
\newblock \url{https://bit.ly/359YSAW}, 2020.
\newblock Accessed: April, 2020.

\bibitem{cisco20}
Cisco.
\newblock Cisco annual internet report.
\newblock
  \url{https://www.cisco.com/c/en/us/solutions/executive-perspectives/annual-internet-report/index.html},
  2020.
\newblock Accessed: April, 2020.

\bibitem{aguzzi2013definition}
S~Aguzzi, D~Bradshaw, M~Canning, M~Cansfield, P~Carter, G~Cattaneo,
  S~Gusmeroli, G~Micheletti, D~Rotondi, and R~Stevens.
\newblock Definition of a research and innovation policy leveraging cloud
  computing and iot combination.
\newblock {\em Final Report, European Commission, SMART}, 37:2013, 2013.

\bibitem{tikunov2007traffic}
Denis Tikunov and Toshikazu Nishimura.
\newblock Traffic prediction for mobile network using holt-winter’s
  exponential smoothing.
\newblock In {\em 2007 15th International Conference on Software,
  Telecommunications and Computer Networks}, pages 1--5. IEEE, 2007.

\bibitem{ma2020survey}
Bo~Ma, Weisi Guo, and Jie Zhang.
\newblock A survey of online data-driven proactive 5g network optimisation
  using machine learning.
\newblock {\em IEEE Access}, 8:35606--35637, 2020.

\bibitem{narejo2018an}
Sanam {Narejo} and Eros Gian~Alessandro {Pasero}.
\newblock An application of internet traffic prediction with deep neural
  network.
\newblock 55:139--149, 2018.

\bibitem{bui2017survey}
Nicola Bui, Matteo Cesana, S~Amir Hosseini, Qi~Liao, Ilaria Malanchini, and
  Joerg Widmer.
\newblock A survey of anticipatory mobile networking: Context-based
  classification, prediction methodologies, and optimization techniques.
\newblock {\em IEEE Communications Surveys \& Tutorials}, 19(3):1790--1821,
  2017.

\bibitem{li2017learning}
Rongpeng Li, Zhifeng Zhao, Jianchao Zheng, Chengli Mei, Yueming Cai, and
  Honggang Zhang.
\newblock The learning and prediction of application-level traffic data in
  cellular networks.
\newblock {\em IEEE Transactions on Wireless Communications}, 16(6):3899--3912,
  2017.

\bibitem{tran2019cellular}
Quang~Thanh {Tran}, Li~{Hao}, and Quang~Khai {Trinh}.
\newblock Cellular network traffic prediction using exponential smoothing
  methods.
\newblock {\em Journal of Information and Communication Technology},
  18(1):1--18, 2019.

\bibitem{huang2017study}
Chih-Wei Huang, Chiu-Ti Chiang, and Qiuhui Li.
\newblock A study of deep learning networks on mobile traffic forecasting.
\newblock In {\em 2017 IEEE 28th Annual International Symposium on Personal,
  Indoor, and Mobile Radio Communications (PIMRC)}, pages 1--6. IEEE, 2017.

\bibitem{samulevicius2015most}
Saulius {Samulevicius}, Torben~Bach {Pedersen}, and Troels~Bundgaard
  {Sorensen}.
\newblock Most: Mobile broadband network optimization using planned
  spatio-temporal events.
\newblock In {\em 2015 IEEE 81st Vehicular Technology Conference (VTC Spring)},
  pages 1--5, 2015.

\bibitem{pollakis2016anticipatory}
Emmanuel {Pollakis} and Slawomir {Stanczak}.
\newblock Anticipatory networking for energy savings in 5g systems.
\newblock {\em WSA}, pages 1--7, 2016.

\bibitem{hua2018traffic}
Yuxiu {Hua}, Zhifeng {Zhao}, Zhiming {Liu}, Xianfu {Chen}, Rongpeng {Li}, and
  Honggang {Zhang}.
\newblock Traffic prediction based on random connectivity in deep learning with
  long short-term memory.
\newblock In {\em 2018 IEEE 88th Vehicular Technology Conference (VTC-Fall)},
  page 8690851, 2018.

\bibitem{qiu2018spatio}
Chen {Qiu}, Yanyan {Zhang}, Zhiyong {Feng}, Ping {Zhang}, and Shuguang {Cui}.
\newblock Spatio-temporal wireless traffic prediction with recurrent neural
  network.
\newblock {\em IEEE Wireless Communications Letters}, 7(4):554--557, 2018.

\bibitem{syarif2016svm}
Iwan Syarif, Adam Prugel-Bennett, and Gary Wills.
\newblock Svm parameter optimization using grid search and genetic algorithm to
  improve classification performance.
\newblock {\em Telkomnika}, 14(4):1502, 2016.

\bibitem{chen2018deep}
Longbiao Chen, Dingqi Yang, Daqing Zhang, Cheng Wang, Jonathan Li, et~al.
\newblock Deep mobile traffic forecast and complementary base station
  clustering for c-ran optimization.
\newblock {\em Journal of Network and Computer Applications}, 121:59--69, 2018.

\bibitem{zhang2019deep}
Chaoyun {Zhang}, Paul {Patras}, and Hamed {Haddadi}.
\newblock Deep learning in mobile and wireless networking: A survey.
\newblock {\em IEEE Communications Surveys and Tutorials}, 21(3):2224--2287,
  2019.

\bibitem{kraus2020deep}
Mathias {Kraus}, Stefan {Feuerriegel}, and Asil {Oztekin}.
\newblock Deep learning in business analytics and operations research: Models,
  applications and managerial implications.
\newblock {\em European Journal of Operational Research}, 281(3):628--641,
  2020.

\bibitem{guidotti2019a}
Riccardo {Guidotti}, Anna {Monreale}, Salvatore {Ruggieri}, Franco {Turini},
  Fosca {Giannotti}, and Dino {Pedreschi}.
\newblock A survey of methods for explaining black box models.
\newblock {\em ACM Computing Surveys}, 51(5):1--42, 2019.

\bibitem{Ansari2020}
Mohammed Ansari, Saeed Alsamhi, Yuansong Qiao, Yuhang Ye, and Brian Lee.
\newblock Security of distributed intelligence in edge computing: Threats and
  countermeasures.
\newblock In Patricia~Endo Theo~Lynn, John~Mooney and Brian Lee, editors, {\em
  The Cloud-to-Thing Continuum - Opportunities and Challenges in Cloud, Fog and
  Edge Computing}, chapter~6. Palgrave Macmillan, Cham, Switzerland, 2020.

\bibitem{ordonez2016deep}
Francisco~Javier Ord{\'o}{\~n}ez and Daniel Roggen.
\newblock Deep convolutional and lstm recurrent neural networks for multimodal
  wearable activity recognition.
\newblock {\em Sensors}, 16(1):115, 2016.

\bibitem{hochreiter1998vanishing}
Sepp Hochreiter.
\newblock The vanishing gradient problem during learning recurrent neural nets
  and problem solutions.
\newblock {\em International Journal of Uncertainty, Fuzziness and
  Knowledge-Based Systems}, 6(02):107--116, 1998.

\bibitem{jozefowicz2015empirical}
Rafal Jozefowicz, Wojciech Zaremba, and Ilya Sutskever.
\newblock An empirical exploration of recurrent network architectures.
\newblock In {\em International Conference on Machine Learning}, pages
  2342--2350, 2015.

\bibitem{greff2017lstm}
Klaus Greff, Rupesh~K Srivastava, Jan Koutn{\'\i}k, Bas~R Steunebrink, and
  J{\"u}rgen Schmidhuber.
\newblock Lstm: A search space odyssey.
\newblock {\em IEEE transactions on neural networks and learning systems},
  28(10):2222--2232, 2017.

\bibitem{lstmExample2016block}
Shi Yan.
\newblock Understanding lstm and its diagrams.
\newblock \url{https://bit.ly/2JvRwhr}, 2016.
\newblock Accessed: August, 2018.

\bibitem{lin2016abnormal}
Hsiu-Yu Lin, Yu-Ling Hsueh, and Wen-Nung Lie.
\newblock Abnormal event detection using microsoft kinect in a smart home.
\newblock In {\em Computer Symposium (ICS), 2016 International}, pages
  285--289. IEEE, 2016.

\bibitem{chung2014empirical}
Junyoung Chung, Caglar Gulcehre, KyungHyun Cho, and Yoshua Bengio.
\newblock Empirical evaluation of gated recurrent neural networks on sequence
  modeling.
\newblock {\em arXiv preprint arXiv:1412.3555}, 2014.

\bibitem{chung2015gated}
Junyoung Chung, Caglar Gulcehre, Kyunghyun Cho, and Yoshua Bengio.
\newblock Gated feedback recurrent neural networks.
\newblock In {\em International Conference on Machine Learning}, pages
  2067--2075, 2015.

\bibitem{kaiser2015neural}
Łukasz {Kaiser} and Ilya {Sutskever}.
\newblock Neural gpus learn algorithms.
\newblock {\em arXiv preprint arXiv:1511.08228}, 2015.

\bibitem{arkhipenko2016comparison}
K.~{Arkhipenko}, I.~{Kozlov}, J.~{Trofimovich}, K.~{Skorniakov}, A.~{Gomzin},
  and D.~{Turdakov}.
\newblock Comparison of neural network architectures for sentiment analysis of
  russian tweets.
\newblock {\em Computational Linguistics and Intellectual Technologies.
  International Conference Dialog 2016 Proceedings}, 15, 2016.

\bibitem{arora2016analysis}
Preeti Arora, Shipra Varshney, et~al.
\newblock Analysis of k-means and k-medoids algorithm for big data.
\newblock {\em Procedia Computer Science}, 78:507--512, 2016.

\bibitem{rai2010survey}
Pradeep Rai and Shubha Singh.
\newblock A survey of clustering techniques.
\newblock {\em International Journal of Computer Applications}, 7(12):1--5,
  2010.

\bibitem{zhang2018citywide}
Chuanting Zhang, Haixia Zhang, Dongfeng Yuan, and Minggao Zhang.
\newblock Citywide cellular traffic prediction based on densely connected
  convolutional neural networks.
\newblock {\em IEEE Communications Letters}, 22(8):1656--1659, 2018.

\bibitem{wang2017spatiotemporal}
Jing {Wang}, Jian {Tang}, Zhiyuan {Xu}, Yanzhi {Wang}, Guoliang {Xue}, Xing
  {Zhang}, and Dejun {Yang}.
\newblock Spatiotemporal modeling and prediction in cellular networks: A big
  data enabled deep learning approach.
\newblock In {\em IEEE INFOCOM 2017 - IEEE Conference on Computer
  Communications}, pages 1--9, 2017.

\bibitem{alawe2018improving}
Imad {Alawe}, Adlen {Ksentini}, Yassine {Hadjadj-Aoul}, and Philippe {Bertin}.
\newblock Improving traffic forecasting for 5g core network scalability: A
  machine learning approach.
\newblock {\em IEEE Network}, 32(6):42--49, 2018.

\bibitem{zhang2018long}
Chaoyun {Zhang} and Paul {Patras}.
\newblock Long-term mobile traffic forecasting using deep spatio-temporal
  neural networks.
\newblock In {\em Proceedings of the Eighteenth ACM International Symposium on
  Mobile Ad Hoc Networking and Computing}, pages 231--240, 2018.

\bibitem{feng2018deeptp}
Jie {Feng}, Xinlei {Chen}, Rundong {Gao}, Ming {Zeng}, and Yong {Li}.
\newblock Deeptp: An end-to-end neural network for mobile cellular traffic
  prediction.
\newblock {\em IEEE Network}, 32(6):108--115, 2018.

\bibitem{eurostat20}
Eurostat.
\newblock Population on 1 january by age groups and sex - cities and greater
  cities.
\newblock \url{https://bit.ly/2YqqJeW}, 2020.
\newblock Accessed: April, 2020.

\bibitem{barlacchi2015multi}
Gianni Barlacchi, Marco De~Nadai, Roberto Larcher, Antonio Casella, Cristiana
  Chitic, Giovanni Torrisi, Fabrizio Antonelli, Alessandro Vespignani, Alex
  Pentland, and Bruno Lepri.
\newblock A multi-source dataset of urban life in the city of milan and the
  province of trentino.
\newblock {\em Scientific data}, 2:150055, 2015.

\bibitem{hussain2019mobile}
Bilal Hussain, Qinghe Du, Sihai Zhang, Ali Imran, and Muhammad~Ali Imran.
\newblock Mobile edge computing-based data-driven deep learning framework for
  anomaly detection.
\newblock {\em IEEE Access}, 7:137656--137667, 2019.

\bibitem{mededovic2019node}
Emil Mededovic, Vaggelis~G Douros, and Petri M{\"a}h{\"o}nen.
\newblock Node centrality metrics for hotspots analysis in telecom big data.
\newblock In {\em IEEE INFOCOM 2019-IEEE Conference on Computer Communications
  Workshops (INFOCOM WKSHPS)}, pages 417--422. IEEE, 2019.

\bibitem{zhang2020citywide}
Dehai Zhang, Linan Liu, Cheng Xie, Bing Yang, and Qing Liu.
\newblock Citywide cellular traffic prediction based on a hybrid spatiotemporal
  network.
\newblock {\em Algorithms}, 13(1):20, 2020.

\bibitem{amin2020hotspots}
Farhan Amin and Gyu~Sang Choi.
\newblock Hotspots analysis using cyber-physical-social system for a smart
  city.
\newblock {\em IEEE Access}, 8:122197--122209, 2020.

\bibitem{yuan2019research}
Chunhui Yuan and Haitao Yang.
\newblock Research on k-value selection method of k-means clustering algorithm.
\newblock {\em J—Multidisciplinary Scientific Journal}, 2(2):226--235, 2019.

\bibitem{syakur2018integration}
MA~Syakur, BK~Khotimah, EMS Rochman, and BD~Satoto.
\newblock Integration k-means clustering method and elbow method for
  identification of the best customer profile cluster.
\newblock In {\em IOP Conference Series: Materials Science and Engineering},
  volume 336, page 012017. IOP Publishing, 2018.

\bibitem{wang2018analysis}
Weijie Wang and Yanmin Lu.
\newblock Analysis of the mean absolute error (mae) and the root mean square
  error (rmse) in assessing rounding model.
\newblock In {\em IOP Conference Series: Materials Science and Engineering},
  volume 324, page 012049. IOP Publishing, 2018.

\bibitem{kingma2014adam}
Diederik~P Kingma and Jimmy Ba.
\newblock Adam: A method for stochastic optimization.
\newblock {\em arXiv preprint arXiv:1412.6980}, 2014.

\bibitem{elliott2011sas}
Alan~C Elliott and Linda~S Hynan.
\newblock A sas{\textregistered} macro implementation of a multiple comparison
  post hoc test for a kruskal--wallis analysis.
\newblock {\em Computer methods and programs in biomedicine}, 102(1):75--80,
  2011.

\bibitem{ericsson2019}
Ericsson.
\newblock Ericsson mobility report: 5g uptake even faster than expected.
\newblock \url{https://mb.cision.com/Main/15448/2836925/1060118.pdf}, 2019.
\newblock Accessed: April, 2020.

\end{thebibliography}






\end{document}